\input form
%DEFINIZIONI LOCALI
% beginning of gothic
\font\teneufm=eufm10
\font\seveneufm=eufm7
\font\fiveeufm=eufm5
\newfam\eufmfam
\textfont\eufmfam=\teneufm
\scriptfont\eufmfam=\seveneufm
\scriptscriptfont\eufmfam=\fiveeufm

%\def\upsilon{{\eufm z}}
%\def\Pset{\hbox{\eulfnt P}}

% end of gothic

\def\rmI{{\rm I}}
\def\rmII{{\rm II}}
\def\rmIII{{\rm III}}
\def\ee{{e}}
\def\AAA{{\cal A}}
\def\BBB{{\cal B}}
\def\CCC{{\cal C}}
\def\DDD{{\cal D}}
\def\EEE{{\cal E}}
\def\Ebarra{{\bar E}}
\def\FFF{{\cal F}}
\def\GGG{{\cal G}}
\def\HHH{{\cal H}}
\def\III{{\cal I}}

\def\MMM{{\cal M}}

\def\OOO{{\cal O}}

\def\TTT{{\cal T}}
\def\UUU{{\cal U}}

\def\XXX{{\cal X}}
\def\WWW{{\cal W}}

\def\Pset{{\cal P}}
\def\PPset{\widehat{\cal P}}

\def\realpart{\mathop{\rm Re}\nolimits}
\def\imaginary{\mathop{\rm Im}\nolimits}
\def\imunit{{\bf i}}
\def\der{{\rm d}}
\def\Lie{{\cal L}}
\def\telchi{{\TTT}}
\def\frac#1#2{{#1 \over #2}}
\def\parder#1#2{{\partial#1 \over \partial#2}}
\def\scalprod#1#2{#1\cdot#2}
\def\build#1_#2^#3{\mathrel{
\mathop{\kern 0pt#1}\limits_{#2}^{#3}}}

%FINE DEFINIZIONI LOCALI
%BIBLIOGRAFIA
\cita{Ber-Bia-2011}{M. Berti and L. Biasco: {\sl Branching of            
   Cantor manifolds of elliptic tori and applications to PDEs},
   Commun. Math. Phys., {\bf 305}, 741--796 (2011).}
\cita{Bia-Chi-Val-2003}{L. Biasco, L. Chierchia and E. Valdinoci: {\sl
    Elliptic two-dimensional invariant tori for the planetary
    three-body problem}, Arch. Rational Mech. Anal., {\bf 170},
  91--135 (2003).}
\cita{Bia-Chi-Val-2006}{L. Biasco, L. Chierchia and E. Valdinoci: {\sl
    N-dimensional elliptic invariant tori for the planar (N+1)-body
    problem}, SIAM J. Math. Anal., {\bf 37}, n.5,
  1560--1588 (2006).}
\cita{Eliasson-1988}{L.H. Eliasson: {\sl Perturbations of stable invariant
    tori for Hamiltonian systems}, Ann. Scuola Norm. Sup. Pisa,
  Cl. Sci., IV Serie, {\bf 15}, 115--147 (1988).}
\cita{Giorgilli-2003}{A. Giorgilli: {\sl Notes on exponential
    stability of Hamiltonian systems}, in {\sl Dynamical Systems, Part
    I\/}. Pubbl. Cent. Ric. Mat. Ennio De Giorgi, Sc. Norm. Sup.
    Pisa, 87--198 (2003).}
\cita{Gio-Loc-1997}{Giorgilli, A. and Locatelli, U., {\it Kolmogorov
    theorem and classical perturbation theory}, J. of App. Math. and
   Phys. (ZAMP), {\bf 48}, 220--261 (1997).}
\cita{Gio-Loc-1997.1}{A. Giorgilli and U. Locatelli: {\sl On classical
    series expansion for quasi-periodic motions}, MPEJ, {\bf 3}, n.5,
  1--25 (1997).}
\cita{Gio-Mar-2010}{A. Giorgilli and S. Marmi: {\sl Convergence radius
    in the Poincar\'e-Siegel problem}, DCDS Series S, {\bf 3}, 601--621
    (2010).}
\cita{Grobner-60}{Gr\"obner, W.: {\it Die  Lie-Reihen und Ihre 
  Anwendungen}, Springer Verlag, Berlin (1960); Italian transl.: {\it Le
  serie di Lie e le loro applicazioni}, Cremonese, Roma (1973).}
\cita{Jef-Mos-1966}{W.H. Jefferys and J. Moser: {\sl Quasi-periodic
    solutions for the three-body problem}, Astronom. J., {\bf 71},
  568--578 (1966).}
\cita{Kolmogorov-1954}{A.N. Kolmogorov: {\sl Preservation of
    conditionally periodic movements with small change in the Hamilton
    function}, Dokl. Akad. Nauk SSSR, {\bf 98}, 527--530
    (1954). English transl. in: Los Alamos Scientific Laboratory
    translation LA-TR-71-67.}
\cita{Kuksin-1988}{S.B. Kuksin: {\sl The perturbation theory for the
    quasi-periodic solutions of infinite-dimensional Hamiltonian
    systems and its applications to the Korteweg de Vries equation},
    Matem. Sbornik {\bf 136} (1988); English transl. in: Math.  USSR
    Sbornik {\bf 64}, 397--413 (1989).}
\cita{Laskar-1989b}{J. Laskar: {\sl Syst\`emes de variables et
  \'el\'ements}, in D. Benest, C. Froeschl\'e (eds.), {\sl Les
  M\'ethodes modernes de la M\'ecanique C\'eleste (Goutelas 89)},
  63--87 (1989).}
\cita{Lieberman-1971}{B.B. Lieberman: {\sl Existence of quasi-periodic
    solutions to the three-body problem}, Celestial Mechanics, {\bf
  3}, 408--426 (1971).}
\cita{Loc-Gio-2000}{U. Locatelli and A. Giorgilli: {\sl Invariant
    tori in the secular motions of the three--body planetary systems},
  Celest. Mech. Dyn. Astr., {\bf 78}, 47--74 (2000).}
\cita{Melnikov-1965}{V.K. Melnikov: {\sl On some cases of                
   conservation of almost periodic motions with a small change of the
   Hamiltonian function}, Dokl. Akad. Nauk SSSR, {\bf 165}, 1245--1248
   (1965).}
\cita{Poschel-1989}{J. P{\"o}schel: {\sl On elliptic lower dimensional
    tori in Hamiltonian systems}, Math. Z., {\bf 202}, 559--608
  (1989).}
\cita{Poschel-1996}{J. P{\"o}schel: {\sl A KAM-theorem for some
    nonlinear PDEs}, Ann. Sc. Norm. Sup. Pisa Cl. Sci., {\bf 23},
  119--148 (1996).}
\cita{San-Loc-Gio-2011}{M. Sansottera, U. Locatelli and A. Giorgilli:
  {\sl A semi-analytic algorithm for constructing lower dimensional
  elliptic tori in planetary systems}, Celest. Mech. Dyn. Astr., {\bf
  111}, 337--361 (2011).}

%FINE BIBLIOGRAFIA

\def\testos{\hfill A. Giorgilli, U. Locatelli, M. Sansottera}
\def\testod{On the convergence of an algorithm constructing the normal form for
       lower $\ldots$\hfill}

\title{On the convergence of an algorithm\riga
       constructing the normal form for\riga
       lower dimensional elliptic tori in\riga
       planetary systems}

\author{\it ANTONIO GIORGILLI\hfil\break
Dipartimento di Matematica, Universit\`a degli Studi di Milano,\hfil\break 
via Saldini 50, 20133\ ---\ Milano, Italy.}

\author{\it UGO LOCATELLI\hfil\break
Dipartimento di Matematica, 
Universit\`a degli Studi di Roma ``Tor Vergata'',\hfil\break 
Via della Ricerca Scientifica 1, 00133\ ---\ Roma, Italy.}

\author{\it MARCO SANSOTTERA\hfil\break 
D\'epartement de Math\'ematique \& NAXYS, Universit\'e de Namur ,\hfil\break
Rempart de la Vierge 8, B-5000\ ---\ Namur, Belgium.}

\abstract{We give a constructive proof of the existence of lower
dimensional elliptic tori in nearly integrable Hamiltonian systems.
In particular we adapt the classical Kolmogorov's normalization
algorithm to the case of planetary systems, for which elliptic tori
may be used as replacements of elliptic keplerian orbits in
Lagrange-Laplace theory.  With this paper we support with rigorous
convergence estimates the semi-analytical work in our previous
article~\dbiref{San-Loc-Gio-2011}, where an explicit calculation of an
invariant torus for a planar model of the Sun-Jupiter-Saturn-Uranus
system has been made.  With respect to previous works on the same
subject we exploit the characteristic of Lie series giving a precise
control of all terms generated by our algorithm.  This allows us to
slightly relax the non-resonance conditions on the frequencies.}

\section{sec:intro}{Introduction}
Lower dimensional invariant tori in weakly perturbed Hamiltonian
system, as opposed to periodic orbits and full dimensional tori, are a
natural class of objects which deserve to be investigated.  We are
particularly interested in the existence of such invariant tori for
the planetary problem.  We refer in particular to the theory of
secular motions originally developed by Lagrange and Laplace.  In that
theory one investigates the motion near to keplerian orbits, with
inclinations and eccentricities equal to zero.  We point out that in
the neighborhood of these orbits there are elliptic invariant tori
that may be used in order to generalize and hopefully improve the
previous theory.

As a general fact, the existence of elliptic lower dimensional
invariant tori was first stated by Melnikov\bibref{Melnikov-1965}
and, more than 20 years later, proved independently by
Eliasson\bibref{Eliasson-1988} and Kuksin\bibref{Kuksin-1988}.
Their results have also been extended to Hamiltonian PDEs (see,
again,~\dbiref{Kuksin-1988}, \dbiref{Poschel-1989},~\dbiref{Poschel-1996}
and, for more recent results, \dbiref{Ber-Bia-2011} and references
therein).

Concerning the planetary problem, the existence of lower dimensional
tori for the three-body system has been proven by Jefferys and
Moser\bibref{Jef-Mos-1966} and Lieberman\bibref{Lieberman-1971}.
However the configurations considered in those papers are quite far
from the ones in the original Lagrange-Laplace theory.
In~\dbiref{Jef-Mos-1966} the case of large mutual inclinations is
investigated, so that the lower dimensional tori are partially
hyperbolic.  In~\dbiref{Lieberman-1971} the ratio of the semi-major
axes of the planets is assumed to be small enough and the perihelia
are locked in phase.  An application of P{\"o}schel's method to the
Solar System has been produced by Biasco, Chierchia and Valdinoci in
two different cases, namely the spatial three-body planetary problem
and a planar system with a central star and $n$ planets
(see~\dbiref{Bia-Chi-Val-2003} and~\dbiref{Bia-Chi-Val-2006},
respectively).  However, as often happens in the framework of KAM
theory, their approach is deep from a theoretical point of view,
but seems not to be suitable for explicit calculations, even if one is
just interested in finding the locations of an elliptic invariant
torus.

A constructive algorithm for elliptic tori has been produced by the
authors in a previous paper~\dbiref{San-Loc-Gio-2011}, where the
formal procedure is described in detail.  Furthermore an explicit
calculation for a planar model of the Sun-Jupiter-Saturn-Uranus system
has been performed using algebraic manipulation on a computer, and the
resulting orbits on an elliptic torus have been found to be in
agreement with those obtained by direct numerical integration.  The
construction of an elliptic torus is performed by giving the
Hamiltonian a suitable normal form using an infinite sequence of near
the identity canonical transformations defined by Lie series.
However, a rigorous proof of the convergence of the whole procedure
does not follow from previous ones, and is still lacking.  In the
present paper we publish such a proof.

From a technical point the main difference with respect to the
previous works is that our algorithm is fully constructive, and
specially designed to deal with the Hamiltonian of a planetary system.
Moreover we transport in the KAM framework a non-resonance condition
that has been introduced in \dbiref{Gio-Mar-2010} for the
Poincar\'e-Siegel problem.  Actually that condition turns out to be
equivalent to the Bruno's one, but it produces better analytical
estimates.  The perturbation procedure is followed by a geometric
argument concerning the estimate of the measure of a suitable set of
non-resonant frequencies, which is basically an adaptation of the
approach described in~\dbiref{Poschel-1989}.

We state here our main result.  Let us consider a
$2(n_1+n_2)$-dimensional phase space endowed with $n_1$ pairs of
action-angle coordinates $(p,q)\in\OOO_1\times\toro^{n_1}$ and other
$n_2$ pairwise conjugated canonical variables
$(x,y)\in\OOO_2\subseteq\reali^{2n_2}$, where both
$\OOO_1\subseteq\reali^{n_1}$ and $\OOO_2$ are open sets including the
origin.  We also introduce an open set $\UUU\subset\reali^{n_1}$ and
the frequency vector $\omega^{(0)}\in\UUU$ which plays the role of a
parameter.

\theorem{teor:enunciato-nonquantitativo}{Consider the following
family of real Hamiltonians, parameterized by the $n_1$-dimensional
frequency vector $\omega^{(0)}$,
$$
\eqalign{
\HHH^{(0)}(p,q,x,y;\omega^{(0)}) =
\omega^{(0)}\cdot p+
\epsilon\sum_{j=1}^{n_2}\left[
\frac{\Omega_j^{(0)}(\omega^{(0)})}{2}\left(x_j^2+y_j^2\right)\right]+
\epsilon\FFF_0(q;\omega^{(0)} &)
\cr
+\epsilon\FFF_1(q,x,y;\omega^{(0)})+\epsilon\FFF_2(p,q,x,y;\omega^{(0)})+
\FFF_{\rm h.o.t.}(p,q,x,y;\omega^{(0)} &)\ ,
\cr
}
\formula{def:H0-espansione-rozza}
$$
with $\epsilon$ playing the usual role of small parameter.
Let us assume that

\item{(a)} the frequencies $\Omega_j^{(0)}:\UUU\to\reali$ are analytic functions of $\omega^{(0)}\in\UUU\,$; similarly $\FFF_0\,$, $\FFF_1\,$, $\FFF_2$ and $\FFF_{\rm h.o.t.}$ are analytic functions of $(p,q,x,y;\omega^{(0)})\in\OOO_1\times\toro^{n_1}\times\OOO_2\times\UUU\,$;

\item{(b)} one has $\Omega_i^{(0)}(\omega^{(0)})\neq\Omega_j^{(0)}(\omega^{(0)})$
  for $\omega^{(0)}\in\UUU$ and $1\le i<j\le n_2\,$;

\item{(c)} the function $\FFF_0$ is independent of $p$ and $(x,y)\,$;
  $\FFF_1$ is independent of $p$ and linear in $(x,y)\,$; $\FFF_2$ is
  either linear in $p$ or quadratic in $(x,y)\,$; $\FFF_{\rm h.o.t.}$
  is of higher order in $p$ and $(x,y)$, i.e., $\FFF_{\rm
  h.o.t.}\,=o\big(\|p\|+\|(x,y)\|^2\big)$;

\item{(d)} $\FFF_{\rm h.o.t.}$ splits as
  $\FFF_{\rm h.o.t.}(p,q,x,y;\omega^{(0)})=\FFF_{\rm
  int}(p;\omega^{(0)})+ \epsilon\FFF_{\rm
  n.i.}(p,q,x,y;\omega^{(0)})\,$; moreover, the average of $\FFF_2$
  over the angles is equal to zero;

\item{(e)} $\HHH^{(0)}$ is invariant with respect to the $\theta$-family of canonical diffeomorphisms
  $$
  \eqalign{
  \big(p_1,&\ldots,p_{n_1},q_1,\ldots,q_{n_1},x_1,\ldots,x_{n_2},
  y_1,\ldots,y_{n_2}\big)
  \mapsto
  \cr
  &\big(p_1,\ldots,p_{n_1},q_1+\vartheta,\ldots,q_{n_1}+\vartheta,
  x_1\cos\vartheta+y_1\sin\vartheta,\ldots,
  x_{n_2}\cos\vartheta+y_{n_2}\sin\vartheta,
  \cr
  &\phantom{\big(}\quad y_1\cos\vartheta-x_1\sin\vartheta,\ldots,
  y_{n_2}\cos\vartheta-x_{n_2}\sin\vartheta\big)\ 
  \cr
  }
  $$
  where $\vartheta\in\toro\,$;
\item{(f)} one has
  $$
    \eqalign{
    \sup_{(p,q,x,y;\omega^{(0)})\in\OOO_1\times\toro^{n_1}\times\OOO_2\times\UUU}
    \big|\FFF_j(p,q,x,y;\omega^{(0)})\big| &\leq E
    \qquad\hbox{for } j=0,\,1,\,2\ ,
    \cr
    \sup_{(p,q,x,y;\omega^{(0)})\in\OOO_1\times\toro^{n_1}\times\OOO_2\times\UUU}
    \big|\FFF_{\rm h.o.t.}(p,q,x,y;\omega^{(0)})\big| &\leq E\ ,
    \cr
    }
  $$
  for some $E>0$.  

\noindent
Then, there is a positive $\epsilon^{\star}$ such that for
$0\le \epsilon<\epsilon^{\star}$ the following statement holds true:
there exists a non-resonant set $\UUU^{(\infty)}\subset\UUU$ of positive
Lebesgue measure, such that for each $\omega^{(0)}\in\UUU^{(\infty)}$
there exists an analytic canonical transformation
$(p,q,x,y)=\psi_{\omega^{(0)}}^{(\infty)}(P,Q,X,Y)$ leading the
Hamiltonian in the normal form
$$
\vcenter{\openup1\jot\halign{
 \hbox {\hfil $\displaystyle {#}$}
&\hbox {\hfil $\displaystyle {#}$\hfil}
&\hbox {$\displaystyle {#}$\hfil}\cr
\HHH^{(\infty)}(P,Q,X,Y;\omega^{(0)}) &=
&\omega^{(\infty)}\cdot P+
\epsilon\sum_{j=1}^{n_2}\frac{\Omega_j^{(\infty)}\left(X_j^2+Y_j^2\right)}{2}+
o\big(\|P\|+\|(X,Y)\|^2\big)\ ,
\cr
}}
$$
where $\omega^{(\infty)}=\omega^{(\infty)}(\omega^{(0)})$ and
$\Omega^{(\infty)}=\Omega^{(\infty)}(\omega^{(0)})\,$.}\endclaim

\noindent
The existence of an elliptic invariant torus is a straightforward
consequence of the normal form above.  Indeed the torus $P=0$, $X=Y=0$
is clearly invariant and elliptic, and carries a quasi-periodic motion
with frequencies $\omega^{(\infty)}$.  This is the natural
adaptation of the original scheme of Kolmogorov.

Actually, we shall prove two more quantitative statements, i.e.,
propositions~\proref{analitico} and~\proref{prop:geometrico}.  Let us
highlight some points related to the theorem above.

If $\FFF_0 = \FFF_1 = \FFF_2 = 0$ then the
Hamiltonian~\frmref{def:H0-espansione-rozza} is clearly in normal
form.  Thus, we collect in $\FFF_0$, $\FFF_1$ and $\FFF_2$ all
coupling terms that should be removed in order to prove that an
elliptic torus, possibly with different frequencies, persists under the
perturbation.  The request that $\FFF_2$ has zero average is not
restrictive.

A relevant characteristic of the
Hamiltonians~\frmref{def:H0-espansione-rozza} is that setting
$\epsilon=0$ one is left with the so-called ``keplerian
approximation'', depending on the actions $p$ only.  This is very
typical for the application of KAM theory to a planetary system.
Furthermore, the frequencies of the oscillations transversal to an
elliptic torus are $\Oscr(\epsilon)$ with respect to those related to
the quasi-periodic motion on the invariant lower dimensional
torus. Thus a natural distinction arises between the fast variables
$(p,q)$ and the slow {\it secular variables}, according to the common
language in Celestial Mechanics.  Hypothesis~(e) turns out to be
natural if one is interested in planetary systems. Actually, it means
that $\HHH^{(0)}$ is invariant with respect to rotations around the
direction of the total angular momentum. This symmetry is equivalent
to assume the so-called ``d'Alembert rules'' in Celestial Mechanics.
Let us emphasize that hypothesis~(e) allows us to deal with both the
{\it spatial} planetary problem, where
$\Omega_{n_2}^{(0)}(\omega^{(0)})=0\,$, and the {\it planar} one. The
previous mentioned papers by Biasco, Chierchia and Valdinoci are
restricted either to the spatial three-body problem (after the
reduction of the angular momentum) or to the planar case, because they
also require that $\Omega_{j}^{(0)}(\omega^{(0)})\neq 0$ for
$j=1,\,\ldots,\,n_2\,$.  Actually our proof can be modified by
replacing~(e) with the weaker assumption that $\HHH^{(0)}$ is
invariant with respect to the diffeomorphism
$(p,q_1,\ldots,q_{n_1},x,y)\mapsto(p,q_1+\pi,\ldots,q_{n_1}+\pi,-x,-y)\,$.
This could be interesting in order to state a theorem that applies to
a planetary system after the reduction of the angular momentum.  In
our opinion, a further modification of the proof could also cover the
case of restricted problems with three or more bodies, where the
symmetries are lost.

Let us highlight that our statement does not assume the usual
non-degeneracy hypothesis on the $p$-dependence of the Hamiltonian,
which is required in the classical framework of KAM-like
theorems.  Actually, we just use the non-degeneracy property of the
keplerian approximation so as to {\it preliminarly} give the
Hamiltonian the form~\frmref{def:H0-espansione-rozza}.

The paper is organized as follows. In section~\secref{sec:ellformale}
we recall the formal algorithm, also introducing the peculiar
property that must be satisfied by the expansions of our Hamiltonians,
so as to fit with the aforementioned d'Alembert rules.  A full
justification of the algorithm may be found
in~\dbiref{San-Loc-Gio-2011}. In section~\secref{sec:stime}, we
introduce some unavoidable analytical settings. In
section~\secref{sec:quantitative} we produce the quantitative
estimates that are necessary in order to prove the convergence.  Most
of these estimates are now standard matter, so we skip some
calculations that may be easily reconstructed by the reader.  Instead, a
special emphasis is given to the control of small divisors,
since this is new in KAM theory (see subsection~\sbsref{sbs:indexes}).
In section~\secref{measure} we prove that our procedure applies to a
set of initial frequencies of large measure.  In this part we simplify
the discussion by using the Diophantine condition.  However this is
legitimate, because our non-resonance conditions imply the Diophantine
ones, so that the conclusion concerning the measure remains
valid. Finally, in section~\secref{sec:fine-dim-teor-quantitativo}, we
give the proof of the theorem~\thrref{teor:enunciato-nonquantitativo}.
An appendix containing the technical calculations is included at the
end.

\section{sec:ellformale}{Formal algorithm}
This section is devoted to the algorithm leading in normal form a
Hamiltonian~\frmref{def:H0-espansione-rozza} of the family
$\HHH^{(0)}$ that is parameterized with respect to the frequency vector
$\omega^{(0)}$.  Our constructive procedure is described here from a
purely formal point of view, by including all the (sometimes tedious)
formul{\ae} that will be necessary to analyze the convergence of such
an algorithm in the next sections. Let us recall that our procedure
can be effectively implemented with the aid of manipulations made by
computer algebra (see~\dbiref{San-Loc-Gio-2011}).  As a very minor
difference with respect to~\dbiref{San-Loc-Gio-2011}, here we have
found convenient to use the complex variables $z=(x+\imunit
y)/\sqrt{2}$ in order to deal with the transversal directions with
respect to the elliptic tori (as is usually done). One can immediately
verify that the transformation $(p,q,z,\imunit\bar z)\mapsto(p,q,x,y)$
is canonical.

\subsection{sbs:settings-formal-algorithm}{Initial settings and strategy
of the formal algorithm} 
For some fixed positive integer $K$ we introduce the classes of
functions $\PPset_{\hat m,\hat l,sK}$ with integers $\hat m,\,\hat
l,\,s\ge 0\,$, which can be written as
$$
g(p,q,z,\imunit\bar z) =
\sum_{{\scriptstyle{m\in\naturali^{n_1}}}\atop{\scriptstyle{|m|=\hat m}}}
\,\sum_{{\scriptstyle{(l,\bar l)\in\naturali^{2n_2}}}\atop{\scriptstyle{|l|+|\bar l|=\hat l}}}
\,\sum_{{\scriptstyle{{k\in\interi^{n_1}}}\atop{\scriptstyle{|k|\le sK}}}} c_{m,l,\bar l,k}
\,p^{m} z^{l}(\imunit\bar z)^{\bar l} \exp(\imunit k\cdot q)\ ,
\formula{frm:esempio-g-in-PPset}
$$
with coefficients $c_{m,l,\bar l,k}\in\complessi$.  Here we denote by
$|\cdot|$ the $l_1$-norm and we adopt use the multi-index notation,
i.e., $p^{m}=\prod_{j=1}^{n_1} p_j^{m_j}$.

Furthermore we say that $g\in\Pset_{\ell,sK}$ in case
$$
g\in\bigcup_{{\scriptstyle {\hat m\ge 0\,,\,\hat l\ge 0}}\atop
{\scriptstyle {2\hat m+\hat l=\ell}}}\PPset_{\hat m,\hat l,sK}
$$
and the Taylor-Fourier expansion of $g$ satisfies the following property:
setting 
$$
\CCC_{\MMM}(l,\bar l)=\sum_{j=1}^{n_2}(l_j-{\bar l}_j)\ ,
\qquad
\CCC_{\III}(k)=\sum_{j=1}^{n_1} k_j\ ,
\formula{def:characteristics}
$$
one has $c_{m,l,\bar l,k}=0$ for $\CCC_{\MMM}(l,\bar
l)\neq\CCC_{\III}(k)$.  We also set
$\Pset_{-2,sK}=\Pset_{-1,sK}=\{0\}$ for $s\ge 0$, $K>0\,$.

\noindent
The latter definition is equivalent to hypothesis~(e) of
theorem~\thrref{teor:enunciato-nonquantitativo} and includes also the
d'Alembert rules, mentioned in the introduction.  In Celestial
Mechanics these rules are usually stated by saying that all terms
appearing in the expansions have the ``monomial characteristic''
$\CCC_{\MMM}(l,\bar l)$ equal to the ``characteristic of the
inequality'' $\CCC_{\III}(k)\,$.

Finally we shall denote by $\langle g\rangle_{\vartheta}
= \int_{\toro^n}\diff\vartheta_1\ldots\diff\vartheta_n \,g/(2\pi)^n$
the average of a function $g$ with respect to the angles
$\vartheta\,$.  We shall also omit the dependence of the function from
the variables, unless it has some special meaning.

The relevant algebraic property is stated by the following
\lemma{lem:proprieta-classi-funz}{Let $g\in\Pset_{\ell,sK}$
and $g^{\prime}\in\Pset_{\ell^{\prime},s^{\prime}K}$ for some
$\ell,\,s,\,\ell^{\prime},\,s^{\prime}\ge 0$ and $K>0\,$. Then
$\{g,g^{\prime}\}\in\Pset_{\ell+\ell^{\prime}-2,(s+s^{\prime})K}\,$.
}\endclaim
\noindent
The proof of the lemma above is left to the reader being a
straightforward consequence of the definition of the Poisson bracket.

We start with the Hamiltonian in the form
$$
\eqalign{
H^{(0)} &= \scalprod{\omega^{(0)}}{p} +
\epsilon\sum_{j=1}^{n_2}\Omega^{(0)}_{j}z_{j}{\bar z}_{j}
+\sum_{\ell>2}\sum_{s\geq 0} \epsilon^{s} f_{\ell}^{(0,s)}
\cr
&\quad+\sum_{s\geq 1} \epsilon^{s} f_{0}^{(0,s)}+
\sum_{s\geq 1} \epsilon^{s} f_{1}^{(0,s)}+
\sum_{s\geq 1} \epsilon^{s} f_{2}^{(0,s)}\ ,
\cr
}
\formula{frm:H(0)}
$$
where $f_{\ell}^{(0,s)}\in\Pset_{\ell,sK}$; moreover, as in
hypothesis~(c) of theorem~\thrref{teor:enunciato-nonquantitativo},
$f_{\ell}^{(0,0)}=f_{\ell}^{(0,0)}(p;\omega^{(0)})$ for $\ell\ge 3$
and $\langle f_{2}^{(0,1)}\rangle_{q}=0\,$.  The
Hamiltonian~\frmref{def:H0-espansione-rozza} may be written in the
form~\frmref{frm:H(0)} (see section~\secref{sec:stime}).

In the spirit of the original Kolmogorov's proof scheme, starting from
$H^{(0)}$, we construct an infinite sequence of Hamiltonians
$\left\{H^{(r)}\right\}_{r\geq 0}$ with the request that each
$H^{(r)}$ is in normal form up to order $r\,$, in a sense to be
defined below.  To this aim, we perform a sequence of normalization
steps, transforming the Hamiltonian $H^{(r-1)}$ into $H^{(r)}$ via a
near the identity canonical transformation.  The canonical
transformation at order $r$ is generated by a composition of four Lie
series/transforms of the form
$$
\telchi_{\epsilon^{r-1}\Dscr_2^{(r)}}\circ
\exp\left( \epsilon^r\Lie_{\chi_2^{(r)}} \right)\circ
\exp\left( \epsilon^r\Lie_{\chi_1^{(r)}} \right)\circ
\exp\left( \epsilon^r\Lie_{\chi_0^{(r)}} \right)
\formula{serie-Lie-passo-r}
$$
where $\Lie_g\cdot=\{\cdot,g\}$ is the Lie derivative operator and
$\chi_0^{(r)}(q)\in\Pset_{0,rK}\,$, $\chi_1^{(r)}(q,z,\imunit\bar
z)\in\Pset_{1,rK}\,$, $\chi_2^{(r)}(p,q,z,\imunit\bar
z)\in\Pset_{2,rK}$.  The Lie transform operator
$\telchi_{\epsilon^{r-1}\Dscr_2^{(r)}}$, with a sequence of functions
$\big\{\epsilon^{j(r-1)}\Dscr_2^{(r;j)}(z,\imunit\bar
z)\in\Pset_{2,0}\big\}_{j\ge 1}\,$ actually induces a canonical linear
change of the coordinates $(z,\imunit\bar z)$ (see
subsection~\sssref{sss:diagonal}).  Self-consistent introductions to
the Lie series formalism can be found, e.g., in~\dbiref{Grobner-60}
and in~\dbiref{Giorgilli-2003}.

The generating functions $\chi_0^{(r)}$, $\chi_1^{(r)}$, $\chi_2^{(r)}$
and $\Dscr_2^{(r)}$ are determined by some homological equations.  The
main difference with respect to the original Kolmogorov's algorithm is
that the frequencies $\omega^{(r)}$ and $\Omega^{(r)}$ may change
at every normalization step by a small quantity (see
formul{\ae}~\frmref{chgfreq.veloci} and~\frmref{chgfreq.lente}).

In order to control the small divisors, we need to introduce at each
$r$-th step two non-resonance conditions up to a finite order
$rK$, namely
$$
\,\min_{\scriptstyle{k\in\interi^{n_1}\,,\,0<|k|\leq
rK}\atop\scriptstyle{l\in\interi^{n_2}\,,\,0\leq
|l|\leq2}}\big| \scalprod{k}{\omega^{(r-1)}(\omega^{(0)})}
+ \epsilon\,\scalprod{l}{\Omega^{(r-1)}(\omega^{(0)})}\big| \ge a_r\ ,
\formula{nonres}
$$
and
$$
\,\min_{1\le i<j\le n_2}\big|\Omega_i^{(r-1)}(\omega^{(0)})
-\Omega_j^{(r-1)}(\omega^{(0)})\big|\ge b_r\ ,
\formula{nonres.a}
$$
where $\{a_r\}_{r\ge 1}$ and $\{b_r\}_{r\ge 1}$ are two monotonically
decreasing sequences such that $a_r\to 0$ and $b_r\to b_{\infty}>0$
when $r\to+\infty\,$.  For $|l|= 1,2$ condition~\frmref{nonres} is
usually referred to as the first and second Melnikov condition,
respectively, while for $|l|=0$ it is the usual condition of strong
non-resonance.

In the rest of this section we provide a detailed description of the
generic $r$-th normalization step.  Let us write the Hamiltonian
$H^{(r-1)}$, which is in normal form up to order $r-1$, as
$$
\eqalign{
H^{(r-1)} &= \scalprod{\omega^{(r-1)}}{p} +
\epsilon\sum_{j=1}^{n_2}\Omega^{(r-1)}_{j}z_{j}{\bar z}_{j}
+\sum_{\ell>2}\sum_{s\geq 0} \epsilon^{s} f_{\ell}^{(r-1,s)}
\cr
&\quad+\sum_{s\geq r} \epsilon^{s} f_{0}^{(r-1,s)}+
\sum_{s\geq r} \epsilon^{s} f_{1}^{(r-1,s)}+
\sum_{s\geq r} \epsilon^{s} f_{2}^{(r-1,s)}\ ,
\cr
}
\formula{frm:H(r-1)}
$$
where $f_{\ell}^{(r-1,s)}\in\Pset_{\ell,sK}$; moreover, we have
$f_{\ell}^{(r-1,0)}=f_{\ell}^{(r-1,0)}(p;\omega^{(0)})$
for $\ell\ge 3\,$ and, just for $r=1\,$, $\langle
f_{2}^{(0,1)}\rangle_{q}=0\,$. 
In the expansion above, the functions $f_{\ell}^{(r-1,s)}$ may depend
analytically on $\epsilon$, the relevant information being that they
carry a common factor $\epsilon^s$.  Such an expansion is clearly not
unique, but this is harmless.

\subsection{sbs:firststep}{First stage of the normalization step}
Our aim is to remove the term
$f_{0}^{(r-1,r)}$.  Thus, we determine the generating function
$\chi^{(r)}_{0}$ by solving the homological equation
$$
\Lie_{\chi^{(r)}_{0}} \left(\scalprod{\omega^{(r-1)}}{p}\right)
+ f_{0}^{(r-1,r)} - \langle f_{0}^{(r-1,r)}\rangle_{q} = 0\ .
\formula{frm:chi0r}
$$
This equation admits a solution in view of the non-resonance
condition~\frmref{nonres} with $|l|=0\,$. Indeed, considering the
Taylor-Fourier expansion
$$
f_{0}^{(r-1,r)}(q)=
\sum_{{\scriptstyle{|k|\le rK}}\atop{\scriptstyle{\CCC_{\III}(k)=0}}}
c_{0,0,0,k}^{(r-1)}\exp(\imunit k\cdot q)
\formula{frm:espansione-f_0^(r-1,r)}
$$
one can easily check that
$$
\chi^{(r)}_{0}(q)=\sum_{{\scriptstyle{0<|k|\le rK}}\atop{\scriptstyle{\CCC_{\III}(k)=0}}}
\frac{c_{0,0,0,k}^{(r-1)}}{\imunit\scalprod{k}{\omega^{(r-1)}}}
\exp(\imunit k\cdot q)\ .
\formula{frm:espansione-chi0r}
$$

The new Hamiltonian is determined as the Lie series with generating
function $\epsilon^r \chi^{(r)}_{0}$, namely
$$
\eqalign{
H^{(\rmI;r)} &= \exp\left(\epsilon^{r}\Lie_{\chi^{(r)}_{0}}\right)H^{(r-1)}\cr
&= \scalprod{\omega^{(r-1)}}{p} +
\epsilon\sum_{j=1}^{n_2}\Omega^{(r-1)}_{j}z_{j}{\bar z}_{j}
+\sum_{\ell>2}\sum_{s\geq 0} \epsilon^{s} f_{\ell}^{(\rmI;r,s)}
\cr
&\quad+\sum_{s\geq r} \epsilon^{s} f_{0}^{(\rmI;r,s)}+
\sum_{s\geq r} \epsilon^{s} f_{1}^{(\rmI;r,s)}+
\sum_{s\geq r} \epsilon^{s} f_{2}^{(\rmI;r,s)}\ .
\cr
}
\formula{frm:H(I;r)-espansione}
$$
The functions $f_{l}^{(\rmI;r,s)}$ are recursively defined as
$$
\vcenter{\openup1\jot 
\halign{
$\displaystyle\hfil#$&$\displaystyle{}#\hfil$&$\displaystyle#\hfil$\cr
f_{0}^{(\rmI;r,r)} &= 0\ ,&\cr
\noalign{\smallskip}
f_{0}^{(\rmI;r,r+m)} &= f_{0}^{(r-1,r+m)} &\quad\hbox{for } 0<m<r\,,\cr
\noalign{\smallskip}
f_{\ell}^{(\rmI;r,s)} &=\sum_{j=0}^{\lfloor s/r\rfloor}
\frac{1}{j!} \Lie^{j}_{\chi^{(r)}_{0}}f^{(r-1,s-jr)}_{\ell+2j}&
\quad\hbox{for }{\vtop{\hbox{${\ell =0\,,\ s\ge 2r\ \hbox{ or } \ \ell=1,\,2\,,\ s\ge r}$}
\vskip-2pt\hbox{\hskip-5pt$\hbox{or }\ \ell\ge 3\,,\ s\ge 0\ ,$}}}\cr
}}
\formula{frm:fI}
$$
with $f_{\ell}^{(\rmI;r,s)}\in\Pset_{\ell,sK}$.  The constant term
$c_{0,0,0,0}^{(r-1)}= \langle f_{0}^{(r-1,r)}\rangle_{q}$ has been
omitted.

\subsection{sbs:second}{Second stage of the normalization step}
We now remove $f_{1}^{(\rmI;r,r)}$ appearing in the
expansion~\frmref{frm:H(I;r)-espansione}.  To this aim, we
determine a new generating function $\chi^{(r)}_{1}$ by solving the
homological equation
$$
\Lie_{\chi^{(r)}_{1}}\biggl(\scalprod{\omega^{(r-1)}}{p}+
\epsilon\sum_{j=1}^{n_2}\Omega^{(r-1)}_{j}z_{j}{\bar z}_{j}\biggr)
+f_{1}^{(\rmI;r,r)}= 0\ .
\formula{frm:chi1r}
$$
Again, we consider the Taylor-Fourier expansion
$$
f_{1}^{(\rmI;r,r)}(q,z,\imunit\bar z)=\sum_{|l|+|\bar l|=1}\,
\sum_{{\scriptstyle{0<|k|\le rK}}\atop{\scriptstyle{\CCC_{\III}(k)=\CCC_{\MMM}(l,\bar l)}}}\,
c_{0,l,\bar l,k}^{(\rmI;r)}z^{l}(\imunit\bar z)^{\bar l}\exp(\imunit\scalprod{k}{q})\ .
\formula{frm:espansione-f_1^(I;r,r)}\ ,
$$
where the average $\langle f_{1}^{(\rmI;r,r)}\rangle_{q}$ is zero in
view of the d'Alembert rules.  In more detail, since each term
appearing in expansion~\frmref{frm:espansione-f_1^(I;r,r)} must
satisfy $\CCC_{\III}(k)=\CCC_{\MMM}(l,\bar l)=\pm 1$ (recall their
definitions in~\frmref{def:characteristics}), then all the
coefficients $c_{0,l,\bar l,k}^{(\rmI;r)}$ with even $|k|$ must be
zero.  The solution of the homological equation~\frmref{frm:chi1r} is
given by
$$
\chi^{(r)}_{1}(q,z,\imunit\bar z)=\sum_{|l|+|\bar l|=1}\,
\sum_{{\scriptstyle{0<|k|\le rK}}\atop{\scriptstyle{\CCC_{\III}(k)=\CCC_{\MMM}(l,\bar l)}}}
\,\frac{c_{0,l,{\bar l},k}^{(\rmI;r)}\,z^{l}(\imunit\bar z)^{\bar l}
\exp(\imunit k\cdot q)}{\imunit\big[\scalprod{k}{\omega^{(r-1)}}+
\epsilon\scalprod{(l-{\bar l})}{\Omega^{(r-1)}}\big]}\ ,
\formula{frm:espansione-chi1r}
$$
where the divisors cannot vanish in view of condition~\frmref{nonres}
with $|l|=1\,$.

We emphasize that the condition $\Omega_{j}^{(0)}\neq 0$ for
$j=1,\,\ldots,\,n_2\,$, which has been required in previous theorems,
is not requested here thanks to the d'Alembert rules.

Again, the new Hamiltonian is calculated as
$$
H^{(\rmII;r)} = \exp\left(\epsilon^{r}\Lie_{\chi^{(r)}_{1}}\right) H^{(\rmI;r)}\ ,
\formula{def:H(II;r)}
$$
and may be given the form~\frmref{frm:H(I;r)-espansione}, replacing
the upper index ${\rm I}$ by ${\rm II}\,$, with
$$
\vcenter{\openup1\jot 
\halign{
$\displaystyle\hfil#$&$\displaystyle{}#\hfil$&$\displaystyle#\hfil$\cr
f_{\ell}^{(\rmII;r,r)} &= 0 &\quad\hbox{for } \ell=0,1\,,\cr
\noalign{\smallskip}
f_{\ell}^{(\rmII;r,r+m)} &= f_{\ell}^{(\rmI;r,r+m)}
&\quad\hbox{for } \ell=0,1\,,\ 0<m<r\,,\cr
\noalign{\smallskip}
f_{0}^{(\rmII;r,2r)} &= f_{0}^{(\rmI;r,2r)} +
\frac{1}{2} \Lie_{\chi_{1}^{(r)}}f_{1}^{(\rmI;r,r)}\ ,&\cr
\noalign{\smallskip}
f_{0}^{(\rmII;r,2r+m)} &= f_{0}^{(\rmI;r,2r+m)} + \Lie_{\chi_{1}^{(r)}}f_{1}^{(\rmI;r,r+m)}
&\quad\hbox{for } 0<m<r\,,\cr
\noalign{\smallskip}
f_{\ell}^{(\rmII;r,s)} &= \sum_{j=0}^{\lfloor s/r\rfloor} \frac{1}{j!} \Lie^{j}_{\chi^{(r)}_{1}}f^{(\rmI;r,s-jr)}_{\ell+j}&
\quad\hbox{for }{\vtop{\hbox{${\ell =0\,,\ s\ge 3r\ \hbox{ or } \ \ell=1\,,\ s\ge 2r}$}
\vskip-2pt\hbox{\hskip-5pt$\hbox{or }\ \ell=2\,,\ s\ge r\ \hbox{ or }\ \ell\ge 3\,,\ s\ge 0\,.$}}}\cr
}}
\formula{frm:fII}
$$

\subsection{sbs:thirdstep}{Third stage of the normalization step}
In order to remove $f_{2}^{(\rmII;r,r)}$ we proceed in two steps.
First we remove the $q$-dependent part and then, in the
next section, the average one.

We determine the generating function $\chi^{(r)}_2$ by solving the
homological equation
$$
\Lie_{\chi^{(r)}_{2}} \biggl(\scalprod{\omega^{(r-1)}}{p}+
\epsilon\sum_{j=1}^{n_2}\Omega^{(r-1)}_{j}z_{j}{\bar z}_{j}\biggr)
+ f_{2}^{(\rmII;r,r)} - \langle f_{2}^{(\rmII;r,r)}\rangle_{q}= 0\ .
\formula{frm:chi2r}
$$
Again, considering the Taylor-Fourier expansion
$$
\eqalign{
f_{2}^{(\rmII;r,r)}(p,q,z,\imunit\bar z)
&=\sum_{|m|=1}\,
\sum_{{\scriptstyle{|k|\le rK}}\atop{\scriptstyle{\CCC_{\III}(k)=0}}}\,
c_{m,0,0,k}^{(\rmII;r)}p^{m}\exp(\imunit\scalprod{k}{q})
\cr
&\phantom{=}+\sum_{|l|+|\bar l|=2}\,
\sum_{{\scriptstyle{|k|\le rK}}\atop{\scriptstyle{\CCC_{\III}(k)=\CCC_{\MMM}(l,\bar l)}}}\,
c_{0,l,\bar l,k}^{(\rmII;r)}z^{l}(\imunit\bar z)^{\bar l}
\exp(\imunit\scalprod{k}{q})\ ,
\cr
}
\formula{frm:espansione-f_2^(II;r,r)}
$$
one can easily check that
$$
\eqalign{
\chi^{(r)}_{2}(p,q,z,\imunit\bar z)
&=\sum_{|m|=1}\,\sum_{{\scriptstyle{0<|k|\le rK}}\atop{\scriptstyle{\CCC_{\III}(k)=0}}}
\,\frac{c_{m,0,0,k}^{(\rmII;r)}p^{m}\exp(\imunit k\cdot q)}
{\imunit\scalprod{k}{\omega^{(r-1)}}}
\cr
&\phantom{=}+\sum_{|l|+|\bar l|=2}\,
\sum_{{\scriptstyle{0<|k|\le rK}}\atop{\scriptstyle{\CCC_{\III}(k)=\CCC_{\MMM}(l,\bar l)}}}
\,\frac{c_{0,l,{\bar l},k}^{(\rmII;r)}z^{l}(\imunit\bar z)^{\bar l}
\exp(\imunit k\cdot q)}{\imunit\big[\scalprod{k}{\omega^{(r-1)}}+
\epsilon\scalprod{(l-{\bar l})}{\Omega^{(r-1)}}\big]}
\ ,
\cr
}
\formula{frm:espansione-chi2r}
$$
where the divisors cannot vanish in view of condition~\frmref{nonres}
with $|l|=0,\,2\,$.

The transformed Hamiltonian is calculated as
$$
H^{(\rmIII;r)} = \exp\left(\epsilon^{r}\Lie_{\chi_{2}^{(r)}}\right)H^{(\rmII;r)}
\formula{def:H(III;r)}
$$
and may be given the form~\frmref{frm:H(I;r)-espansione}, replacing
the upper index ${\rm I}$ by ${\rm III}\,$, with 
$$
\vcenter{\openup1\jot 
\halign{
$\displaystyle\hfil#$&$\displaystyle{}#\hfil$&$\displaystyle#\hfil$\cr
f_{\ell}^{(\rmIII;r,r)} &= 0 &\quad\hbox{for } \ell=0,1\,,\cr
\noalign{\smallskip}
f_{\ell}^{(\rmIII;r,s)} &=
\sum_{j=0}^{\lfloor s/r\rfloor-1}\frac{1}{j!} \Lie^{j}_{\chi^{(r)}_{2}}
f^{(\rmII;r,s-jr)}_{\ell}&\quad\hbox{for } \ell=0,1
\,,\ s>r\,,\cr
\noalign{\smallskip}
f_{2}^{(\rmIII;r,r)} &= \langle f_{2}^{(\rmII;r,r)}\rangle_{q}\ ,
\cr
\noalign{\smallskip}
f_{2}^{(\rmIII;r,jr)} &= \frac{j-1}{j!}\Lie^{j-1}_{\chi^{(r)}_{2}}
f^{(\rmII;r,r)}_{2}+\sum_{i=0}^{j-2}\frac{1}{i!} \Lie^{i}_{\chi^{(r)}_{2}}
f^{(\rmII;r,(j-i)r)}_{2}&\quad\hbox{for } j\geq2\,,\cr
\noalign{\smallskip}
f_{2}^{(\rmIII;r,jr+m)} &= \sum_{i=0}^{j-1}\frac{1}{i!}
\Lie^{i}_{\chi^{(r)}_{2}} f^{(\rmII;r,(j-i)r+m)}_{2}&\quad\hbox{for } j\geq 1
\,,\ 0<m<r\,,\cr
\noalign{\smallskip}
f_{\ell}^{(\rmIII;r,s)} &= \sum_{j=0}^{\lfloor s/r\rfloor}
\frac{1}{j!} \Lie^{j}_{\chi^{(r)}_{2}} f^{(\rmII;r,s-jr)}_{\ell}&
\quad\hbox{for } \ell\geq 3\,,\ s\geq 0\,.
\cr
}}
\formula{frm:fIII}
$$

\subsubsection{sss:diagonal}{Diagonalization of the quadratic normal 
form part in $(z,\imunit\bar z)$}
The last term to deal with is
$$
\langle f_{2}^{(\rmII;r,r)}\rangle_{q}=
\sum_{|m|=1}c_{m,0,0,0}^{(\rmII;r)}p^{m}+
\sum_{|l|=|\bar l|=1}c_{0,l,\bar l,0}^{(\rmII;r)}z^{l}(\imunit\bar z)^{\bar l}\ .
$$
We should remove the non diagonal terms in the latter expansion,
namely the terms with $l\neq\bar l$.  This could be done with standard
algebraic methods.  However, in order to construct a coherent scheme of
estimates, we found it convenient to proceed with a Lie transform
operator $\telchi_{\XXX}\,$, with a sequence of generating functions
$\{\XXX_j\}_{j\ge 1}\,$.  We recall that the Lie transform of a
generic function $g$ is defined as
$$
\telchi_{\XXX}g=\sum_{j=0}^{\infty}\EEE_{j}\,g
\qquad
{\rm with}
\qquad
\EEE_{0}\,g=g\ ,
\qquad
\EEE_{j}\,g=\sum_{i=1}^{j}\frac{i}{j}\Lie_{\XXX_i}\EEE_{j-i}\,g
\ .
\formula{frm:def-telchi}
$$

We look for a sequence of functions
$\Dscr_2^{(r)}= \big\{\epsilon^{j(r-1)}\Dscr_2^{(r;j)}\big\}_{j\ge 1}$
such that
$$
\telchi_{\Dscr_2^{(r)}}\left(\epsilon Z_0^{(r)}\right)+
\telchi_{\Dscr_2^{(r)}}\left(\epsilon^r g_{1}^{(r)}\right)=
\sum_{j=0}^{+\infty}\epsilon^{j(r-1)+1} Z_j^{(r)}\ ,
\formula{frm:diagonalizza-con-telchi}
$$
where
$$
Z_0^{(r)}=\sum_{j=1}^{n_2}\Omega^{(r-1)}_{j}z_{j}{\bar z}_{j}\ ,
\qquad
g_{1}^{(r)}(z,\imunit\bar z)=
f_{2}^{(\rmIII;r,r)}(0,z,\imunit\bar z)
\formula{def:Z0-g1}
$$
and $Z_j^{(r)}$, for $j\ge 1\,$, is the polynomial
$$
Z_j^{(r)}=\sum_{|l|=1}c_{0,l,l,0}^{(r;j)}z^{l}(\imunit\bar z)^{l}\ ,
\formula{frm:struttura-Z_j^(r)}
$$
with coefficients $c_{0,l,l,0}^{(r;j)}\,$ to be found.  The functions
$\Dscr_2^{(r;j)}$ are recursively defined so that
$$
\EEE_{j}^{(r)}Z_0^{(r)}+\EEE_{j-1}^{(r)}\,g_1^{(r)}=Z_j^{(r)}\ .
\formula{frm:diagonalizza-con-telchi-di-passo-j}
$$
The latter equation is solved by rearranging it as
$$
\Lie_{\Dscr_2^{(r;j)}}Z_0^{(r)}+\Psi_{j}^{(r)}=Z_j^{(r)}\ ,
\formula{frm:D2r-di-passo-j}
$$
with
$$
\Psi_{j}^{(r)}=
\sum_{i=1}^{j-1}\left[\frac{i}{j}\Lie_{\Dscr_2^{(r;i)}}
\left(Z_{j-i}^{(r)}-\EEE_{j-i-1}^{(r)}\,g_1^{(r)}\right)\right]
+\EEE_{j-1}^{(r)}\,g_1^{(r)}\ .
\formula{frm:def-Psi_j^(r)}
$$
Let us give some more details.  Proceeding by induction, assume that
$\Psi_{j}^{(r)}\in\Pset_{2,0}$ and depends only on $(z,\imunit\bar z)$; this is true for
$j=1$.  Thus we can write
$$
\Psi_{j}^{(r)}=\sum_{|l|=|\bar l|=1}\,c_{0,l,{\bar l},0}^{(r;j)}
z^{l}(\imunit\bar z)^{\bar l}
\formula{frm:espansione-Psi_j^(r)}
$$
and the homological equation~\frmref{frm:D2r-di-passo-j} is solved with
$$
\Dscr_2^{(r;j)}=\sum_{{\scriptstyle{|l|=|\bar l|=1}}\atop{\scriptstyle{l\neq\bar l}}}
\,\frac{c_{0,l,{\bar l},0}^{(r;j)}}{\imunit\scalprod{(l-{\bar l})}{\Omega^{(r-1)}}}
z^{l}(\imunit\bar z)^{\bar l}\ ,
\formula{frm:espansione-D2r}
$$
where $Z_j^{(r)}$ has the form~\frmref{frm:struttura-Z_j^(r)}.  The
divisors cannot vanish in view of condition~\frmref{nonres.a}; the
cases $|l|=2$ or $|\bar l|=2$ cannot occur in view of d'Alembert rules
for terms independent of the angles $q$
(see~\frmref{def:characteristics}).  Again, let us emphasize that 
condition $\Omega_{i}^{(r)}\neq 0$ for $i=1,\,\ldots,\,n_2\,$ is not
needed here.  By lemma~\lemref{lem:proprieta-classi-funz} all
functions so constructed depend just on $(z,\imunit\bar z)$ and belong
to $\Pset_{2,0}\,$. This ensures the formal
consistency of the whole procedure.

Applying the Lie transform operator $\telchi_{\Dscr_2^{(r)}}$ we
finally get the Hamiltonian in normal form up to order $r$ as
$$
H^{(r)} = \telchi_{\epsilon^{r-1}\Dscr_2^{(r)}}H^{(\rmIII;r)}\ .
\formula{def:H(r)}
$$
The transformed Hamiltonian may be given the
form~\frmref{frm:H(I;r)-espansione}, replacing $r-1$ with $r$, namely
$$
\eqalign{
H^{(r)} &= \scalprod{\omega^{(r)}}{p} +
\epsilon\sum_{j=1}^{n_2}\Omega^{(r)}_{j}z_{j}{\bar z}_{j}
+\sum_{\ell>2}\sum_{s\geq 0} \epsilon^{s} f_{\ell}^{(r,s)}
\cr
&\quad+\sum_{s\geq r+1} \epsilon^{s} f_{0}^{(r,s)}+
\sum_{s\geq r+1} \epsilon^{s} f_{1}^{(r,s)}+
\sum_{s\geq r+1} \epsilon^{s} f_{2}^{(r,s)}\ ,
\cr
}
\formula{frm:H(r)-espansione}
$$
possibly with a change of the frequencies $\omega^{(r)}$ and
$\Omega^{(r)}$, that we briefly discuss in the next section.

\subsubsection{sss:frequencies}{Change of frequencies and transformed Hamiltonian}
The key remark here is that the function $f_{2}^{(\rmIII;r,r)}$ still
contains a part that is $\Oscr(\epsilon^r)$ and belongs to
$\Pset_{2,0}\,$, i.e., $\sum_{|m|=1}c_{m,0,0,0}^{(\rmII;r)} p^{m}+
\sum_{i\ge 1}\epsilon^{i(r-1)}\sum_{|l|=1}c_{0,l,l,0}^{(r;i)}
z^{l}(\imunit\bar z)^{l}\,$.  This kind of terms cannot be eliminated,
then they must be added to the normal form part, producing small
corrections of the frequencies so that
$$
\omega_{j}^{(r)}=
\omega_{j}^{(r-1)}+\epsilon^{r}\parder{f_{2}^{(\rmIII;r,r)}}{p_j}
\qquad
\hbox{for } j=1,\,\ldots,\,n_1
\formula{chgfreq.veloci}
$$
and
$$
\Omega_{j}^{(r)}=\Omega_{j}^{(r-1)}+\sum_{i=1}^{+\infty}\left[\epsilon^{i(r-1)}
\frac{\partial^2 Z_i^{(r)}}{\partial z_j\partial (\imunit{\bar z}_j)}\right]
\qquad
\hbox{for } j=1,\,\ldots,\,n_2\ .
\formula{chgfreq.lente}
$$

Recalling that in view of lemma~\lemref{lem:proprieta-classi-funz}
each class $\Pset_{\ell,sK}$ is invariant under the
action of the operator
$\EEE_{j}^{(r)}=\sum_{i=1}^{j}\frac{i}{j}\Lie_{\Dscr_2^{(r;i)}}\EEE_{j-i}^{(r)}$
 with $j\ge 1\,$, we get the explicit expressions
$$
\vcenter{\openup1\jot 
\halign{
$\displaystyle\hfil#$&$\displaystyle{}#\hfil$&$\displaystyle#\hfil$\cr
f_{\ell}^{(r,r)} &= 0 &\quad\hbox{for } \ell=0,1,2\,,\cr
\noalign{\smallskip}
f_{\ell}^{(r,0)} &= f_{\ell}^{(\rmIII;r,0)}
&\quad\hbox{for } \ell\ge 3\,,\cr
\noalign{\smallskip}
f_{\ell}^{(r,s)} &= 
\sum_{j\geq0} \epsilon^{j(r-1)}\EEE_{j}^{(r)} f_{\ell}^{(\rmIII;r,s)}&
\quad\hbox{for }{\vtop{\hbox{${0\le\ell\le 2\,,\ s>r}$}
\vskip-2pt\hbox{\hskip-5pt$\hbox{or }\ \ell\ge 3\,,\ s\ge 1\ ,$}}}\cr
\cr
}}
\formula{frm:fr}
$$
Remark that the second equation means that $f_{\ell}^{(\rmIII;r,0)}$
remains unchanged under the whole normalization step, since one has
$f_{\ell}^{(\rmIII;r,0)}=f_{\ell}^{(\rmII;r,0)}=f_{\ell}^{(\rmI;r,0)}=
f_{\ell}^{(r-1,0)}=f_{\ell}^{(r-1,0)}(p;\omega^{(0)})$ for $\ell\ge
3\,$, recall equation~\frmref{frm:H(r-1)} and
formul{\ae}~\frmref{frm:fI}, \frmref{frm:fII} and \frmref{frm:fIII}.

Let us stress that the first normalization step does not
change the frequencies, namely
$$
\omega^{(1)}=\omega^{(0)}\ ,
\qquad
\Omega^{(1)}=\Omega^{(0)}\ .
\formula{chgfreq-passo1}
$$
This will play a main role in the quantitative scheme.  Moreover it
also remove a natural doubt about the convergence of the Lie series
appearing in the definition~\frmref{def:H(r)} for $r=1\,$, since it
seems that the generating function does not get small when
$\epsilon\to 0\,$. Indeed, the Lie transform operator
$\telchi_{\Dscr_2^{(1)}}$ turns out to be equal to the identity in
view of the assumption $\langle
f_{2}^{(0,1)}\rangle_{q}=0\,$. Actually we get the chain of
inequalities
$$
\eqalign{
f_{2}^{(\rmIII;1,1)} &= \langle f_{2}^{(\rmII;1,1)}\rangle_{q}=
\langle f^{(\rmI;1,1)}_{2}\rangle_{q}+
\langle \Lie_{\chi^{(1)}_{1}}f^{(\rmI;1,0)}_{3}\rangle_{q}=
\langle f^{(\rmI;1,1)}_{2}\rangle_{q}=
\cr
&=\langle f^{(0,1)}_{2}\rangle_{q}+
\langle \Lie_{\chi^{(1)}_{0}}f^{(0,0)}_{4}\rangle_{q}=
\langle f^{(0,1)}_{2}\rangle_{q}=0\ .
\cr
}
\formula{eq:f_2^(III;1,1)=0}
$$
This in view of the recursive formul{\ae}~\frmref{frm:fI},
\frmref{frm:fII}, \frmref{frm:fIII} and taking into account that
$\Lie_{\chi^{(1)}_{1}}f^{(\rmI;1,0)}_{3}=\Lie_{\chi^{(1)}_{0}}f^{(0,0)}_{4}=0\,$,
as both the generating functions $\chi^{(1)}_{0}$ and
$\chi^{(1)}_{1}$ have zero angular average, while
$f^{(0,0)}_{4}\in\Pset_{4,0}$ and $f^{(\rmI;1,0)}_{3}\in\Pset_{3,0}$
do not depend on the angles.

Finally, let us remark that the Hamiltonian $H^{(r)}$
in~\frmref{frm:H(r)-espansione} has the same form of $H^{(r-1)}$, so
that the induction step can be iterated provided the
conditions~\frmref{nonres} and~\frmref{nonres.a} hold true with $r+1$
in place of $r\,$.

Let us emphasize that our formulation of the algorithm works both for
real and complex Hamiltonians.  This is useful because all the
analytical estimates will be worked out in the framework of complex
functions, as it is usual in perturbation theory.  However, if the
expansion~\frmref{frm:H(0)} contains only real functions, then all
terms of type $\scalprod{\omega^{(r)}}{p}\,$,
$\sum_{j=1}^{n_2}\Omega^{(r)}_{j}z_{j}{\bar z}_{j}$ and
$f_{\ell}^{(r,s)}$ generated by the algorithm are real too, as easily
checked.

\section{sec:stime}{Analytical Settings}
We introduce the complex domains $\DDD_{\rho,R,\sigma,h}
= \GGG_{\rho} \times \toro^{n_1}_{\sigma} \times \BBB_{R} \times \WWW_h\,$,
where $\GGG_{\rho}\subset\complessi^{n_1}$ and
$\BBB_R\subset\complessi^{n_2}\times\complessi^{n_2}$ are open balls
centered at the origin with radii $\rho$ and $R$,
respectively, $\WWW$ is a subset of $\reali^{n_1}$ while the
subscripts $\sigma$ and $h$ denote the usual complex
extensions\footnote{\dag}{Precisely,
$\GGG_{\rho}=\big\{z\in\complessi^{n_1}:\max_{1\le j\le
n_1}|z_j|<\rho\big\}$,
$\toro^{n_1}_{\sigma}=\big\{q\in\complessi^{n_1}:\realpart
q_j\in\toro,\break\ \max_{1\le j\le n_1}|\imaginary q_j|<\sigma\big\}\,$,
$\BBB_R=\{z\in\complessi^{2n_2}: \max_{1\le j\le 2n_2} |z_j|<R\,\}$
and\break $\WWW_h = \big\{z\in\complessi^{n_1}: \exists\ \omega\in\WWW\,,
\ \max_{1\le j\le n_1}|z_j-\omega_j|<h\big\}$.} of real domains
(see~\dbiref{Giorgilli-2003}).

Let us consider a generic analytic function $g:\DDD_{\rho,
R, \sigma, h}\to\complessi$,
$$
g(p,q,z,\imunit\bar z;\omega) =
\sum_{{\scriptstyle{k\in\interi^{n_1}}}} g_{k}(p,z,\imunit\bar z;\omega)
\exp(\imunit k\cdot q)\ ,
\formula{frm:funz}
$$
where $g_{k}:\GGG_{\rho}\times \BBB_{R} \times \WWW_h\to\complessi\,$.
We define the weighted Fourier norm
$$
\|g\|_{\rho,R,\sigma,h}=\sum_{{\scriptstyle{k\in\interi^{n_1}}}}
\big|g_{k}\big|_{\rho,R,h}\exp(|k|\sigma)\ ,
$$
where
$$
\left|g_{k}\right|_{\rho,R,h}=
\sup_{{\scriptstyle{p\in\GGG_{\rho}}}
      \atop{{\scriptstyle{(z,\imunit\bar z)\in \BBB_{R}}}
      \atop{\scriptstyle{\omega\in \WWW_h}}}}
\big|g_{k}(p,z,\imunit\bar z;\omega)\big|\ .
\formula{def:norma}
$$
It is also convenient to introduce the Lipschitz constant
related to the Jacobian of the function
$\Omega^{(0)}:\,\WWW_{h_0}\to\complessi^{n_2}$ as follows
$$
\left| \frac{\partial\Omega^{(0)}}{\partial\omega^{(0)}} \right|_{\infty;\WWW_{h_0}}=
\,\sup_{{\scriptstyle{\omega^{(0)}\in \WWW_{h_0}}}}\,
\sup_{{{\scriptstyle{\beta\neq 0}}}\atop{\scriptstyle{\beta+\omega^{(0)}\in \WWW_{h_0}}}}\,
\frac{\max_{1\le i\le n_2}\left|\Omega^{(0)}_i(\omega^{(0)}+\beta)-
\Omega^{(0)}_i(\omega^{(0)})\right|}{\max_{1\le j\le n_1}\left|\beta_j\right|}\ .
\formula{def:cost-Lipschitz}
$$

Let us remark that the dependence on the parameter $\omega$ plays no
role in any of the following statements, so hereafter, we shorten the
notation by ignoring the index $\omega$. We are now ready to claim the
following
\lemma{lem:analiticita-espansione-Ham}{Let us assume the same
hypotheses of theorem~\thrref{teor:enunciato-nonquantitativo} over the
family of Hamiltonians $\HHH^{(0)}$. Then, there exist positive parameters $\rho\,$, $R\,$, $\sigma\,$, $h_0\,$, $\gamma\,$,
$\tau\,$, $\bar b\,$, $J_0$, $\Ebarra\,$, a compact set
$\WWW\subset\reali^{n_1}$ and a positive integer value $K$ such that
the canonical change of coordinates $(p,q,z,\imunit\bar
z)\mapsto(p,q,x,y)$ transforms $\HHH^{(0)}$ in the Hamiltonian
$H^{(0)}:\DDD_{\rho,R,\sigma}\times\WWW_{h_0}\to\complessi$ described by
the expansion~\frmref{frm:H(0)}, where both 
$\Omega^{(0)}(\omega^{(0)})$ and all the terms of
type $f_{\ell}^{(0,s)}$ are real analytic functions of
$\omega^{(0)}\in\WWW_{h_0}\,$. Moreover, the following properties are
satisfied

\item{(a')} the initial set $\WWW$ of frequencies  is
  {\it non-resonant} up to the finite order $2K\,$, namely every
  $\omega^{(0)}\in\WWW$ satisfy
  $$
  \,\min_{\scriptstyle{k\in\interi^{n_1}\,,\,0<|k|\leq
  2K}\atop\scriptstyle{l\in\interi^{n_2}\,,\,0\leq
  |l|\leq2}}\big| \scalprod{k}{\omega^{(0)}}
  + \epsilon\,\scalprod{l}{\Omega^{(0)}(\omega^{(0)})}\big|\,>
  \frac{2\gamma}{K^\tau}
  $$
  and
  $$
  \,\min_{1\le i<j\le n_2}\big|\Omega_i^{(0)}(\omega^{(0)})
  -\Omega_j^{(0)}(\omega^{(0)})\big|\,> 2 \bar b\ ;
  $$

\item{(b')} the Jacobian of
  $\Omega^{(0)}(\omega^{(0)})$ is uniformly bounded in extended domain
  $\WWW_{h_0}\,$, namely
  $\big| \partial\Omega^{(0)}/\partial\omega^{(0)}\big|_{\infty;\WWW_{h_0}} \leq
  J_0 < \infty\,$;

\item{(c')} $f_{\ell}^{(0,s)}\in\Pset_{\ell,sK}$;

\item{(d')} $f_{\ell}^{(0,0)}=f_{\ell}^{(0,0)}(p;\omega^{(0)})$
  for $\ell\ge 3\,$; moreover, $\langle
  f_{2}^{(0,1)}\rangle_{q}=0\,$;

\item{(e')} the following upper bounds hold true
  $$
  \left\|f_{\ell}^{(0,s)}\right\|_{\rho,R,\sigma}\le \frac{\Ebarra}{2^{\ell}}\ .
  $$
}\endclaim
\noindent
We give a sketch of the proof, which is not too difficult.  The
Hamiltonians should be split in many terms such that their
Taylor-Fourier expansions are finite.  {\sl For any fixed value of
index} $\ell\,$, standard arguments on the Fourier decay of the
coefficients allow us to determine a suitable value of the parameters
$K$ and $\sigma\,$, such that the norms of the functions
$f_{\ell}^{(0,s)}$ are bounded by the same constant for $s\ge 0$
(see, e.g., the proof of lemma~5.2 in~\dbiref{Giorgilli-2003}). Having
fixed $K\,$, some classical Diophantine inequalities allow us to
determine $\gamma>0\,$, $\tau> n_1-1\,$, $h_0>0$ and a compact set
$\WWW\subset\UUU$ such that property~(a') is satisfied (being $\UUU$
the initial set of frequency vectors $\omega^{(0)}$ appearing in the
hypotheses of theorem~\thrref{teor:enunciato-nonquantitativo}). Point
(b') is a straightforward consequence of the analyticity of
$\Omega_j^{(0)}$ on the domain $\UUU$.  Property~(c') follows from
hypothesis~(e) of theorem~\thrref{teor:enunciato-nonquantitativo} as
discussed at the beginning of
subsection~\sbsref{sbs:settings-formal-algorithm}. Moreover, (d') is
an immediate consequence of hypotheses~(c)--(d) of
theorem~\thrref{teor:enunciato-nonquantitativo}. Finally, other basic
arguments on the Taylor expansions of homogeneous polynomials allow
us to choose suitable values of $\rho$ and $R\,$, such that the
inequality at point~(e') of
lemma~\lemref{lem:analiticita-espansione-Ham} is satisfied. For
$\epsilon<1$ we have $\sum_{\ell\,,\,s}\epsilon^s
\|f_{\ell}^{(0,s)}\|_{\rho,R,\sigma}\le 2\Ebarra/(1-\epsilon)\,$;
taking into account that the usual sup-norm is bounded by the
weighted Fourier one (defined in~\frmref{def:norma}), this implies
that the Hamiltonian $H^{(0)}$ is {\sl
analytic} in $\DDD_{\rho,R,\sigma}\times\WWW_{h_0}\,$.

The algorithm described in section~\secref{sec:ellformale} clearly
shows that we need some suitable estimates bounding the Lie
series/transforms.  Such estimates are provided by the following
statements.  In order to shorten the notation, hereafter, we will
denote by $\|\cdot\|_{\alpha}\,$ the norm $\|\cdot\|_{\alpha(\rho,
R,\sigma)}$, being $\alpha$ any real positive number.

\lemma{lem:stima-termini-serie-Lie}{Let $d$ and
$d^{\prime}$ be real numbers such that $d>0\,$, $d^{\prime}\ge 0$ and
$d+d^{\prime}<1\,$; let $\XXX$ and $g$ be two analytic functions on
$\DDD_{(1-d^{\prime})(\rho,R,\sigma)}$
having finite norms $\|\XXX\|_{1-d^{\prime}}$ and
$\|g\|_{1-d^{\prime}}\,$, respectively. Then, for $j\ge 1$, we have
$$
\frac{1}{j!}\left\|\Lie^{j}_{\XXX}g\right\|_{1-d-d^{\prime}}
\le\frac{1}{\ee^{2}}
\left(\frac{2e}{\rho\sigma}+\frac{e^2}{R^{2}}\right)^{j}
\frac{1}{d^{2j}}\|\XXX\|^{j}_{1-d^{\prime}}\|g\|_{1-d^{\prime}}\ .
\formula{frm:stimalie}
$$
}\endclaim
\noindent
Actually, similar estimates to~\frmref{frm:stimalie} are included in
some previous papers of the authors. Nevertheless, a little additional
work is needed in order to adapt them to the present context.  The
proof of lemma above is deferred to
appendix~\sbsref{app:stima-serie-di-Lie}.

\lemma{lem:stima-termini-trasformata-Lie}{Let $d$ and
$d^{\prime}$ be real numbers such that $d>0\,$, $d^{\prime}\ge 0$ and
$d+d^{\prime}<1\,$; let the functions $Z_0\,$, $g\,$ and $g^{\prime}$
satisfy

\item{(i)} $Z_0=\sum_{i=1}^{n_2}\Xi_{i}z_{i}{\bar z}_{i}$
with $\min_{1\le i<j\le n_2}\big|\Xi_i-\Xi_j\big|
\,\ge \Xi^*>0\,$;

\item{(ii)} $g^{\prime}=g^{\prime}(z,\imunit\bar z)$
is such that $g^{\prime}\in\Pset_{2,0}$ and it is so small that
$$
\epsilon^{\star}_{{\rm diag}}=
\left(\frac{2e^2}{d^2}+\frac{2^9}{(1-d^{\prime})^2}\right)
\frac{\|g^{\prime}\|_{1-d^{\prime}}}{\Xi^*R^2}\le \frac{1}{2}\ ;
$$

\item{(iii)} $g$ is an analytic function on
$\DDD_{(1-d^{\prime})(\rho,R,\sigma)}$
with finite norm $\|g\|_{1-d^{\prime}}\,$;

Then, there exist a sequence of generating functions $\{\XXX_j\}_{j\ge 1}$ such
that
$\telchi_{\XXX}\,Z_0+\telchi_{\XXX}\,g^{\prime}=\sum_{j=0}^{+\infty}Z_j\,$,
where $\telchi_{\XXX}$ is the Lie transform operator that has been
introduced in~\frmref{frm:def-telchi} and the Taylor expansion of the
``normal form terms'' $Z_j$ is of the same type as that of $Z_0\,$.
Moreover, for $j\ge 1$, the following inequalities hold true:
$$
\|\EEE_{j}\,g\|_{1-d-d^{\prime}}\le
\big(\epsilon^{\star}_{{\rm diag}}\big)^{j}
\left\|g\right\|_{1-d^{\prime}}\ ,
\qquad
\left\|Z_j\right\|_{1-d^{\prime}}\le
\big(\epsilon^{\star}_{{\rm diag}}\big)^{j-1}
\left\|g^{\prime}\right\|_{1-d^{\prime}}\ .
\formula{frm:stima-finale-E_j-Z_j}
$$
}\endclaim
\noindent
Again, the proof of lemma above is deferred to
appendix~\sbsref{app:stima-serie-di-Lie}.

\section{sec:quantitative}{``Purely analytical'' study of the
normalization algorithm}
In this section, we translate our formal algorithm into a recursive
scheme of estimates on the norms of the functions involved in the
normalization process.

\subsection{sbs:indexes}{Small divisors and selection rule}
It is well known that the accumulation of the small divisors can
prevent the convergence of any perturbative proof scheme, that is
designed so as to ensure the existence of invariant tori for
quasi-integrable systems.  In the present subsection, we introduce the
tools that will allow us to keep control of the accumulation of the
small divisors. Here, we follow rather closely~\dbiref{Gio-Mar-2010};
nevertheless, we think it is convenient to adapt that approach to the
present context in a self-consistent way, because it is one of the
most delicate points of the whole proof. The key of our argument is to
focus our attention on the indexes corresponding to the small
denominators, rather than their actual values.

Let $I=\{j_1,\ldots,j_s\}$ and $I'=\{j'_1,\ldots,j'_s\}$ be two sets
of indexes with the same number $s$ of elements.  Let us introduce the
following relation of partial ordering on those sets: we say that
$I\ilt I'$ in case there is a permutation of the indexes such that the
relation $j_m\le j'_m$ holds true for $m=1,\ldots,s\,$.  If two sets
of indexes contain a different number of elements, first, we pad the
shorter one with zeros and, then, we use the same way to compare them.

\definition{def:insiemi-indici}{For all integers $r\geq0$ and
$s\gt 0\,$, let us introduce the family of indexes sets
$$
\Jscr_{r,s} = \bigl\{I=\{j_1,\ldots,j_{s-1}\}\>:\>
 0\leq j_m\leq\min\{r,\lfloor s/2\rfloor\}\,,\>I\ilt I_s^*\bigr\}\ ,
\formula{nrmlie.20}
$$
where
$$
I_s^* = \left\{
\biggl\lfloor\frac{s}{s}\biggr\rfloor,\,
\biggl\lfloor\frac{s}{s-1}\biggr\rfloor,\,\ldots,
\biggl\lfloor\frac{s}{2}\biggr\rfloor
\right\}\ .
\formula{nrmlie.45}
$$
}\endclaim

\noindent
In agreement with~\dbiref{Gio-Mar-2010}, sometimes we will refer to
the condition $I\ilt I^*_s$ as the {\sl selection rule} $\Smat$.  We
are now ready to claim two technical lemmas, which will be useful in
the following.

\lemma{nrmlie.44}{For the set of indexes $I_s^*=\{j_1,\ldots,j_s\}$ the
following statements hold true:
\item{(i)}the maximal index is  $j_{\rm
max}=\bigl\lfloor\frac{s}{2}\bigr\rfloor\,$;
\item{(ii)}for every $k\in\{1,\ldots,j_{\max}\}$ the index $k$
appears exactly
$\bigl\lfloor\frac{s}{k}\bigr\rfloor -\bigl\lfloor\frac{s}{k+1}\bigr\rfloor$
times;
\item{(iii)}for $0\lt r\le s$ one has
$$
\bigl(\{r\}\cup I^*_r\cup I^*_s\bigr) \ilt I^*_{r+s}\ .
$$
}\endclaim

\lemma{nrmlie.23}{For the sets of indexes $\Jscr_{r,s}$ the following
statements hold true:
\item{(i)}$\Jscr_{r,s}=\Jscr_{\min\{r,\lfloor s/2\rfloor\},s}\,$;
\item{(ii)}$\Jscr_{r-1,s}\subseteq\Jscr_{r,s}\,$;
\item{(iii)}if $I\in\Jscr_{r-1,r}$ and $I'\in\Jscr_{r,s}\,$, then
$\bigr(\{\min\{r,s\}\} \cup I \cup
I'\bigl)\in\Jscr_{r,r+s}\,$.}\endclaim
\noindent The proofs of the two lemmas above are deferred to
appendix~\sbsref{app:indici}.

Now, we think it can be useful to describe the mechanism of
accumulation of the small divisors in a rather informal way. Let us
assume some initial upper bounds on the perturbing terms, like those
at point~(e') of lemma~\lemref{lem:analiticita-espansione-Ham} and
focus on the first normalization step.  Looking at the Taylor-Fourier
expansion~\frmref{frm:espansione-chi0r} of the generating function
$\chi^{(1)}_{0}$, it is well expected that the estimate of its norm is
$\Oscr(\Ebarra/a_1)\,$, in view of the non-resonance
condition~\frmref{nonres}. Of course, the recursive
formula~\frmref{frm:fI} propagate the factors $1/a_1$ to the upper
bounds of the terms $f_{\ell}^{(\rmI;1,s)}$ appearing in the
expansion~\frmref{frm:H(I;r)-espansione} of the Hamiltonian
$H^{(\rmI;1)}$. In particular, the third equation in
formula~\frmref{frm:fI} allows us to remark that the ``most
dangerous'' terms in the estimate of $f_{\ell}^{(\rmI;1,s)}$ are
$\Oscr(\Ebarra/a_1^s)$ (when $\ell =0$, $s\ge 2$ or
$\ell\ge 1$, $s\ge 1$). Here, if there is a sum of some upper
bounds containing different small divisors, ``most dangerous'' means
that we just consider the smallest denominator (in
agreement with the definition of partial ordering $\ilt$ among the
sets of indexes).  Since $f_{1}^{(\rmI;1,1)}=\Oscr(\Ebarra/a_1)\,$, from
equations~\frmref{frm:chi1r}--\frmref{frm:espansione-chi1r} and
condition~\frmref{nonres} it follows that the estimate of the norm of
$\chi^{(1)}_{1}$ is $\Oscr(\Ebarra/a_1^2)\,$. The factors $1/a_1^2$ are
newly propagated to the upper bounds on the terms
$f_{\ell}^{(\rmII;1,s)}$ by the recursive definition
in~\frmref{frm:fII}.  In particular, the estimate for
$f_{0}^{(\rmII;1,2)}$ is $\Oscr(\Ebarra/a_1^3)\,$, while it is
$\Oscr(\Ebarra/a_1^{2s})$ in the general case of $f_{\ell}^{(\rmII;1,s)}$
when $\ell+s\ge 3\,$.  At each normalization step, the generating
function containing more divisors is $\chi^{(r)}_{2}$, since
$f_{2}^{(\rmI;1,1)}=\Oscr(\Ebarra/a_1^2)\,$ from
equations~\frmref{frm:chi2r}--\frmref{frm:espansione-chi2r} and
condition~\frmref{nonres} it follows that an upper bound on
$\chi^{(1)}_{2}$ is $\Oscr(\Ebarra/a_1^3)\,$. By analyzing the
accumulation of the divisors due to the recursive definition
in~\frmref{frm:fIII} as we already did for~\frmref{frm:fI}
and~\frmref{frm:fII}, we can claim that the upper bounds on the terms
$f_{\ell}^{(\rmIII;1,s)}$ are $\Oscr(\Ebarra/a_1^{3s-3+\ell})$ for
$0\le\ell\le 2$, $s\ge 2\,$, while they are more simply
$\Oscr(\Ebarra/a_1^{3s-3})$ when $\ell\ge 3$, $s\ge 0\,$. As we
will explain in the next subsection, the estimates of the Poisson
bracket involving the generating function
$\epsilon^{r-1}\Dscr_2^{(r)}$ do not propagate any small divisors;
therefore, we can roughly say that the estimate just stated about
$f_{\ell}^{(\rmIII;1,s)}$ hold true also for $f_{\ell}^{(1,s)}$.

Let us continue to consider the accumulation of the small divisors in
the estimates up to the generic $r$-th normalization step, in the same
way as we did above for the first step; thus, let us imagine to unfold
all the recursive inequalities necessary to provide an upper bound on
the norm of the terms appearing in the expansion of $H^{(r)}$.  In the
following table we summarize all the relevant information about the
set of indexes appearing in the denominators, due to the accumulation
of the small divisors.
$$
{
\vcenter{\tabskip=0pt\offinterlineskip
\def\tablerule{}%\noalign{\hrule}}
\halign{
%\vrule\hfil$\>${#}$\>$\hfil\vrule
%&$\>$\hfil\strut{#}$\>$\hfil\vrule
%&$\>${#}\hfil$\>$\vrule\ignorespaces
\hfil$\>${#}$\>$\hskip25pt\hfil
&$\>$\hfil\strut{#}$\>$\hskip25pt\hfil
&$\>${#}\hfil$\>$\ignorespaces
\cr
\tablerule
\phantom{\vbox to 14pt{\relax}}
 Function & conditions & set of indexes\cr
\tablerule
\phantom{\vbox to 14pt{\relax}}
$f_{0}^{(r,s)}$
& $0\le r < s$
& $\big(\Jscr_{r,s}\big)^3$  \cr
\tablerule
\phantom{\vbox to 14pt{\relax}}
$f_{1}^{(r,s)}$
& $0\le r < s$
& $\big(\Jscr_{r,s}\big)^3\cup\{r\}$  \cr
\tablerule
\phantom{\vbox to 14pt{\relax}}
$f_{2}^{(r,s)}$
& $0\le r < s$
& $\big(\Jscr_{r,s}\big)^3\cup\big(\{r\}\big)^2$  \cr
\tablerule
\phantom{\vbox to 14pt{\relax}}
$f_{\ell{\scriptscriptstyle\ge} 3}^{(r,s)}$
& $r\ge 0\,,\>s\ge 1$
& $\big(\Jscr_{r,s}\cup\{\min\{r,s\}\}\big)^3$  \cr
\tablerule
\phantom{\vbox to 14pt{\relax}}
$\chi_{0}^{(r)}$
& $r\ge 1$
& $\big(\Jscr_{r-1,r}\big)^3\cup\{r\}$  \cr
\tablerule
\phantom{\vbox to 14pt{\relax}}
$\chi_{1}^{(r)}$
& $r\ge 1$
& $\big(\Jscr_{r-1,r}\big)^3\cup\big(\{r\}\big)^2$  \cr
\tablerule
\phantom{\vbox to 14pt{\relax}}
$\chi_{2}^{(r)}$
& $r\ge 1$
& $\big(\Jscr_{r-1,r}\cup\{r\}\big)^3$  \cr
\tablerule
}}
}
\formula{insiemiindici}
$$

\noindent
With a little abuse of notation, in the table above we introduced a
sort of power of a set so that, for instance,
$(\Jscr_{r,s})^3=\Jscr_{r,s}\cup\Jscr_{r,s}\cup\Jscr_{r,s}\,$.  We
think that the rules described in~\frmref{insiemiindici} are really
the keystone of the whole proof; they are implicitly demonstrated in
appendix~\sbsref{app:lemmone} (dealing with the proof of the main
lemma of the ``purely analytic'' part) through some estimates
involving the sequence $\{T_{r,s}\}_{r,s\ge 0}$ that will be
introduced later.  Let us emphasize that the discussion preceding the
table in~\frmref{insiemiindici} can be translated in a formal proof of
those rules for $r=1\,$; moreover, a similar reasoning can be
extended to the general case.

Let us recall that at each normalization step $r$ we need to estimate
multiple Poisson brackets and this requires to restrict the
analyticity domain. Of course, the series taking into account all
these restrictions must converge; therefore, the divisor $d^{2j}$
appearing in lemma~\lemref{lem:stima-termini-serie-Lie} must shrink to
zero when $r\to\infty\,$. This is another possible source of 
divergence of the whole algorithm. Actually, the present discussion
aims at introducing the tools that allow us to control the accumulation of
the factors due to the restriction of the domains, in a simultaneous
way to the small denominators arising from the solution of the
homological equations. Let us define the sequences $\{d_r\}_{r\geq0}$
and $\{\delta_r\}_{r\ge 1}$ as
$$
d_{0}=0\ ,\qquad
d_{r}=d_{r-1}+4\delta_{r}\ ,\qquad
\delta_{r}=\frac{3}{8\pi^{2} r^{2}}\ .
\formula{frm:dr-and-deltar}
$$
At the $r$-th step of the algorithm, it is convenient to make a
restriction $\delta_r$ of the domain for each of the four canonical
transformations prescribed by the normalization procedure, so that the
Hamiltonian $H^{(r)}$ will be analytic on
$\DDD_{(1-d_r)(\rho,R,\sigma)}\,$. Let us remark that
$\lim_{r\to\infty}d_{r}=1/4$ in view of the definition
in~\frmref{frm:dr-and-deltar}; thus, the previous sequences of domains
converge to a compact set whose interior is not empty.

In order to translate the accumulation rules for small divisors into
quantitative estimates, it is useful to introduce the sequence of
positive real numbers $T_{r,s}$ that are associated to the sets of
indexes $\Jscr_{r,s}$ so that
$$
T_{0,s} = T_{s,0} = 1\ \hbox{ for } s\ge 0\ ,
\qquad
T_{r,s} = \max_{I\in\Jscr_{r,s}}\, \prod_{j\in I\,,\,j\ge 1}
\frac{1}{a_j\delta_j^2}
\ \hbox{ for } r\ge 1\,,\ s\ge 1\ .
\formula{nrmlie.26}
$$
For consistency reasons, of course, the product in the definition
above is put to be equal to $1$ when any factor of type
$(a_j\delta_j^2)^{-1}$ does not occur.

\lemma{nrmlie.27}{The sequence $T_{r,s}$ satisfies the following
properties for all $r,s\ge 1\,$:
\item{(i)}$T_{r-1,s} \le T_{r,s}$ and $T_{r^{\prime},s} = T_{s,s}$
for $r^{\prime}>s\,$;
\item{(ii)}$T_{r-1,r} T_{r,s} / (a_m\delta_m^2) \le
T_{r,r+s}\,$, where $m=\min\{r,s\}\,$.
}\endclaim
\noindent
In the present case too, the proof of lemma above is deferred to
appendix~\sbsref{app:indici}.

\subsection{sec:stime}{Convergence of the algorithm under
non-resonance conditions}
The estimates of the norms of the functions must take into account
many contributions of different type. In the previous subsection, we
provided the tools to control the accumulation of the small divisors,
now we need some suitable definitions to evaluate other contributions.
First, it is convenient to introduce the constant
$$
M=\max\left\{1\,,\,\Ebarra\left(\frac{2e}{\rho\sigma}+\frac{e^2}{R^{2}}\right)
\right\}\ ,
\formula{eq:def-M}
$$
so that many parameters can be considered all together in the
estimates.  Moreover, in order to bound the effects due to the
generating functions $\epsilon^{r-1}\Dscr_{2}^{(r)}$, which remove the
non-diagonal terms depending on $(z,\imunit\bar z)\,$, we define the
sequence $\{\zeta_{r}\}_{r\ge 0}$ as
$$
\zeta_{0} =0\ ,\qquad\zeta_{1} =0\ ,
\qquad\zeta_{r}=\zeta_{r-1}+\frac{2^{-(r+6)}}{1-2^{-(r+6)}}
\quad\hbox{for } r\ge 2\ .
\formula{eq:def-succ-zetagreco}
$$
Since $\Dscr_{2}^{(1)}=0$ (as shown in
subsection~\sssref{sss:diagonal}), the first value of
index $r$ for which $\zeta_{r}\neq 0$ refers to the second step of
normalization. Furthermore, the scheme of estimates also requires to
control the number of summands involved in the recursive
formul{\ae}~\frmref{frm:fI}, \frmref{frm:fII} and~\frmref{frm:fIII}.
For this purpose, we introduce three sequences of integer numbers
$\{\nu_{r,s}\}_{r\ge 0\,,\,s\ge 0}$, $\{\nu_{r,s}^{(\rmI)}\}_{r\ge
1\,,\,s\ge 0}$ and $\{\nu_{r,s}^{(\rmII)}\}_{r\ge 1\,,\,s\ge 0}$
defined as
$$
\vcenter{\openup1\jot 
\halign{
$\displaystyle\hfil#$&$\displaystyle{}#\hfil$&$\displaystyle#\hfil$\cr
\nu_{0,s} &= 1
&\quad\hbox{for } s\ge 0\,,
\cr
\nu_{r,s}^{(\rmI)} &= \sum_{j=0}^{\lfloor s/r \rfloor} \nu_{r-1,r}^{j}\nu_{r-1,s-jr}
&\quad\hbox{for } r\ge 1\,,\ s\ge 0\,,
\cr
\nu_{r,s}^{(\rmII)} &= \sum_{j=0}^{\lfloor s/r \rfloor}
(\nu_{r,r}^{(\rmI)})^{j}\nu_{r,s-jr}^{(\rmI)}
&\quad\hbox{for } r\ge 1\,,\ s\ge 0\,,
\cr
\nu_{r,s} &= \sum_{j=0}^{\lfloor s/r \rfloor}
(\nu_{r,r}^{(\rmII)})^{j}\nu_{r,s-jr}^{(\rmII)}
&\quad\hbox{for } r\ge 1\,,\ s\ge 0\,.
\cr
}}
\formula{frm:seqnu}
$$

We are now ready to claim the main lemma of the ``purely analytic''
part.
\lemma{lem:lemmone}{Let us consider a Hamiltonian
$H^{(0)}$ expanded as in~\frmref{frm:H(0)} and satisfying points~(c')--(e') of
lemma~\lemref{lem:analiticita-espansione-Ham}. Let us assume that on
$H^{(0)}$ we can perform at least the first $r\ge 1$ normalization
steps of the formal algorithm described in
section~\secref{sec:ellformale} and
$$
\epsilon^{i-1}\left[\frac{M^{3i}}{b_i}
\,\frac{T_{i,i}^3}{\left(a_i\delta_i^2\right)^2}\,
\nu_{i,i}\exp(i\zeta_{i-1})\right] \le \frac{1}{2^{i+6}}
\qquad
\hbox{for } 2\le i\le r\ .
\formula{frm:cond-eps-trasf-Diag}
$$

Then, the following upper bounds on the generating functions hold true: 
$$
\vcenter{\openup1\jot 
\halign{
$\displaystyle\hfil#$&$\displaystyle{}#\hfil$&$\displaystyle#\hfil$\cr
\left(\frac{2e}{\rho\sigma}+\frac{e^2}{R^{2}}\right)\frac{1}{\delta_r^2}
\,\|\chi_{0}^{(r)}\|_{1-d_{r-1}}&\leq M^{3r-2}\,
\frac{T_{r-1,r}^3}{a_r\delta_r^2}\,\nu_{r-1,r}\exp(r\zeta_{r-1})\ ,
\cr
\left(\frac{2e}{\rho\sigma}+\frac{e^2}{R^{2}}\right)\frac{1}{\delta_r^2}
\,\|\chi_{1}^{(r)}\|_{1-d_{r-1}-\delta_{r}}&\leq M^{3r-1}\,
\frac{T_{r,r}^3}{\left(a_r\delta_r^2\right)^2}
\,\nu_{r,r}^{(\rmI)}\exp(r\zeta_{r-1})\ ,
\cr
\left(\frac{2e}{\rho\sigma}+\frac{e^2}{R^{2}}\right)\frac{1}{\delta_r^2}
\,\|\chi_{2}^{(r)}\|_{1-d_{r-1}-2\delta_{r}}&\leq M^{3r} 
\left(\frac{T_{r,r}}{a_r\delta_r^2}\right)^3
\nu_{r,r}^{(\rmII)}\exp(r\zeta_{r-1})\ ,
\cr
\epsilon^{j(r-1)}\|\EEE_{j}^{(r)}g\|_{1-d_{r}}&\leq 2^{-j(r+6)}\|g\|_{1-d_{r-1}-3\delta_{r}}
\ ,
\cr
}}
\formula{stime:generatrici-lemmone}
$$
where the latter inequality is satisfied for all $j\ge 0$ and any
analytic function $g$ with finite norm
$\|g\|_{1-d_{r-1}-3\delta_{r}}\,$, being $\{\EEE_{j}^{(r)}\}_{j\ge 0}$
the sequence of operators defined in~\frmref{frm:def-telchi}
and related to the Lie transform operator
$\telchi_{\epsilon^{r-1}\Dscr_2^{(r)}}\,$. Furthermore, the terms
appearing in the expansion of the new Hamiltonian
$H^{(r)}$ in~\frmref{frm:H(r)-espansione} are bounded by
$$
\vcenter{\openup1\jot 
\halign{
$\displaystyle\hfil#$&$\displaystyle{}#\hfil$&$\displaystyle#\hfil$\cr
\|f_{\ell}^{(r,s)}\|_{1-d_{r}}&\leq \frac{\Ebarra M^{3s-3+\ell}}{2^{\ell}}\,
\frac{T_{r,s}^3}{\left(a_r\delta_r^2\right)^{\ell}}
\,\nu_{r,s}\exp(s\zeta_{r})
&\quad\,{\hbox{for }0\le\ell\le 2\,,\ s>r\,,}
\cr
\|f_{\ell}^{(r,0)}\|_{1-d_{r}}&\leq \frac{\Ebarra}{2^{\ell}}\nu_{r,0}
&\quad\,{{\hbox{\rm for}\ \ell\ge 3\,,}}
\cr
\|f_{\ell}^{(r,s)}\|_{1-d_{r}}&\leq \frac{\Ebarra M^{3s}}{2^{\ell}}
\left(\frac{T_{r,s}}{a_m\delta_m^2}\right)^3 \nu_{r,s}\exp(s\zeta_{r})&
\quad\hbox{for }{\vtop{\hbox{${\ell\ge 3\,,\ s\ge 1\,,}$}
\vskip-2pt\hbox{\hskip-5pt$\hbox{with }\ m=\min\{r,s\}\,.$}}}\cr
}}
\formula{stime:f_l^(r,s)-lemmone}
$$
Finally, for $r\ge 2\,$, the variations of the frequencies, induced by
the $r$-th normalization step, are bounded by
$$
\max_{{1\le i\le n_1}\atop{1\le j\le n_2}}
\left\{\frac{1}{\sigma}\big|\omega_{i}^{(r)}-\omega_{i}^{(r-1)}\big|\,,\,
\epsilon\big|\Omega_{j}^{(r)}-\Omega_{j}^{(r-1)}\big|\right\}\le
\epsilon^{r}M^{3r} \left(\frac{T_{r,r}}{a_r\delta_r^2}\right)^3
\nu_{r,r}\exp(r\zeta_{r})\ .
\formula{frm:stima-variazione-frequenze}
$$
}\endclaim
\noindent
The proof of lemma above
needs many essentially trivial computations, nevertheless it is
outlined in an exhaustive way in appendix~\sbsref{app:lemmone}.

In the statement of lemma~\lemref{lem:lemmone}, the hypothesis
requiring that the first $r$ normalization steps can be formally
performed essentially means that the non-resonance
conditions~\frmref{nonres} and~\frmref{nonres.a} are satisfied, so that
all $a_1,\,\ldots,\,a_r$ and $b_1,\,\ldots,\,b_r$ are
positive. However, in order to ensure the uniform convergence of the
Hamiltonian $H^{(r)}$ to the wanted normal form, we need some stronger
assumptions like, for instance, the following one (that has been
adopted also in~\dbiref{Gio-Mar-2010}).

\definition{def:Ctau}{We say that the sequence $\{a_r\}_{r\ge 1}$,
introduced in~\frmref{nonres}, satisfies the {\bf condition
$\tauvet$}, if
$$
-\sum_{r\ge 1} \frac{\log a_r}{r(r+1)} = \Gamma \lt \infty\ .
\formula{frm:Ctau}
$$
}\endclaim

We can now provide an estimate for the quantities $T_{r,s}$
appearing in lemma~\lemref{lem:lemmone}.
\lemma{nrmlie.42}{Let the sequence $\{a_r\}_{r\ge 1}$ satisfy
condition~$\tauvet$ and the sequence $\{\delta_r\}_{r\ge 1}$ be
defined as in~\frmref{frm:dr-and-deltar}. Then, the sequence
$\{T_{r,s}\}_{r\ge 0\,,\,s\ge 0}$ is bounded by
$$
T_{r,s} \le \frac{1}{a_s\delta_s^2} T_{r,s} \le
\left(2^{15}e^{\Gamma}\right)^s
\qquad
\hbox{for } r\ge 1\,,\ s\ge 1\ .
$$
}\endclaim

Also the number of summands involved in the recursive
formul{\ae}~\frmref{frm:fI}, \frmref{frm:fII} and~\frmref{frm:fIII}
needs to be controlled by some special estimates.
\lemma{lem:nuest}{The sequence of positive integer numbers
$\{\nu_{r,s}\}_{r\ge 0\,,\,s\ge 0}$ defined
in~\frmref{frm:seqnu} is bounded by
$$
\nu_{r,s}\le\nu_{s,s}\leq 2^{8s}
\qquad
\hbox{for } r\ge 0\,,\ s\ge 0\ .
$$
}\endclaim
\noindent The proofs of the two lemmas above are deferred to
appendix~\sbsref{app:seq}.

All the
formul{\ae}~\frmref{frm:cond-eps-trasf-Diag}--\frmref{frm:stima-variazione-frequenze}
appearing in the main lemma~\lemref{lem:lemmone} look extremely
complex, as they are written in a suitable way to prove the lemma by
induction.  Therefore, we summarize in a more complete and compact way
all the ``purely analytical'' part studying the convergence of our
algorithm.

\proposition{analitico}{Let us consider an analytic Hamiltonian
$H^{(0)}:\DDD_{\rho,R,\sigma}\to\complessi\,$ expanded as in~\frmref{frm:H(0)},
that satisfies hypotheses~(c')--(e') of
lemma~\lemref{lem:analiticita-espansione-Ham} and is such that it is
possible to perform infinitely many normalization steps of the formal
algorithm described in section~\secref{sec:ellformale}. Moreover,
assume that

\item{(f')} the sequences of frequency vectors $\{\omega^{(r)}\}_{r\ge 0}$
and $\{\Omega^{(r)}\}_{r\ge 0}$ (defined in
subsection~\sssref{sss:frequencies}) fulfill the non-resonance
conditions~$\tauvet$ and~\frmref{nonres.a} with $b_{r}\ge\bar b>0$ for
$r\ge 1\,$, respectively;

\item{(g')} the parameter $\epsilon$ is smaller than the ``analytic
threshold value'' $\epsilon^{\star}_{{\rm an}}\,$, being
$$
\epsilon^{\star}_{{\rm an}}=
\frac{1}{2^8}\left(\frac{\min\left\{1\,,\,\bar b\right\}}{\AAA}\right)^2
\qquad \hbox{with} \qquad
\AAA=\big(2^{18}M\,e^{\Gamma}\big)^3\ ,
\formula{def:soglia-analitica}
$$
where $M$ and $\Gamma$ are defined in~\frmref{eq:def-M}
and~\frmref{frm:Ctau}, respectively.

\noindent
Then, there exists an analytic canonical transformation
$\Phi_{\omega^{(0)}}^{(\infty)}:\DDD_{1/2(\rho,R,\sigma)}\to
\DDD_{3/4(\rho,R,\sigma)}$ such that the
Hamiltonian $H^{(\infty)}=H^{(0)}\circ\Phi_{\omega^{(0)}}^{(\infty)}$ is in
normal form, i.e.,
$$
H^{(\infty)} = \scalprod{\omega^{(\infty)}}{p} +
\epsilon\sum_{j=1}^{n_2}\Omega^{(\infty)}_{j}z_{j}{\bar z}_{j}
+\sum_{\ell>2}\sum_{s\geq 0} \epsilon^{s} f_{\ell}^{(\infty,s)}\ .
\formula{frm:H(infty)-espansione}
$$
Furthermore, the norms of the functions
$f_{\ell}^{(\infty,s)}\in\Pset_{\ell,sK}$ are bounded by
$$
\left\|f_{\ell}^{(\infty,s)}\right\|_{3/4}\le
\frac{\Ebarra}{2^{\ell}}\AAA^s
\qquad\hbox{for } \ell\ge 3\,,\ s\ge 0
\formula{stime:f_l^(infty,s)}
$$
and both the limit values of the frequency vectors
$\omega^{(\infty)}$ and $\Omega^{(\infty)}$ are well defined, being
$\{\omega^{(r)}\}_{r\ge 0}$ and $\{\Omega^{(r)}\}_{r\ge 0}$ Cauchy
sequences, as their $r$-th variations are such that
$$
\max_{1\le i\le n_1}
\left\{\big|\omega_{i}^{(r)}-\omega_{i}^{(r-1)}\big|\right\}\le
\sigma\big(\epsilon\AAA\big)^r\ ,
\qquad
\max_{1\le j\le n_2}
\left\{\big|\Omega_{j}^{(r)}-\Omega_{j}^{(r-1)}\big|\right\}\le
\epsilon^{r-1}\AAA^r\ ,
\formula{frm:stima-variazione-frequenze-+-esplicita}
$$
for $r\ge 2\,$, while at the first normalization step
the equations in~\frmref{chgfreq-passo1} hold true.
}\endclaim
\noindent
Since most of the preliminary work has been previously carried out
through all the present section, the proof of the proposition above is
now rather easy and can be sketched as follows. First, let us remark
that we can apply lemma~\lemref{lem:lemmone}, since
condition~\frmref{frm:cond-eps-trasf-Diag} is always satisfied under
the hypotheses of proposition~\proref{analitico}, as it can be easily
verified using lemmas~\lemref{nrmlie.42}--\lemref{lem:nuest} and the
elementary inequality $\exp(\zeta_{s})<2$ for $s\ge 0$
(see~\frmref{eq:def-succ-zetagreco}). Thus, starting from the
inequalities in formula~\frmref{stime:f_l^(r,s)-lemmone} and using
property~(i) of lemma~\lemref{nrmlie.27}, some trivial calculations
allow us to ensure that
$$
\|f_{\ell}^{(r,s)}\|_{1-d_{r}}\le \frac{\Ebarra}{2^{\ell}}\AAA^s
\qquad{\hbox{\rm for }0\le\ell\le 2\,,\ s>r
\ \hbox{ or }\ \ell\ge 3\,,\ s\ge 0\ .}
\formula{stime:f_l^(r,s)-intermedie}
$$
Since $\epsilon\AAA<1$ (actually $\epsilon\AAA<1/\AAA\ll 1\,$, in view
of condition~(g') combined with the definitions in~\frmref{eq:def-M}
and \frmref{frm:Ctau}), as an immediate consequence of the
estimate~\frmref{stime:f_l^(r,s)-intermedie}, we can deduce that
$H^{(r)}$, written in~\frmref{frm:H(r)-espansione}, is analytic on
$\DDD_{(1-d_r)(\rho,R,\sigma)}\supset\DDD_{3/4(\rho,R,\sigma)}\,$.
Starting from~\frmref{frm:stima-variazione-frequenze} and using
condition~(g'), some calculations analogous to those required
by~\frmref{stime:f_l^(r,s)-intermedie} allow us to verify the
inequalities in~\frmref{frm:stima-variazione-frequenze-+-esplicita}.

Let us now focus on the difference $H^{(r)}-H^{(r-1)}$. Starting
from equations~\frmref{frm:H(r)-espansione}
and~\frmref{frm:H(r-1)}, we can write
$$
\eqalign{
&H^{(r)}-H^{(r-1)}=\scalprod{\big(\omega^{(r)}-\omega^{(r-1)}\big)}{p} +
\epsilon\sum_{j=1}^{n_2}\big(\Omega^{(r)}_{j}-\Omega^{(r-1)}_{j}\big)
z_{j}{\bar z}_{j}
\cr
&\phantom{H^{(r)}} +\sum_{\ell\ge 0}\sum_{s\geq r} \epsilon^{s}
\left(f_{\ell}^{(r,s)}-f_{\ell}^{(r-1,s)}\right)
+\sum_{\ell>2}\sum_{s=1}^{r-1}\sum_{j\ge 1}
\epsilon^{s+j(r-1)}\EEE_{j}^{(r)} f_{\ell}^{(r-1,s)}\ ,
}
\formula{frm:H(r)menoH(r-1)-espansione}
$$
where we used the recursive definitions in~\frmref{frm:fr}, the fact
that $\EEE_{0}^{(r)}$ is equal to the identity and the equation
$f_{\ell}^{(\rmIII;r,s)}=f_{\ell}^{(r-1,s)}$
for $\ell\ge 3\,,\ 1\le s\le r-1$ (because of
formul{\ae}~\frmref{frm:fI}, \frmref{frm:fII}, \frmref{frm:fIII}).
Therefore, we have that
$$
\left\|H^{(r)}-H^{(r-1)}\right\|_{3/4}\le
\left(n_1\sigma\rho+\frac{n_2R^2}{\epsilon}
+\frac{4\Ebarra}{1-\epsilon\AAA}\right)\big(\epsilon\AAA\big)^r
+\frac{2^{-(r+6)}}{1-2^{-(r+6)}}\frac{\Ebarra\,\epsilon\AAA}{1-\epsilon\AAA}\ ,
\formula{frm:stima-H(r)menoH(r-1)}
$$
where we used the fourth inequality
in~\frmref{stime:generatrici-lemmone} and those
in~\frmref{frm:stima-variazione-frequenze-+-esplicita}--\frmref{stime:f_l^(r,s)-intermedie}. Since
the r.h.s. of the estimate above tends to zero for $r\to\infty$ and that
the sup-norm is bounded by the weighted Fourier one, then
$\{H^{(r)}\}_{r\ge 0}$ is a {\sl Cauchy sequence} of analytic
Hamiltonians and it admits a limit $H^{(\infty)}$. Moreover,
formul{\ae}~\frmref{frm:H(r)menoH(r-1)-espansione}--\frmref{frm:stima-H(r)menoH(r-1)}
imply that also $\{\omega^{(r)}\}_{r\ge 0}\,$, $\{\Omega^{(r)}\}_{r\ge
0}$ and $\{f_{\ell}^{(r,s)}\}_{r\ge 0}$ for $\ell\ge 3\,,\ s\ge 0$ are
{\sl Cauchy sequences}, thus the expansion of the Hamiltonian
$H^{(\infty)}=\lim_{r\to\infty}H^{(r)}$ cannot differ from that
written in~\frmref{frm:H(infty)-espansione}, where the terms
$f_{\ell}^{(\infty,s)}\in\Pset_{\ell,sK}$ are bounded as
in~\frmref{stime:f_l^(infty,s)}, in view of
inequality~\frmref{stime:f_l^(r,s)-intermedie}.

In order to conclude the proof of proposition~\proref{analitico} we
need some arguments which are often used in the framework of a KAM
theorem based on canonical transformations performed by Lie series.
For sake of completeness, we sketch here the essential ideas; more
details are available, for instance, in subsection~4.3
of~\dbiref{Gio-Loc-1997}.  Let us denote with $\phi^{(r)}$ the
canonical change of coordinates induced by the $r$-th normalization
step of the formal algorithm described in
section~\secref{sec:ellformale}, i.e.,
$$
\phi^{(r)}(p,q,z,\imunit\bar z)=
\exp\left( \epsilon^r\Lie_{\chi_0^{(r)}} \right)
\exp\left( \epsilon^r\Lie_{\chi_1^{(r)}} \right)
\exp\left( \epsilon^r\Lie_{\chi_2^{(r)}} \right)
\telchi_{\epsilon^{r-1}\Dscr_2^{(r)}}(p,q,z,\imunit\bar z)\ .
\formula{frm:def-phi(r)}
$$
Using the fourth inequality in~\frmref{stime:generatrici-lemmone},
one can easily verify that
$$
\max_{1\le j\le n_2}\left\{
\Big\|\telchi_{\epsilon^{r-1}\Dscr_2^{(r)}}z_j-z_j\Big\|_{3/4}\right\}
<\delta_r R\ .
\formula{frm:scartamento-dovuto-al-telchi}
$$
Analogous estimates can be deduced for both the other Lie series
appearing in~\frmref{frm:def-phi(r)} and all the canonical variables,
using again the inequalities in~\frmref{stime:generatrici-lemmone}.
Therefore, one has that
$\phi^{(r)}(\DDD_{(1/2+d_{r-1})(\rho,R,\sigma)})\subset
\DDD_{(1/2+d_{r})(\rho,R,\sigma)}$ and
$\Phi^{(r)}(\DDD_{1/2(\rho,R,\sigma)})\subset\DDD_{3/4(\rho,R,\sigma)}\,$,
being $\Phi^{(r)}=\phi^{(1)}\circ\ldots\circ\phi^{(r)}\,$. By
repeatedly using the so-called exchange theorem for Lie series (and
Lie transforms), one immediately obtains that
$H^{(r)}=H^{(0)}\circ\Phi^{(r)}$. By using
estimate~\frmref{frm:scartamento-dovuto-al-telchi} and the 
ones related to the other generating functions, we can prove that the
canonical transformation
$\Phi_{\omega^{(0)}}^{(\infty)}=\lim_{r\to\infty}\Phi^{(r)}$ is well
defined in $\DDD_{1/2(\rho,R,\sigma)}\,$. Finally, we have that
$H^{(0)}\circ\Phi_{\omega^{(0)}}^{(\infty)}=\lim_{r\to\infty}H^{(0)}\circ\Phi^{(r)}=
\lim_{r\to\infty}H^{(r)}=H^{(\infty)}\,$.

Actually, with some additional effort, we could prove that
$\Phi_{\omega^{(0)}}^{(\infty)}$ differs from the identity just for
terms of order $\Oscr(\epsilon)\,$. As a final comment, let us remark that
in the symbol $\Phi_{\omega^{(0)}}^{(\infty)}\,$, we emphasized the
parametric dependence of that canonical transformation on the initial
frequency $\omega^{(0)}\,$, as it is in the spirit of the next
section.

\section{measure}{Measure of the resonant regions}
The aim of this section is to show that the set of frequencies to
which our algorithm applies has relative big measure.  To this end we
must exploit the dependence of the whole procedure on the frequency
vector $\omega^{(0)}$, that has been neglected in the analytic
construction.  Thus we focus here on the sequence
$\big\{(\omega^{(r)}\,,\,\epsilon\Omega^{(r)})\big\}_{r\ge 0}$ and on
its dependence on $\omega^{(0)}$, with the aim of selecting the set of
frequencies $\omega^{(0)}$ for which the sequence
$(\omega^{(r)}\,,\,\epsilon\Omega^{(r)})$ satisfies all the
non-resonant conditions that are requested so as to ensure the
convergence.  Therefore we extract from the discussion of
section~\secref{sec:quantitative} just the essential information
concerning the shift of the frequencies at each normalization step,
i.e., the
estimates~\frmref{frm:stima-variazione-frequenze-+-esplicita}.

Let us recall that both $\omega^{(r)}$ and $\Omega^{(r)}$ are
iteratively defined according to the prescriptions given in
section~\secref{sec:ellformale} and their values actually depend on
all terms appearing in the expansion~\frmref{frm:H(0)} of the initial
Hamiltonian~$H^{(0)}$. Nevertheless, with a little abuse of notation,
in the present section and in appendix~\sbsref{app:measure}, the
quantities $\omega^{(r)}$ and $\Omega^{(r)}$ are regarded as
analytic functions of the frequencies $\omega^{(0)}$ only, on some open
domain that can be defined as follows. We start from the compact
set $\WWW^{(0)}\subset\reali^{n_1}$ and its complex extension
$\Wscr^{(0)}_{h_0}\,$, where the Hamiltonian~$H^{(0)}$ is well defined
and the Jacobian of $\Omega^{(0)}(\omega^{(0)})$ is bounded.  We
consider a sequence of complex extended domains
$\Wscr^{(0)}_{h_0}\supseteq\Wscr^{(1)}_{h_1}\supseteq\Wscr^{(2)}_{h_2}
\supseteq\ldots\,$, where $\{h_{r}\}_{r\ge 0}$ is a positive non-increasing sequence of real numbers such that $\omega^{(r)}(\omega^{(0)})$ admits an inverse function $\phi^{(r)}$ well
defined on $\Wscr^{(r)}_{h_{r}}\,$.  In detail, let us start by
setting $\Wscr^{(1)}=\Wscr^{(0)}$.  For $r\ge 2$ and some fixed
positive values of the parameters $\gamma\, ,\tau\in\reali$ and
$K\in\naturali\,$, we define the sequence of real domains
$\{\Wscr^{(r)}\}_{r\ge 0}\,$, so that at each step $r$, we remove from
$\Wscr^{(r-1)}$ all the resonant regions related to the new
small divisors appearing in the formal algorithm (see
section~\secref{sec:ellformale}); therefore,
$$
\Wscr^{(r)}=\Wscr^{(r-1)}\,\backslash\,\Rscr^{(r)}
\ ,
\qquad{\rm with}\quad
\Rscr^{(r)} = \bigcup_{{rK<|k|\leq (r+1)K}\atop{|l|\leq 2}}\Rscr^{(r)}_{k,l}\ ,
\formula{def:WWWr}
$$
being
$$
\Rscr_{k,l}^{(r)}=\left\{\omega\in\Wscr^{(r-1)}:
\left| k\cdot\omega+
\epsilon l\,\cdot\Omega^{(r)}\circ\phi^{(r)}(\omega)\right|\lt
\frac{2\gamma}{\big((r+1)K)^{\tau}}\right\}\ .
\formula{def:strisciarisonantekl}
$$
Moreover, it is also convenient to introduce the functions
$\delta\omega^{(r)}$ and $\Delta\Omega^{(r)}$ defined as
$$
\delta\omega^{(r)}=\omega^{(r)}\circ\phi^{(r-1)}-{\rm Id}\ ,
\qquad
\Delta\Omega^{(r)}=\Omega^{(r)}\circ\phi^{(r-1)}-\Omega^{(r-1)}\circ\phi^{(r-1)}\ .
\formula{def:delta-Delta-frequenze}
$$
By the way, let us remark that from equations
in~\frmref{chgfreq-passo1} we have $\delta\omega^{(1)}=0$ and
$\Delta\Omega^{(1)}=0\,$.

We now adapt the approach by P\"oschel\bibref{Poschel-1989} to our
context in the following
\proposition{prop:geometrico}{Let us consider 
the family~\frmref{frm:H(0)} of Hamiltonians $H^{(0)}$ parameterized
by the $n_1$-dimensional frequency vector $\omega^{(0)}$. Assume that
there exist positive parameters $\gamma\,$, $\tau\,$, $\bar b\,$, $J_0$,
a positive integer $K$ and a compact set $\WWW\subset\reali^{n_1}$
such that the function $\Omega^{(0)}:\WWW_{h_0}\to\complessi^{n_2}$ is
analytic and satisfies the properties~(a')--(b') of
lemma~\lemref{lem:analiticita-espansione-Ham}.  Define the sequence
$\{h_{r}\}_{r\ge 0}$ of complex extensions as
$$
h_0 = \min\left\{\frac{\gamma\eta}{4K^{\tau}}\,,
\ \frac{\bar b}{4 J_0}\right\}
\qquad
\hbox{and}
\qquad
h_r = \frac{h_{r-1}}{2^{\tau+2}}
\ \ \hbox{for } r\ge 1\ ,
\formula{def:hr}
$$
where $\eta=\min\{1/K\,,\,\sigma\}\,$.

\noindent
Considering the sequence of Hamiltonians $\{H^{(r)}\}_{r\ge 0}\,$,
formally defined by the algorithm in section~\secref{sec:ellformale},
let us assume that the functions
$\omega^{(1)},\,\Omega^{(1)},\,\ldots\,,\,\omega^{(r)},\,\Omega^{(r)}$
satisfy the following hypotheses up to a fixed normalization step $r\ge 0$

\item{(h')} the function $\omega^{(s)}(\omega^{(0)})$ has analytic inverse
$\phi^{(s)}$ on $\Wscr^{(s)}_{h_{s}}\,$, for $0\le s \le r-1$, where
$\phi^{(0)}={\rm Id}$ and the domains are recursively defined as
extensions of those given by
formul{\ae}~\frmref{def:WWWr}--\frmref{def:strisciarisonantekl},
starting from $\WWW^{(0)}=\WWW\,$;

\item{(i')} both
$\omega^{(s)}\circ\phi^{(s-1)}:\,\Wscr^{(s-1)}_{h_{s-1}}\to\complessi^{n_1}$
and
$\Omega^{(s)}\circ\phi^{(s-1)}:\,\Wscr^{(s-1)}_{h_{s-1}}\to\complessi^{n_2}$
are analytic functions, for $1\le s \le r$ ;

\item{(j')} there exist positive
parameters $\epsilon\,$, $\sigma$ and $\Ascr\ge 1$ satisfying

$$
\max_{1\le j\le n_1}\sup_{\omega\in \WWW^{(s-1)}_{h_{s-1}}}
\big|\delta\omega^{(s)}_j(\omega)\big|\le
\sigma\big(\epsilon\AAA\big)^s\ ,
\quad
\max_{1\le j\le n_2}\sup_{\omega\in \WWW^{(s-1)}_{h_{s-1}}}
\big|\Delta\Omega^{(s)}_j(\omega)\big|\le
\epsilon^{s-1}\AAA^s\ ,
\formula{frm:stima-variazione-frequenze-per-parte-geom}
$$
\item{} for $2\le s \le r\,$, where $\delta\omega^{(s)}$ and $\Delta\Omega^{(s)}$ are defined
as in~\frmref{def:delta-Delta-frequenze}, with $\delta\omega^{(1)}=0$
and $\Delta\Omega^{(1)}=0\,$;

\item{(k')} the parameter $\epsilon$ is smaller than the ``geometric
threshold value''

$$
\epsilon^{*}_{{\rm ge}} = \min\left\{\frac{1}{(2J_0+1)\eta}\,,
\ \frac{1}{2^{\tau+3}\Ascr}\,
\min\left\{1\,,\,\frac{h_0}{8\Ascr}\,,\,\frac{\bar b}{8\Ascr}\right\}
\right\}\ ;
\formula{def:soglia-geometrica}
$$

\noindent
Then, the function $\omega^{(r)}(\omega^{(0)})$ admits an analytic
inverse
$\phi^{(r)}:\WWW^{(r)}_{h_{r}} \to \WWW^{(0)}_{h_{0}}$
on its domain of definition and satisfies the inclusion relation
$\phi^{(r)}\big(\WWW^{(r)}_{h_{r}}\big)
\subset\phi^{(r-1)}\big(\WWW^{(r-1)}_{h_{r-1}}\big)\,$. Moreover, the following
non-resonance inequalities hold true:
$$
\eqalign{
\min_{\scriptstyle{k\in\interi^{n_1}\,,\,0<|k|\leq (r+1)K}
\atop\scriptstyle{l\in\interi^{n_2}\,,\,0\leq |l|\leq2}}
\,\inf_{\omega\in\Wscr^{(r)}_{h_{r}}}
\left| k\cdot\omega
+\epsilon l\,\cdot\Omega^{(r)}\big(\phi^{(r)}(\omega)\big)\right|
&\ge
\frac{\gamma}{\big((r+1)K\big)^{\tau}}\ ,
\cr
\min_{1\le i<j\le n_2}\,\inf_{\omega\in\Wscr^{(r)}_{h_{r}}}
\left| \Omega^{(r)}_i\big(\phi^{(r)}(\omega)\big)-
\Omega^{(r)}_j\big(\phi^{(r)}(\omega)\big)\right|
&\ge
\bar b\ .
}
\formula{nonres-cond-diof}
$$
Finally, the Lipschitz constant related to the Jacobian of the
functions $\phi^{(r)}$ and $\Omega^{(r)}\circ\phi^{(r)}$ are uniformly
bounded as
$$
\left|\frac{\partial\big(\phi^{(r)}-{\rm Id}\big)}{\partial\omega}
\right|_{\infty;\WWW^{(r)}_{h_r}}\leq \epsilon\sigma\ ,
\qquad
\left|\frac{\partial\big(\Omega^{(r)}\circ\phi^{(r)}\big)}{\partial\omega}
\right|_{\infty;\WWW^{(r)}_{h_r}}\leq 2 J_0 + 1\ .
\formula{diseq:Lipschitz-definitive}
$$

}\endclaim

\noindent
The proof is deferred to appendix~\sbsref{app:measure}.

The proposition above allows us to prove the persistence of a set of
tori characterized by ``Diophantine'' frequencies, according to the
following
\definition{def:dioph-freq}{For any fixed $\omega^{(0)}\,$, we say that
the sequence of frequency vectors $\big\{\big(\omega^{(r)}(\omega^{(0)})\,,\,
\epsilon\Omega^{(r)}(\omega^{(0)})\big)\big\}_{r\ge 0}$ is
Diophantine, if there are three positive constants $\gamma\,$, $\tau$
and $\bar b$ such that, for all $r\ge 1$, \frmref{nonres}
and~\frmref{nonres.a} are satisfied with $a_r=\gamma/(rK)^{\tau}$ and
$b_r\ge\bar b\,$.}\endclaim

\noindent
We emphasize that the non-resonance condition given by the definition
above is stronger than the condition~$\tauvet$ in~\frmref{frm:Ctau}
considered in section~\secref{sec:quantitative}.

We denote by $\Kscr_l^{(r)}$ the closed convex hull of the gradient set
$$
\Gscr_l^{(r)}=
\left\{\partial_{\omega}\big[\epsilon
\scalprod{l}{\Omega^{(r)}\big(\phi^{(r)}(\omega)\big)}\big]
\>:\ \omega\in\Wscr^{(r)}\right\}\ .
\formula{frm:gradient-set-passo0}
$$
Due to the fact that the transversal frequencies are
$\Oscr(\epsilon)\,$, we can ensure that the closed convex hull does not
contain any integer vector except the zero vector.  To this aim, we must
ensure the further smallness condition
$$
\epsilon < \frac{1}{4 (2J_0+1)}\ ,
\formula{diseq:eps-piccolo-per-guscio-convesso}
$$
so that ${\rm dist}(k,\Kscr_l^{(r)})\geq 1/2$ for
$k\in\interi^{n_1}\setminus\{0\}\,$ (with respect to the euclidean
norm).

We now need to estimate the volume of the resonant zones that we must
remove at each step of the procedure and show that the final ``good
domain'', $\lim_{r\to\infty}\phi^{(r)}(\Wscr^{(r)})\,$, has positive
Lebesgue measure.  To this aim, we report lemma~8.1
of P\"oschel\bibref{Poschel-1989}.

\lemma{lem:Poescel-dist}{If ${\rm dist}(k,\Kscr_l^{(r)})=s>0$ then
$$
{\rm m}(\Rscr^{(r)}_{k,l})\leq \frac{4\gamma}{((r+1)K)^\tau}
\frac{D^{n_1-1}}{s}\ ,
$$
where $D$ is the diameter of $\Wscr^{(0)}$ with respect to the
sup-norm.}
\endclaim
The volume of the resonant regions must be compared with respect to
the initial set $\Wscr^{(0)}$.  Therefore, it is convenient to
estimate the measure of $\phi^{(r)}(\Rscr^{(r)}_{k,l})\,$ in the
original coordinates $\omega^{(0)}$. Using
lemma~\lemref{lem:Poescel-dist} and
assumption~\frmref{diseq:eps-piccolo-per-guscio-convesso}, for
$k\in\interi^{n_1}\setminus\{0\}$ we have
$$
{\rm m}\Big(\phi^{(r)}\big(\Rscr^{(r)}_{k,l}\big)\Big)\le
\frac{8\gamma D^{n_1-1}}{((r+1)K)^\tau}\,\sup_{\omega\in\Wscr^{(r)}}
\det\left(\frac{\partial\phi^{(r)}}{\partial \omega}\right)\ .
\formula{stima-zona-risonante-in-omeghini-iniz}
$$
Starting from the first inequality
in~\frmref{diseq:Lipschitz-definitive} and using the well known
Gershgorin circle theorem, one can easily control the actual
expansion of the resonant zones due to the stretching of the
frequencies, by verifying that
$$
\sup_{\omega\in\Wscr^{(r)}}
\det\left(\frac{\partial\phi^{(r)}}{\partial \omega}\right)
\le 2
\qquad{\rm when}\quad
\epsilon\le\frac{\log 2}{\sigma n_1^2}\ .
\formula{diseq:stima-dilatazione-frequenze-veloci}
$$
Using the
inequalities~\frmref{stima-zona-risonante-in-omeghini-iniz}--\frmref{diseq:stima-dilatazione-frequenze-veloci},
we can easily obtain a final estimate of the total volume of the
resonant regions included in $\WWW^{(0)}$:
$$
\eqalign{
\sum_{r=2}^{\infty}\,\sum_{{rK<|k|\leq (r+1)K}\atop{|l|\leq 2}}
{\rm m}\Big(\phi^{(r)}\big(\Rscr^{(r)}_{k,l}\big)\Big) &\leq
c_{n_2}\sum_{r=2}^{\infty}\,\sum_{(r-1)K<|k|\leq rK}
\frac{16\gamma D^{n_1-1}}{((r+1)K)^\tau}
\cr
&\leq \gamma\ \frac{2^{n_1+4} c_{n_2} D^{n_1-1}}{K^{\tau-n_1}}
\sum_{r=3}^{\infty}\frac{1}{r^{\tau-n_1+1}}\ ,
\cr
}
\formula{eq:measure}
$$
where $c_{n_2}=(2n_2+2)(2n_2+1)/2$ is the maximum number of polynomial
terms having degree $\le 2$ in the transversal variables
$(z,\imunit\bar z)\,$.  The last series is convergent if $\tau>n_1$
and it is of order $\Oscr(\gamma)$.

\section{sec:fine-dim-teor-quantitativo}{Proof of theorem~\thrref{teor:enunciato-nonquantitativo}} 
\noindent
The proof is a straightforward combination of
propositions~\proref{analitico} and~\proref{prop:geometrico}, which
summarize the ``purely analytical'' study of the convergence of our
algorithm and the more ``geometrical part'', respectively.
We sketch the argument.

According to lemma~\lemref{lem:analiticita-espansione-Ham} the family
of Hamiltonians $\HHH^{(0)}$, parameterized by the frequency vectors
$\omega^{(0)}$ and defined on a real domain, can be extended to a
complex domain $\DDD_{\rho,R,\sigma}\times\WWW_{h_0}$ with suitable
parameters; moreover, their expansions can be written as $H^{(0)}$
in~\frmref{frm:H(0)}.  Possibly modifying the values of parameters
$\gamma$ and $\tau\,$, we can choose $\gamma$ and $\tau>n_1$ such that
the estimate of the resonant volume in the last row
of~\frmref{eq:measure} is smaller than ${\rm m}(\WWW_{h_0})\,$ and
property~(a') of lemma~\lemref{lem:analiticita-espansione-Ham} is
still satisfied.  Let us consider values of the small parameter
$\epsilon$ such that $\epsilon<\epsilon^{\star}$, with
$$
\epsilon^{\star}=
\min\left\{\epsilon^{\star}_{{\rm an}}\,,\,\epsilon^{\star}_{{\rm ge}}
\,,\,\frac{1}{4 (2J_0+1)}\,,\,\frac{\log 2}{\sigma n_1^2}\right\}\ ,
\formula{def:soglia-finale}
$$
where $\epsilon^{\star}_{{\rm an}}$ and $\epsilon^{\star}_{{\rm ge}}$
are defined in~\frmref{def:soglia-analitica}
and~\frmref{def:soglia-geometrica}, respectively.

Recall that the $r$-th normalization step of the formal algorithm
described in section~\secref{sec:ellformale} can be performed if the
non-resonance conditions~\frmref{nonres}--\frmref{nonres.a} are
satisfied.  Assuming the threshold value $\epsilon^{\star}$ as
in~\frmref{def:soglia-finale}
lemma~\lemref{lem:analiticita-espansione-Ham} and
proposition~\proref{prop:geometrico} ensures that the first step can
be performed for every frequency vector $\omega^{(0)}\in\WWW^{(0)}_{h_0}\,$.

We now proceed by induction.  Let us suppose that $r-1$ steps have
been performed and proposition~\proref{prop:geometrico} applies.  In
view of the non-resonance condition~\frmref{nonres-cond-diof} the
$r$-th normalization step can be performed.  We now check that
proposition~\proref{prop:geometrico} applies again. By construction,
both $\omega^{(r)}(\omega^{(0)})$ and $\Omega^{(r)}(\omega^{(0)})$ are
analytic functions on
$\phi^{(r-1)}\big(\Wscr^{(r-1)}_{h_{r-1}}\big)\,$. In view of
$\epsilon<\epsilon^{\star}$, then hypothesis~(g') of
proposition~\proref{analitico} is satisfied, so
lemma~\lemref{lem:lemmone} applies and the
estimate~\frmref{frm:stima-variazione-frequenze-+-esplicita} on the
shift of the frequencies holds true.  Thus
proposition~\proref{prop:geometrico} can be applied at the $r$-th step
which complete the induction.

We conclude that the non-resonance
conditions~\frmref{nonres-cond-diof} hold true for $r\ge 0$ and
that the sequence of frequency vectors
$\big\{\big(\omega^{(r)}(\omega^{(0)})\,,\,
\epsilon\Omega^{(r)}(\omega^{(0)})\big)\big\}_{r\ge 0}$ is
Diophantine for
 $\omega^{(0)}\in\lim_{r\to\infty}\phi^{(r)}\big(\WWW^{(r)}_{h_r}\big)=
\bigcap_{r=0}^{\infty}\phi^{(r)}\big(\WWW^{(r)}_{h_r}\big)\,$, where we
used the inclusion relation $\phi^{(r)}\big(\WWW^{(r)}_{h_{r}}\big)
\subset\phi^{(r-1)}\big(\WWW^{(r-1)}_{h_{r-1}}\big)\,$ between open sets.
Finally, also hypothesis~(f') of proposition~\proref{analitico} is
satisfied and so, for
$\omega^{(0)}\in\bigcap_{r=0}^{\infty}\phi^{(r)}\big(\WWW^{(r)}_{h_r}\big)$,
there exists an analytic canonical transformation
$\Phi_{\omega^{(0)}}^{(\infty)}$ which gives the initial Hamiltonian
$H^{(0)}$ the normal form~\frmref{frm:H(infty)-espansione}.  Since
$\bigcap_{r=0}^{\infty}\phi^{(r)}\big(\WWW^{(r)}_{h_r}\big)$ is a
countable intersection of open sets, it is measurable and
$$
{\rm m}\bigg(\bigcap_{r=0}^{\infty}\phi^{(r)}\big(\WWW^{(r)}_{h_r}\big)\bigg)
\ge {\rm m}\big(\WWW^{(0)}\big)-
\sum_{r=2}^{\infty}\,\sum_{{rK<|k|\leq (r+1)K}\atop{|l|\leq 2}}
{\rm m}\Big(\phi^{(r)}\big(\Rscr^{(r)}_{k,l}\big)\Big)>0\ ,
$$
where we have taken into account the fact that the complex extension
radius $h_r\to 0$ for $r\to\infty$, the estimate~\frmref{eq:measure}
and the initial choice of the parameters $\gamma$ and $\tau$ at the
beginning of the present section. This concludes the argument proving
theorem~\thrref{teor:enunciato-nonquantitativo}.

We now add a short remark concerning the comparison with the estimates
in previous works.  The threshold value $\epsilon^{\star}$ on the
small parameter $\epsilon$ is explicitly defined
in~\frmref{def:soglia-finale}. Although the definition involves many
parameters, we might produce an asymptotic estimate of the volume of
the resonant regions for $\epsilon\to0$.  In view of
inequality~\frmref{eq:measure} it is $\Oscr(\gamma)$. Moreover, one
can easily show\footnote{\dag}{First, let us remark that
$\epsilon^{\star}_{{\rm an}}=\Oscr(e^{-6\Gamma})$ in view of the
definitions in~\frmref{def:soglia-analitica}. By
definitions~\defref{def:Ctau} and~\defref{def:dioph-freq} we get that
$\Gamma=\sum_{r\ge 1}
[-\log \gamma+\tau\log(rK)]/[r(r+1)]\,$. Therefore,
$e^{-\Gamma}=\Oscr(\gamma)\,$, as one can easily verify that
$\sum_{r\ge 1} 1/[r(r+1)]=1\,$. For this purpose, it is enough to
check by induction that $\sum_{r=1}^{s} 1/[r(r+1)]=s/(s+1)$ for $s\ge
1\,$.} that $\epsilon^{\star}\le\epsilon^{\star}_{{\rm an}}=
\Oscr(\gamma^6)\,$. Thus the complement of the set of the invariant elliptic tori, i.e.,
$\WWW^{(0)}\setminus\bigcap_{r=0}^{\infty}\phi^{(r)}\big(\WWW^{(r)}_{h_r}\big)\,$,
has a measure estimated by $\Oscr(\epsilon^{1/6})\,$. This is
definitely worse than the results obtained
in~\dbiref{Bia-Chi-Val-2003} and~\dbiref{Bia-Chi-Val-2006}, where this
same quantity has been proven to be smaller than a bound
$\Oscr(\epsilon^{b_1})\,$, with $b_1<1/2\,$.  However, let us
emphasize that our main interest is to establish the convergence of a
constructive algorithm suitable for computer assisted applications.
As a matter of facts, by explicitly performing a number of
perturbation steps, both the applicability threshold and the estimate
of the measure can be significantly improved, possibly giving
realistic estimates for physical systems.  On the other hand it is
well known that purely analytical estimates are usually
unrealistically small.  For this reason we did not pay attention in
producing optimal estimates.

\appendix{app1}{Technicalities}
The appendix is devoted to technical details and proofs which have
been moved here in order to avoid the overloading of the text.

\subsection{app:stima-serie-di-Lie}{Estimates for multiple Poisson
brackets}
In the present subsection, we will replace $|\cdot|_{\alpha\rho,\alpha
R}$ with $|\cdot|_{\alpha}\,$, being $\alpha$ any real positive
number, so as to shorten the notation in an analogous way to what is
done for $\|\cdot\|_{\alpha\rho,\alpha R,\alpha\sigma}$ and
$\|\cdot\|_{\alpha}\,$. Let us recall that both the norms
$\|\cdot\|_{\alpha}$ and $|\cdot|_{\alpha}$ are defined
in~\frmref{def:norma}.

Some Cauchy's estimates on the derivatives in the restricted domains
will be useful during the following proof. Let us recall them by
referring to any function $g$ satisfying the hypotheses of
lemma~\lemref{lem:stima-termini-serie-Lie}:
$$
\left|\parder{g}{p_j}\right|_{1-d-d^{\prime}}\leq
\frac{\left|g\right|_{1-d^{\prime}}}{d\rho}\ ,
\qquad
\left|\parder{g}{z_j}\right|_{1-d-d^{\prime}}\leq
\frac{\left|g\right|_{1-d^{\prime}}}{d R}\ .
\formula{frm:stime-Cauchy}
$$
Of course, the latter inequality holds true also by replacing $z_j$
with $\bar z_j\,$.

Before considering the multiple Lie derivatives for the flow along a
generating function, it is convenient to provide a suitable estimate
for a single Poisson bracket.

\lemma{lem:stima-sigola-Pois-par}{Let $d,\,d^{\prime}\in\reali_{+}$ such that
$d+d^{\prime}<1\,$ and $g,\,g^{\prime}$ be two analytic functions such
that their corresponding norms $\|g\|_{1-d-d^{\prime}}$ and
$\|g^{\prime}\|_{1-d^{\prime}}$ are finite. Then, for all
$\delta\in\reali_{+}$ such that $d+d^{\prime}+\delta<1$ the following
inequality holds true:
$$
\left\|\{g,g^{\prime}\}\right\|_{1-d-d^{\prime}-\delta}\leq
\left(\frac{2}{e\rho\sigma}+
\frac{1}{R^{2}}\right)\frac{1}{(d+\delta)\delta}
\left\|g\right\|_{1-d-d^{\prime}}\left\|g^{\prime}\right\|_{1-d^{\prime}}\ .
\formula{frm:stimapois}
$$
}\endclaim
\proof
It is convenient to separate the contributions given by the
derivatives with respect to the conjugate pairs of variables $(p,q)$
and $(z,\imunit\bar z)\,$. Thus, let us first write the following
chain of inequalities:
$$
\eqalign{
&\Bigg\|\sum_{j=1}^{n_1}\left(
\frac{\partial g}{\partial q_j}\frac{\partial g^{\prime}}{\partial p_j}-
\frac{\partial g}{\partial p_j}\frac{\partial g^{\prime}}{\partial q_j}
\right)\Bigg\|_{1-d-d^{\prime}-\delta}
\leq
\cr
&\qquad\qquad\qquad
\sum_{k\in\interi^{n_1}}\sum_{k^{\prime}\in\interi^{n_1}}\Bigg[\Bigg(
\frac{|k|\left|g_{k}\right|_{1-d-d^{\prime}}
\left|g^{\prime}_{k^{\prime}}\right|_{1-d^{\prime}}}{(d+\delta)\rho}
\cr
&\qquad\qquad\qquad
\phantom{\sum_{k\in\interi^{n_1}}\sum_{k^{\prime}\in\interi^{n_1}}\Bigg[\Bigg(}
+\frac{\left|g_{k}\right|_{1-d-d^{\prime}}
|k^{\prime}|\left|g^{\prime}_{k^{\prime}}\right|_{1-d^{\prime}}}{\delta\rho}
\Bigg)\ee^{(|k|+|k^{\prime}|)(1-d-d^{\prime}-\delta)\sigma}\Bigg]
\cr
&\qquad\qquad\qquad
\le\frac{2}{e\rho\sigma}\frac{1}{(d+\delta)\delta}
\left\|g\right\|_{1-d-d^{\prime}}\left\|g^{\prime}\right\|_{1-d^{\prime}}\ ,
}
\formula{frm:stimavel}
$$
being $g_{k^{\prime}}^{\prime}=g_{k^{\prime}}^{\prime}(p,z,\imunit\bar
z)$ the terms appearing in the expansion for $g^{\prime}$ analogous
to~\frmref{frm:funz}; moreover, in the inequalities above we used the
first estimate in~\frmref{frm:stime-Cauchy} and the elementary one
$a\ee^{-ab}\le 1/(\ee b)\,$, holding true for any positive real values
of $a$ and $b\,$.

Let us now focus on the remaining terms of the Poisson bracket.  For
each point $(p,z,\imunit\bar z)\in\GGG_{(1-d-d^{\prime}-\delta)\rho}
\times \BBB_{(1-d-d^{\prime}-\delta)R}$ and for all pairs of vectors
$k,\,k^{\prime}\in\interi^{n_1}$, we introduce an auxiliary function
$$
G_{(p,z,\imunit\bar z);k,k^{\prime}}(t) =
g_k\Bigl(p,\,z-t\parder{g_{k^{\prime}}^{\prime}}{(\imunit{\bar z})},
\,\imunit {\bar z}+t\parder{g_{k^{\prime}}^{\prime}}{z}\Bigr)\ ,
\formula{def:aux-fun-G}
$$
where we use the multi-index notation also for the derivatives,
so that for instance ${\partial g_{k^{\prime}}^{\prime}}/{\partial z}=
\big({\partial g_{k^{\prime}}^{\prime}}/{\partial z_1},\,
\ldots,\,{\partial g^{\prime}}/{\partial z_{n_2}}\big)\,$. Since $g_k$
is analytic on $\GGG_{(1-d-d^{\prime})\rho}\times
\BBB_{(1-d-d^{\prime})R}\,$, then the auxiliary function
$G_{(p,z,\imunit\bar z);k,k^{\prime}}$ is certainly analytic for
$|t|\le \bar t\,$, with
$$
{\bar t}=\frac{\delta R}{{\displaystyle \max_{1\le j\le n_2}}
\left\{\left|\parder{g_{k^{\prime}}^{\prime}}{z_j}\right|_{1-d-d^{\prime}-\delta}\,,\,
\left|\parder{g_{k^{\prime}}^{\prime}}{(\imunit{\bar z}_j)}
\right|_{1-d-d^{\prime}-\delta}\right\}\ . }
\formula{def:disco-analit-G}
$$
Thus, the definitions~\frmref{def:aux-fun-G}--\frmref{def:disco-analit-G} and the Cauchy's
estimate for the derivative of the auxiliary function ensure that
$$
\big|\left\{g_{k},g_{k^{\prime}}^{\prime}\right\}\big|_{1-d-d^{\prime}-\delta}\le
\left| \frac{\der}{\der t} G_{(p,z,\imunit\bar z);k,k^{\prime}}(t)\Bigm|_{t=0}
\right|_{1-d-d^{\prime}-\delta}
\le
\frac{\left|g_{k}\right|_{1-d-d^{\prime}}}{\bar t}\ .
\formula{frm:stima-par-Pois-due-singoli-modi-Fourier}
$$
Using the definition of the norm~\frmref{def:norma}, the previous
inequalities~\frmref{def:disco-analit-G}--\frmref{frm:stima-par-Pois-due-singoli-modi-Fourier}
and the second one in~\frmref{frm:stime-Cauchy}, we get
$$
\Bigg\|\sum_{j=1}^{n_2}\left(
\frac{\partial g}{\partial (\imunit{\bar z}_j)}
\frac{\partial g^{\prime}}{\partial z_j}-
\frac{\partial g}{\partial z_j}
\frac{\partial g^{\prime}}{\partial (\imunit{\bar z}_j)}
\right)\Bigg\|_{1-d-d^{\prime}-\delta}\le
\frac{1}{R^2}
\frac{\left\|g\right\|_{1-d-d^{\prime}}\left\|g^{\prime}\right\|_{1-d^{\prime}}}
{(d+\delta)\delta}
\formula{frm:stimasec}
$$
By joining \frmref{frm:stimavel} and \frmref{frm:stimasec}, one
immediately obtains inequality~\frmref{frm:stimapois}.
\endproof

The estimate~\frmref{frm:stimalie} on the multiple Poisson brackets
can be now easily verified following, for instance, the proof scheme
of lemma~4.2 in~\dbiref{Giorgilli-2003}.
\prooftx{of lemma~\lemref{lem:stima-termini-serie-Lie}}
For $j\ge 1$ let us choose $\delta=d/j$ as small step-size
restriction of the analyticity domain. Thus, we
obtain~\frmref{frm:stimalie} by writing the following chain of
inequalities:
$$
\eqalign{
\left\|\Lie^{j}_{\XXX}g\right\|_{1-d-d^{\prime}}
&\leq\left(\frac{2}{e\rho\sigma}+\frac{1}{R^{2}}\right)
\frac{1}{j\delta^{2}}\|\XXX\|_{1-d^{\prime}}
\left\|\Lie^{j-1}_{\XXX}g\right\|_{1-d^{\prime}-(j-1)\delta}
\cr
&\leq\ldots
\cr
&\leq\frac{j!}{\ee^{2}}\left(\frac{2e}{\rho\sigma}+\frac{e^2}{R^{2}}\right)^{j}\frac{1}{(d^2)^j}\|\XXX\|^{j}_{1-d^{\prime}}\|g\|_{1-d^{\prime}}\ ,
\cr
}
\formula{frm:prova-stimalie}
$$
where we repeatedly applied lemma~\lemref{lem:stima-sigola-Pois-par};
in particular, in the first row, we used it by replacing $g\,$,
$g^{\prime}$ and $d$ with $\Lie^{j-1}_{\XXX}g\,$, $\XXX$ and
$(j-1)\delta\,$, respectively; in the last row
of~\frmref{frm:prova-stimalie}, we used also the trivial inequality
$j^{j}\leq j!\ee^{j-1}$, holding true for $j\ge 1\,$.
\endproof

\prooftx{of lemma~\lemref{lem:stima-termini-trasformata-Lie}}
The formal procedure defining all the sequence of generating functions
$\XXX=\{\XXX_{j}\}_{j\ge 1}$ is described in
subsection~\sssref{sss:diagonal}. In the framework of the present
lemma, let us consider
equations~\frmref{frm:diagonalizza-con-telchi}--\frmref{frm:espansione-D2r},
replacing the symbols $\epsilon\,$, $\Dscr_2^{(r;j)}$, $Z_j^{(r)}$,
$g_{1}^{(r)}$, $\Omega^{(r-1)}_{i}$, $\EEE_{j}^{(r)}$,
$\Psi_{j}^{(r)}$ with $1\,$, $\XXX_j\,$, $Z_j\,$, $g^{\prime}\,$,
$\Xi_{i}\,$, $\EEE_{j}\,$, $\Psi_{j}\,$, respectively. Let us recall
that  by construction $\XXX_j\,$, $Z_j\,$, and
$\Psi_{j}$, for $j\ge 1\,$, belong to $\Pset_{2,0}\cap\PPset_{0,2,0}$ (i.e., the set of
functions that are in $\Pset_{2,0}$ and depend just on the conjugate
canonical coordinates $(z,\imunit\bar z)$). Let us consider a generic
function $g^{\prime\prime}\in\PPset_{0,2,0}\,$; since it is a
homogeneous quadratic polynomial, its norm is well defined on any
domain of type $\DDD_{\rho,R,\sigma}$ and it is ``scale invariant'',
i.e.,
$\|g^{\prime\prime}\|_{d^{\prime\prime}}=(d^{\prime\prime})^2\|g^{\prime\prime}\|_{1}$
for $d^{\prime\prime}\in\reali_{+}\,$. It is convenient to rewrite the
estimate of the Poisson bracket for the special case of functions
belonging to $\PPset_{0,2,0}$ as follows:
$$
\left\|\Lie_{\XXX_j}g^{\prime\prime}\right\|_{1-d^{\prime}}
\le 16\,
\frac{\left\|\XXX_j\right\|_{1-d^{\prime}}\left\|g^{\prime\prime}\right\|_{1-d^{\prime}}}
{(1-d^{\prime})^2 R^2}\qquad\hbox{for }j\ge 1\ .
\formula{frm:stimasec-pol-quadr}
$$
In order to verify the inequality above, it is enough to rewrite the
estimate~\frmref{frm:stimasec} in the special case with $d=0\,$,
$\delta=(1-d^{\prime})/2$ and use the scale invariance of the
norm. Let us stress that the upper bound provided
in~\frmref{frm:stimasec-pol-quadr} does not require any restriction of
the domain. Starting from the homological equation for the
determination of $\XXX_j$ and its solution (see
formul{\ae}~\frmref{frm:D2r-di-passo-j}
and~\frmref{frm:espansione-D2r}, respectively), one has
$$
\left\|\XXX_j\right\|_{1-d^{\prime}}\le
\frac{\left\|\Psi_j\right\|_{1-d^{\prime}}}{\Xi^*}\ ,
\qquad
\left\|Z_j\right\|_{1-d^{\prime}}\le \left\|\Psi_j\right\|_{1-d^{\prime}}
\qquad\hbox{for }j\ge 1\ ,
\formula{frm:stima-elementare-X_j-Z_j}
$$
where we also used hypothesis~(i) of the present lemma.  Using 
inequalities~\frmref{frm:stimasec-pol-quadr}--\frmref{frm:stima-elementare-X_j-Z_j},
the definition of the operator $\EEE_{j}$ in~\frmref{frm:def-telchi}
and the fact that for $\Psi_j$ which is analogous
to~\frmref{frm:def-Psi_j^(r)}, one can easily justify the following
recursive estimates involving just the sequence of functions
$\{\Psi_{j}\}_{j\ge 1}$ and $\{\EEE_{j}\,g^{\prime}\}_{j\ge 0}\,$:
$$
\eqalign{
\left\|\Psi_{j}\right\|_{1-d^{\prime}}
&\le
\left\|\EEE_{j-1}\,g^{\prime}\right\|_{1-d^{\prime}}
\cr
\phantom{\left\|\Psi_{j}\right\|_{1-d^{\prime}}}&\phantom{\le}
+\sum_{i=1}^{j-1}\left[\frac{16\,i\left\|\Psi_i\right\|_{1-d^{\prime}}}
{j(1-d^{\prime})^2 \Xi^* R^2}
\left(\left\|\Psi_{j-i}\right\|_{1-d^{\prime}}+
\left\|\EEE_{j-i-1}\,g^{\prime}\right\|_{1-d^{\prime}}\right)\right]\ ,
\cr
\left\|\EEE_{j}\,g^{\prime}\right\|_{1-d^{\prime}}
&\le \sum_{i=1}^{j}\left[\frac{16\,i\left\|\Psi_i\right\|_{1-d^{\prime}}}
{j(1-d^{\prime})^2 \Xi^* R^2}
\left\|\EEE_{j-i}\,g^{\prime}\right\|_{1-d^{\prime}}\right]\ ,
\cr
}
\formula{frm:stima-ricorsiva-Psi_j-E_j}
$$
holding true for $j\ge 1\,$.  Proceeding by induction, one can easily
get the estimate
$$
\max\left\{
\left\|\EEE_{j-1}\,g^{\prime}\right\|_{1-d^{\prime}}\,,\,
\frac{1}{2}\left\|\Psi_{j}\right\|_{1-d^{\prime}}\right\}
\le
\frac{\lambda_{j}}{j}\left(\frac{2^{7}\left\|g^{\prime}\right\|_{1-d^{\prime}}}
{(1-d^{\prime})^2 \Xi^* R^2}\right)^{j-1}\left\|g^{\prime}\right\|_{1-d^{\prime}}
\ ,
\formula{frm:stima-intermedia-Psi_j-E_j}
$$
for $j\ge 1\,$ where the starting point is given by
$\|\EEE_{0}\,g^{\prime}\|_{1-d^{\prime}}=\|g^{\prime}\|_{1-d^{\prime}}$
and $\{\lambda_j\}_{j\ge 1}$ is the famous Catalan sequence, i.e.,
$$
\lambda_{1}=1\ ,
\qquad
\lambda_{j}=\sum_{i=1}^{j-1}\lambda_{i}\lambda_{j-i}\ .
\formula{frm:def-Catalan}
$$
Using the estimate $\lambda_{j}\le 4^{j-1}$ and
inequalities~\frmref{frm:stima-elementare-X_j-Z_j}
and~\frmref{frm:stima-intermedia-Psi_j-E_j}, for $j\ge 1$ we
obtain
$$
\max\left\{\Xi^*\left\|\XXX_j\right\|_{1-d^{\prime}}\,,\,
\left\|Z_j\right\|_{1-d^{\prime}}\right\}\le
\frac{2\,\left\|g^{\prime}\right\|_{1-d^{\prime}}}{j}
\left(\frac{2^{9}\left\|g^{\prime}\right\|_{1-d^{\prime}}}
{(1-d^{\prime})^2 \Xi^* R^2}\right)^{j-1}\ .
\formula{frm:stima-finale-X_j-Z_j}
$$

The estimate above on the generating function $\XXX_j\,$, the
smallness condition on $\epsilon^{\star}_{{\rm diag}}$ in
hypothesis~(ii) and a straightforward adaptation of proposition~4.3
in~\dbiref{Giorgilli-2003} allow us to prove that the Lie transform
operator $\telchi_{\XXX}$ properly defines a linear canonical
transformation. Actually, such an adaptation is needed because in the
framework of section~4 of~\dbiref{Giorgilli-2003} action-angle
variables have been adopted, while in the present context complex canonical
coordinates of polynomial type are used.%%  This fact just requires to
%% suitably replace the estimates for the Poisson brackets in
%% proposition~4.3 with inequality~\frmref{frm:stimasec}.

The first inequality in~\frmref{frm:stima-finale-E_j-Z_j} is a result
of the modified version of proposition~4.3 in~\dbiref{Giorgilli-2003},
while the second estimate immediately follows
from~\frmref{frm:stima-finale-X_j-Z_j} when $j\ge 2\,$; in the special
case with $j=1\,$, that estimate still holds true because
from~\frmref{frm:stima-elementare-X_j-Z_j}--\frmref{frm:stima-ricorsiva-Psi_j-E_j}
it follows that
$\|Z_1\|_{1-d^{\prime}}\le\|\Psi_1\|_{1-d^{\prime}}\le\|g^{\prime}\|_{1-d^{\prime}}\,$.
\endproof

\subsection{app:indici}{On the sets of indexes}
We report here the proofs of
lemmas~\lemref{nrmlie.44}--\lemref{nrmlie.27}.

\prooftx{of lemma~\lemref{nrmlie.44}}
The claim~(i) is a trivial consequence of the definition.

\noindent
(ii) For each fixed value of $s>0$ and $1\le k\le\lfloor
s/2\rfloor\,$, we have to determine the cardinality of the set
$\MMM_{k,s}=\{m\in\naturali:\ 2\le m\le s\,,\ \lfloor s/m\rfloor=k\}$.
For this purpose, we use the obvious inequalities
$$
\Biggl\lfloor\frac{s}{\lfloor s/k\rfloor}\Biggr\rfloor \geq k
\quad\hbox{and}\quad
\Biggl\lfloor\frac{s}{\lfloor s/k\rfloor+1}\Biggr\rfloor \lt k\ .
$$
After having rewritten the same relations with $k+1$ in place of
$k\,$, one immediately realizes that a index $m\in \MMM_{k,s}$ {\sl if
and only if} $m\le\lfloor s/k\rfloor$ {\sl and} $m\ge \lfloor
s/(k+1)\rfloor+1\,$, therefore
$\#\MMM_{k,s}=\bigl\lfloor\frac{s}{k}\bigr\rfloor
-\bigl\lfloor\frac{s}{k+1}\bigr\rfloor\,$.

\noindent
(iii)~Since $r\le s\,$, the definition in~\frmref{nrmlie.45} implies
that neither $\{r\}\cup I^*_r\cup I^*_s$ nor $I^*_{r+s}\,$ can include
any index exceeding $\bigl\lfloor(r+s)/2\bigr\rfloor\,$.  Thus, let us
define some finite sequences of non-negative integers as follows:
$$
\vcenter{\openup1\jot\halign{
 \hbox {\hfil $\displaystyle {#}$}
&\hbox {\hfil $\displaystyle {#}$\hfil}
&\hbox {$\displaystyle {#}$\hfil}
&\hbox to 2 ex{\hfil$\displaystyle {#}$\hfil}
&\hbox {\hfil $\displaystyle {#}$}
&\hbox {\hfil $\displaystyle {#}$\hfil}
&\hbox {$\displaystyle {#}$\hfil}\cr
R_k &= &\#\bigl\{j\in I^*_{r}\>:\> j\le k\bigr\}\ ,
& &S_k &= &\#\bigl\{j\in I^*_{s}\>:\> j\le k\bigr\}\ ,
\cr
M_k &= &\#\bigl\{j\in \{r\}\cup I^*_r\cup I^*_s\>:\> j\le k\bigr\}\ ,
& &N_k &= &\#\bigl\{j\in I^*_{r+s}\>:\> j\le k\bigr\}\ ,
\cr
}}
$$
where the integer index $k$ ranges in
$\big[1,\,\lfloor(r+s)/2\rfloor\big]\,$. When $k\lt r\,$, the
property~(ii) of the present lemma allows us to
write
$$
R_k = r - \Bigl\lfloor \frac{r}{k+1}\Bigr\rfloor\ ,\quad
S_k = s - \Bigl\lfloor \frac{s}{k+1}\Bigr\rfloor\ ,\quad
N_k = r+s - \Bigl\lfloor \frac{r+s}{k+1}\Bigr\rfloor\ ;
$$
using the elementary estimate $\lfloor x\rfloor + \lfloor
y\rfloor \leq \lfloor x+y\rfloor\,$, from the equations above it
follows that $M_k \geq N_k$ for $1\le k<r\,$.  In the remaining
cases, i.e., when $r\le k\le \lfloor(r+s)/2\rfloor\,$, we have that
$$
R_k = r - 1\ ,\quad
S_k = s - \Bigl\lfloor \frac{s}{k+1}\Bigr\rfloor\ ,\quad
N_k = r+s - \Bigl\lfloor \frac{r+s}{k+1}\Bigr\rfloor\ ;
$$
therefore, $M_k=1+R_k+S_k\ge N_k\,$.
\noindent
Since we have just shown that $M_k \geq N_k$ for $1\le
k\le \lfloor(r+s)/2\rfloor\,$, it is now an easy matter to complete
the proof. Let us first imagine to have reordered both the set of
indexes ${r}\cup I^*_r\cup I^*_s$ and $I^*_{r+s}$ in increasing order;
moreover, let us recall that $\#\big(\{r\}\cup I^*_r\cup I^*_s\big)=\#
I^*_{r+s}=r+s-1\,$, in view of the definition in~\frmref{nrmlie.45}.
Thus, since $M_1\ge N_1\,$, every element equal to $1$ in ${r}\cup
I^*_r\cup I^*_s$ has a corresponding index in $I^*_{r+s}$ the value of
which is at least $1\,$. Analogously, since $M_2\ge N_2\,$, every
index $2$ in ${r}\cup I^*_r\cup I^*_s$ has a corresponding index in
$I^*_{r+s}$ which is at least $2\,$, and so on up to $k
=\lfloor(r+s)/2\rfloor\,$. This allows us to conclude that ${r}\cup
I^*_r\cup I^*_s\ilt I^*_{r+s}\,$.
\endproof

%%%
\prooftx{of lemma~\lemref{nrmlie.23}}
The points~(i) and~(ii) of the statement immediately follow from the
definition of $\Jscr_{r,s}\,$.

\noindent
For what concerns (iii), we first remark that
$\#\bigl(\{\min\{r,s\}\}\cup\, I\cup\, I'\bigr) = 1+\#(I)+\#(I') =
r+s-1\,$.  Moreover, after having recalled the definition
in~\frmref{nrmlie.20}, it is easy to verify that $0\le
j\le \min\{r,\lfloor(r+s)/2\rfloor\}\,$, for
$j\in\bigl(\{\min\{r,s\}\}\cup\, I\cup\, I'\bigr)\,$. In order to
complete the proof, now we have to check that the selection rule
$\Smat$ is satisfied. For this purpose, we first remark that
$\big(\{\min\{r,s\}\}\cup I\cup I'\big)\ilt
\big(\{\min\{r,s\}\}\cup I^*_r\cup I^*_s\big)\,$,
because $I\in\Jscr_{r-1,r}$ and $I'\in\Jscr_{r,s}\,$. Therefore,
property~(iii) of lemma~\lemref{nrmlie.44} allows us to conclude that
$\big(\{\min\{r,s\}\}\cup I\cup I'\big)\ilt I^*_{r+s}\,$.\endproof
%%%

\prooftx{of lemma~\lemref{nrmlie.27}}
The point~(i) of the present lemma immediately follows from
property~(ii) of lemma~\lemref{nrmlie.23}. In fact, we can write
$$
T_{r-1,s}
= \max_{I\in\Jscr_{r-1,s}} \,\prod_{j\in I\,,\,j\ge 1}\frac{1}{a_j\delta_j^2} \le
\max_{I\in\Jscr_{r,s}} \,\prod_{j\in I\,,\,j\ge 1}\frac{1}{a_j\delta_j^2}\ ,
$$
because the maximum is evaluated over a larger set of indexes.
Moreover, the equation $T_{r^{\prime},s} = T_{s,s}$ holds true when
$r^{\prime}>s$, as a trivial consequence of the definition
in~\frmref{nrmlie.26} and property~(i) of lemma~\lemref{nrmlie.23}.

\noindent
Concerning the point~(ii), we can evaluate
$T_{r-1,r} T_{r,s} / (a_m\delta_m^2)\,$, where
$m=\min\{r,s\}\,$, as follows:
$$
\eqalign{
\frac{1}{a_m\delta_m^2} T_{r-1,r} T_{r,s}
&= \frac{1}{a_m\delta_m^2}\,
   \left(\max_{I\in\Jscr_{r-1,r}} \,\prod_{j\in I\,,\,j\ge 1}
   \frac{1}{a_j\delta_j^2}\right)
   \left(\max_{I^{\prime}\in\Jscr_{r,s}}\, \prod_{j^{\prime}\in I^{\prime}\,,\,j^{\prime}\ge 1}
   \frac{1}{a_{j^{\prime}}\delta_{j^{\prime}}^2}\right)
\cr
& = \max_{I\in\Jscr_{r-1,r}\,,\,I^{\prime}\in\Jscr_{r,s}}\,
     \prod_{j\in(\{m\}\cup\,I\cup\,I^{\prime})\,,\,j\ge 1}
     \,\frac{1}{a_j\delta_j^2}
\cr
& \le \max_{J\in\Jscr_{r,r+s}} \prod_{j\in J\,,\,j\ge 1}\frac{1}{a_j\delta_j^2}
= T_{r,r+s}\ ,
\cr
}
$$
where the inequality above holds true in view of property~(iii) of
lemma~\lemref{nrmlie.23}.\endproof

\subsection{app:lemmone}{On the main Estimates of the
``Purely Analytical'' Part}
\prooftx{of lemma~\lemref{lem:lemmone}} We will prove by induction the
upper bounds on the terms appearing in the Hamiltonian, as they are
reported in formula~\frmref{stime:f_l^(r,s)-lemmone}.  By the way,
this will allow us to prove also the
estimates~\frmref{stime:generatrici-lemmone} on the generating
functions.

Let us assume that
inequalities~\frmref{stime:f_l^(r,s)-lemmone} are satisfied by
replacing $r$ with $r-1\ge 1\,$, i.e.,
$$
\vcenter{\openup1\jot 
\halign{
$\displaystyle\hfil#$&$\displaystyle{}#\hfil$&$\displaystyle#\hfil$\cr
\|f_{\ell}^{(r-1,s)}\|_{1-d_{r-1}}&\leq \frac{\Ebarra M^{3s-3+\ell}}{2^{\ell}}\,
\frac{T_{r-1,s}^3}{\left(a_{r-1}\delta_{r-1}^2\right)^{\ell}}
\,\nu_{r-1,s}\exp(s\zeta_{r-1})
&\quad{{\hbox{for }0\le\ell\le 2}\,,\ s\ge r\,,}
\cr
\|f_{\ell}^{(r-1,0)}\|_{1-d_{r-1}}&\leq \frac{\Ebarra}{2^{\ell}}\nu_{r-1,0}
&\quad{{\hbox{\rm for}\ \ell\ge 3\,,}}
\cr
\|f_{\ell}^{(r-1,s)}\|_{1-d_{r-1}}&\leq \frac{\Ebarra M^{3s}}{2^{\ell}}
\left(\frac{T_{r-1,s}}{a_m\delta_m^2}\right)^3 \nu_{r-1,s}\exp(s\zeta_{r-1})&
\quad\hbox{for }{\vtop{\hbox{${\ell\ge 3\,,\ s\ge 1\,,}$}
\vskip-2pt\hbox{\hskip-5pt$m=\min\{r-1,s\}\,.$}}}\cr
\cr
}}
\formula{stime:f_l^(r-1,s)-dim-lemmone}
$$
We remark that for $r=1\,$, the upper bounds
at point~(e') of lemma~\lemref{lem:analiticita-espansione-Ham}
can be written as
$$
\|f_{\ell}^{(0,s)}\|_{1-d_{0}}\leq \frac{\Ebarra M^{3s-3}}{2^{\ell}}\,
T_{0,s}^3\,\nu_{0,s}\exp(s\zeta_{0})
\qquad
{{\rm for}\ \ell\ge 0\,,\ s\ge 1\ ,}
\formula{stime:f_l^(0,s)-dim-lemmone}
$$
where we used the definitions in
formul{\ae}~\frmref{frm:dr-and-deltar}--\frmref{frm:seqnu}.  Actually,
the inequality above is the starting point of the inductive argument
and it slightly differs from~\frmref{stime:f_l^(r-1,s)-dim-lemmone},
because in the r.h.s. of~\frmref{stime:f_l^(0,s)-dim-lemmone} there is
not any divisor of type $a_m\delta_m^2\,$. Let us recall that in those
denominators the index $m$ cannot be equal to zero, because of the
definitions in~\frmref{nonres} and~\frmref{frm:dr-and-deltar}. This
fact will force us to somehow distinguish the special case $r=1$ with
respect to the general one.

Let us deal separately with the easy case with $s=0\,$.  Using the
upper bounds at point~(e') of
lemma~\lemref{lem:analiticita-espansione-Ham}, the estimates in the
second row of~\frmref{stime:f_l^(r,s)-lemmone} can be immediately
proved, because
$f_{\ell}^{(r-1,0)}=f_{\ell}^{(\rmI;r,0)}=f_{\ell}^{(\rmII;r,0)}=
f_{\ell}^{(\rmIII;r,0)}=f_{\ell}^{(r,0)}$, $d_r\ge 0$ and
$\nu_{r-1,0}=\nu_{r,0}=1$ for $r\ge 1\,,\,\ell\ge 3\,$, as one can
easily verify by looking at the recursive
formul{\ae}~\frmref{frm:fI}, \frmref{frm:fII}, \frmref{frm:fIII},
\frmref{frm:fr}, \frmref{frm:dr-and-deltar} and~\frmref{frm:seqnu}.

We consider the first stage of the $r$-th normalization step.
By looking at formul{\ae}~\frmref{frm:chi0r}--\frmref{frm:espansione-chi0r}
and using inequalities~\frmref{nonres}
and~\frmref{stime:f_l^(r-1,s)-dim-lemmone}, we obtain
$$
\|\chi_{0}^{(r)}\|_{1-d_{r-1}}\leq \frac{1}{a_r}\|f_{0}^{(r-1,r)}\|_{1-d_{r-1}}
\le M^{3r-3}\,\frac{T_{r-1,r}^3}{a_r}\,\nu_{r-1,r}\exp(r\zeta_{r-1})\ .
$$
The upper bound on the generating function $\chi_{0}^{(r)}$ in
formula~\frmref{stime:generatrici-lemmone} immediately follows from
the inequality above and the definition in~\frmref{eq:def-M}. In the
inductive argument, let us recall that we must replace the estimate on
$f_{0}^{(r-1,r)}$ in~\frmref{stime:f_l^(r-1,s)-dim-lemmone} with that
on $f_{0}^{(0,1)}$ in~\frmref{stime:f_l^(0,s)-dim-lemmone};
nevertheless, also in the case with $r=1\,$, the estimate for
$\chi_{0}^{(1)}$ in formula~\frmref{stime:generatrici-lemmone} can be
easily justified in the same way as before.

We now aim to prove the following estimates on the terms appearing in
the expansion~\frmref{frm:H(I;r)-espansione} of the Hamiltonian
$H^{(\rmI;r)}$:
$$
\vcenter{\openup1\jot 
\halign{
$\displaystyle\hfil#$&$\displaystyle{}#\hfil$&$\displaystyle#\hfil$\cr
\|f_{\ell}^{(\rmI;r,s)}\|_{1-d_{r-1}-\delta_r}&\leq
\frac{\Ebarra M^{3s-3+\ell}}{2^{\ell}}\,
\frac{T_{r,s}^3}{\left(a_r\delta_r^2\right)^{\ell}}
\,\nu_{r,s}^{(\rmI)}\exp(s\zeta_{r-1})
&\qquad\,{\hbox{\rm for }0\le\ell\le 2\,,\ s\ge r\,,}
\cr
\|f_{\ell}^{(\rmI;r,s)}\|_{1-d_{r-1}-\delta_r}&\leq \frac{\Ebarra M^{3s}}{2^{\ell}}
\left(\frac{T_{r,s}}{a_m\delta_m^2}\right)^3
\nu_{r,s}^{(\rmI)}\exp(s\zeta_{r-1})
&\qquad{{\hbox{\rm for}\ \ell\ge 3\,,\ s\ge 1\,,}\atop
{\quad m=\min\{r,s\}\,.}}
\cr
}}
\formula{stime:f_l^(I;r,s)-lemmone}
$$
In formula above, we omitted the inequality
$\|f_{\ell}^{(\rmI;r,0)}\|_{1-d_{r-1}-\delta_r}\le\Ebarra/2^{\ell}$
for $\ell\ge 3\,$, that has been already proved verifying
the estimates in the second row of~\frmref{stime:f_l^(r,s)-lemmone}.
In order to justify the inequalities
in~\frmref{stime:f_l^(I;r,s)-lemmone}, we have to focus on the
recursive definitions in~\frmref{frm:fI}. For $\ell=0$
and $s=r$ we have nothing to do. When $\ell=0$ and $r<s=r+m<2r\,$,
starting from the corresponding estimate
in~\frmref{stime:f_l^(r-1,s)-dim-lemmone}, we can write
$$
\eqalign{
\|f_{0}^{(\rmI;r,r+m)}\|_{1-d_{r-1}-\delta_r}
&\le \Ebarra M^{3(r+m)-3}\,
T_{r-1,r+m}^3\,\nu_{r-1,r+m}\exp\big((r+m)\zeta_{r-1}\big)
\cr
&\le \Ebarra M^{3(r+m)-3}\,T_{r,r+m}^3
\,\nu_{r,r+m}^{(\rmI)}\exp\big((r+m)\zeta_{r-1}\big)\ ,
\cr
}
\formula{frm:stime-f_0^(I;r,r+m)-deduzione}
$$
where we used property~(i) of lemma~\lemref{nrmlie.27} and the
obvious inequality $\nu_{r-1,r+m}\le\nu_{r,r+m}^{(\rmI)}\,$.

Most of the work to verify the
estimates~\frmref{stime:f_l^(I;r,s)-lemmone} has to be done about the
third definition in~\frmref{frm:fI}. There, it is convenient to
consider separately the cases where $s$ is a multiple of the
normalization step $r\,$. Moreover, it is useful to introduce the
sequence of non-negative integer numbers $\{w_{\ell}\}_{\ell\ge 0}$
defined as
$$
w_{\ell}=3-\ell\quad{\rm for}\ 0\le\ell\le 2\ ,
\qquad
w_{\ell}=0\quad\hbox{for } \ell\ge 3\ .
\formula{eq:def-w}
$$
Thus, for $s=mr$ with $m\ge 2$ when $\ell=0$ or
$m\ge 1$ when $\ell\ge 1\,$, one has

$$
\eqalign{
&\|f_{\ell}^{(\rmI;r,mr)}\|_{1-d_{r-1}-\delta_r}
\le \frac{\Ebarra M^{3mr-w_{\ell}}}{2^{\ell}}\exp(mr\zeta_{r-1})\Bigg\{M^{w_{\ell}-2m}
\,\left(\frac{T_{r-1,r}^3}{a_r\delta_r^2}\right)^m\,\nu_{r-1,r}^{m}\nu_{r-1,0}
\cr
&\qquad\qquad\phantom{\le}+\sum_{j=0}^{m-1}\Bigg[
M^{w_{\ell}-2j-w_{\ell+2j}}\,
\bigg(\frac{T_{r-1,r}^3}{a_r\delta_r^2}\bigg)^j
\frac{T_{r-1,(m-j)r}^3}{\left(a_r\delta_r^2\right)^{3-w_{\ell+2j}}}
\,\nu_{r-1,r}^j\nu_{r-1,(m-j)r}\Bigg]\Bigg\}
\cr
&\qquad\qquad\le \frac{\Ebarra M^{3mr-w_{\ell}}}{2^{\ell}}
\nu_{r,mr}^{(\rmI)}\exp(mr\zeta_{r-1})
\cr
&\qquad\qquad\phantom{\le}\cdot\,
\max\left\{\left(\frac{T_{r-1,r}^3}{a_r\delta_r^2}\right)^m\,,\,
\max_{0\le j\le m-1}\left\{
\bigg(\frac{T_{r-1,r}^3}{a_r\delta_r^2}\bigg)^j
\frac{T_{r-1,(m-j)r}^3}{\left(a_r\delta_r^2\right)^{3-w_{\ell+2j}}}
\right\}\right\}
\cr
&\qquad\qquad\le \frac{\Ebarra M^{3mr-w_{\ell}}}{2^{\ell}}
\,\frac{T_{r,mr}^3}{\left(a_r\delta_r^2\right)^{3-w_{\ell}}}
\,\nu_{r,mr}^{(\rmI)}\exp(mr\zeta_{r-1})
\ ,
\cr
}
\formula{frm:stime-f_l^(I;r,mr)-deduzione}
$$
where we used lemma~\lemref{lem:stima-termini-serie-Lie}, the estimate
for $\chi_{0}^{(1)}$ in formula~\frmref{stime:generatrici-lemmone},
the inductive inequalities in~\frmref{stime:f_l^(r-1,s)-dim-lemmone},
the fact that $\{a_r\delta_r^2\}_{r\ge 1}$ is a non-increasing
sequence and $M\ge 1$ (recall the definitions
in~\frmref{nonres}, \frmref{frm:dr-and-deltar} and~\frmref{eq:def-M}),
some elementary properties\footnote{\dag}{Actually, we used two
elementary inequalities: $w_{\ell}\le 2m$ for $\ell=0$ and $m\ge
2$ or $\ell\ge 1$ and $m\ge 1\,$; $w_{\ell}\le
2j+w_{\ell+2j}$ for $j\ge 0\,$. Both immediately follow
from the definition in~\frmref{eq:def-w}.} of the sequence
$\{w_{\ell}\}_{\ell\ge 0}\,$, the definition of
$\{\nu_{r,s}^{(\rmI)}\}_{r\ge 1\,,\,s\ge 0}$ in~\frmref{frm:seqnu} and
the inequality
$$
\max\left\{\left(\frac{T_{r-1,r}^3}{a_r\delta_r^2}\right)^m\,,\,
\max_{0\le j\le m-1}\left\{
\bigg(\frac{T_{r-1,r}^3}{a_r\delta_r^2}\bigg)^j
\frac{T_{r-1,(m-j)r}^3}{\left(a_r\delta_r^2\right)^{3-w_{\ell+2j}}}
\right\}\right\}
\le
\frac{T_{r,mr}^3}{\left(a_r\delta_r^2\right)^{3-w_{\ell}}}
\formula{frm:stima-denominatori-f_l^(I;r,mr)}
$$
holding true for $r\ge 1\,$, for $\ell=0$, $m\ge 2$ or $\ell\ge 1$,
$m\ge 1\,$. By the way, concerning the starting point of the
induction, let us remark that the
inequality~\frmref{frm:stime-f_l^(I;r,mr)-deduzione} holds true also
for $r=1\,$, we just have to replace the first inductive
estimate~\frmref{stime:f_l^(r-1,s)-dim-lemmone} with that
in~\frmref{stime:f_l^(0,s)-dim-lemmone} and use the fact that
$a_1\delta_1^2<1\,$. Thus, we now have to verify the
inequality~\frmref{frm:stima-denominatori-f_l^(I;r,mr)}, so as to
complete the justification
of~\frmref{frm:stime-f_l^(I;r,mr)-deduzione}. For this purpose, once
again, it is convenient to distinguish some sub-cases.

\item{(i)}{For $m=1$ and, then, $\ell\ge 1$,
inequality~\frmref{frm:stima-denominatori-f_l^(I;r,mr)} is rather
obvious, because $T_{r-1,r}\le T_{r,r}$ (recall property~(i) of
lemma~\lemref{nrmlie.27}), $a_r\delta_r^2<1$ (see the definitions
in~\frmref{nonres} and~\frmref{frm:dr-and-deltar}) and $3-w_{\ell}\ge
1$ for $\ell\ge 1\,$.}

\item{(ii)}{For $m\ge 2$ and, then, $\ell\ge 0$, it is convenient to
separately verify that each term appearing in the
l.h.s. of~\frmref{frm:stima-denominatori-f_l^(I;r,mr)} is not greater
than the one in the r.h.s., as we will do at the following
points~(ii.a) and~(ii.b).}

\item{(ii.a)}{Concerning the first term
in~\frmref{frm:stima-denominatori-f_l^(I;r,mr)}, we can write
$$
\left(\frac{T_{r-1,r}^3}{a_r\delta_r^2}\right)^m
\le
\left[\left(\frac{T_{r-1,r}}{a_r\delta_r^2}\right)^{m-1}T_{r,r}\right]^3
\le
T_{r,mr}^3
\le
\frac{T_{r,mr}^3}{\left(a_r\delta_r^2\right)^{3-w_{\ell}}}
\ ,
$$
where we used lemma~\lemref{nrmlie.27} and the elementary inequalities
$3m-3\ge m$ for $m\ge 2\,$, $a_r\delta_r^2<1$ and
$3-w_{\ell}\ge 0$ for $\ell\ge 0\,$.}

\item{(ii.b)}{For $j=0,\,\ldots,\,m-1\,$, we can estimate the
remaining terms of the
l.h.s. of~\frmref{frm:stima-denominatori-f_l^(I;r,mr)} as
$$
\eqalign{
\bigg(\frac{T_{r-1,r}^3}{a_r\delta_r^2}\bigg)^j
\frac{T_{r-1,(m-j)r}^3}{\left(a_r\delta_r^2\right)^{3-w_{\ell+2j}}}
&\le \left[\bigg(\frac{T_{r-1,r}}{a_r\delta_r^2}\bigg)^jT_{r,(m-j)r}\right]^3
\frac{1}{\left(a_r\delta_r^2\right)^{3-2j-w_{\ell+2j}}}
\cr
&\le \frac{T_{r,mr}^3}{\left(a_r\delta_r^2\right)^{3-w_{\ell}}}\ ,
\cr
}
$$
using again lemma~\lemref{nrmlie.27}, $a_r\delta_r^2<1$ and
$w_{\ell}\le 2j+w_{\ell+2j}$ for $j\ge 0\,$.}

\noindent
This complete the justification
of~\frmref{frm:stima-denominatori-f_l^(I;r,mr)} and,
then, of~\frmref{frm:stime-f_l^(I;r,mr)-deduzione}.

Now, we still have to verify the
estimates~\frmref{stime:f_l^(I;r,s)-lemmone}, starting from the third
definition in~\frmref{frm:fI} when $s$ is not a multiple of the
normalization step $r\,$. The case with $0<s<r$ is trivial and can be
treated in a similar way
to~\frmref{frm:stime-f_0^(I;r,r+m)-deduzione}. For $s>r$ let us put
$m=\lfloor s/r\rfloor$ and $s=mr+i\,$; thus, we focus on the third
definition in~\frmref{frm:fI} with $0<i<r$, for $m\ge 2$,
$\ell=0$ or $m\ge 1$, $\ell\ge 1\,$, for which we can write the
following chain of inequalities

$$
\eqalign{
&\|f_{\ell}^{(\rmI;r,s)}\|_{1-d_{r-1}-\delta_r}
\le \frac{\Ebarra M^{3s-w_{\ell}}}{2^{\ell}}\nu_{r,s}^{(\rmI)}\exp(s\zeta_{r-1})
\max\Bigg\{\left(\frac{T_{r-1,r}^3}{a_r\delta_r^2}\right)^{m}
\frac{T_{r-1,i}^3}{\left(a_{i}\delta_{i}^2\right)^{3-w_{\ell+2m}}}\,,
\cr
&\phantom{\|f_{\ell}^{(\rmI;r,s)}\|_{1-d_{r-1}-\delta_r}}
\qquad\qquad\ \max_{0\le j\le m-1}\left\{
\bigg(\frac{T_{r-1,r}^3}{a_r\delta_r^2}\bigg)^j
\frac{T_{r-1,s-jr}^3}{\left(a_r\delta_r^2\right)^{3-w_{\ell+2j}}}
\right\}\Bigg\}
\cr
&\phantom{\|f_{\ell}^{(\rmI;r,s)}\|_{1-d_{r-1}-\delta_r}}\le
\frac{\Ebarra M^{3s-w_{\ell}}}{2^{\ell}}
\,\frac{T_{r,s}^3}{\left(a_r\delta_r^2\right)^{3-w_{\ell}}}
\,\nu_{r,s}^{(\rmI)}\exp(s\zeta_{r-1})
\ ,
\cr
}
\formula{frm:stime-f_l^(I;r,s)-deduzione}
$$
where we proceeded in a similar way as
for~\frmref{frm:stime-f_l^(I;r,mr)-deduzione},
replacing~\frmref{frm:stima-denominatori-f_l^(I;r,mr)} by
$$
\eqalign{
&\max\Bigg\{\left(\frac{T_{r-1,r}^3}{a_r\delta_r^2}\right)^{m}
\frac{T_{r-1,i}^3}{\left(a_{i}\delta_{i}^2\right)^{3-w_{\ell+2m}}}\,,
\cr
&\phantom{\max\Bigg\{}\ \max_{0\le j\le m-1}\left\{
\bigg(\frac{T_{r-1,r}^3}{a_r\delta_r^2}\bigg)^j
\frac{T_{r-1,s-jr}^3}{\left(a_r\delta_r^2\right)^{3-w_{\ell+2j}}}
\right\}\Bigg\}
\le \frac{T_{r,s}^3}{\left(a_r\delta_r^2\right)^{3-w_{\ell}}}
\ ,
\cr
}
\formula{frm:stima-denominatori-f_l^(I;r,s)}
$$
holding true for $\ell=0$, $m\ge 2$ or $\ell\ge 1$, $m\ge 1\,$.
Therefore, in order to complete the justification
of~\frmref{frm:stime-f_l^(I;r,s)-deduzione}, we now have to verify the
inequality~\frmref{frm:stima-denominatori-f_l^(I;r,s)}.  For this
purpose, it is convenient to separately verify that each term
appearing in the l.h.s. of~\frmref{frm:stima-denominatori-f_l^(I;r,s)}
is not greater than the one in the r.h.s.  First, one can easily verify
that
$$
\eqalign{
\left(\frac{T_{r-1,r}^3}{a_r\delta_r^2}\right)^{m}
\frac{T_{r-1,i}^3}{\big(a_{i}\delta_{i}^2\big)^{3-w_{\ell+2m}}}
&\le \left[T_{r-1,r}\bigg(\frac{T_{r-1,r}}{a_r\delta_r^2}\bigg)^{m-1}
\frac{T_{r,i}}{a_{i}\delta_{i}^2}\right]^3
\frac{1}{\left(a_r\delta_r^2\right)^{3-2m}}
\cr
&\le \frac{T_{r,s}^3}{\left(a_r\delta_r^2\right)^{3-w_{\ell}}}
\frac{1}{\left(a_r\delta_r^2\right)^{w_{\ell}-2m}}
\le \frac{T_{r,s}^3}{\left(a_r\delta_r^2\right)^{3-w_{\ell}}}
\ ,
\cr
}
\formula{frm:stima-denominatori-f_l^(I;r,mr+i)-deduzione}
$$
where we used lemma~\lemref{nrmlie.27}, $a_r\delta_r^2<1$ and $w_{\ell}\le 2m$, both for $\ell=0\,$, $m\ge 2$ and $\ell\ge
1\,$, $m\ge 1$. For $j=0,\,\ldots,\,m-1\,$, we can estimate the
remaining terms of the
l.h.s. of~\frmref{frm:stima-denominatori-f_l^(I;r,s)} in a similar way
to~\frmref{frm:stima-denominatori-f_l^(I;r,mr+i)-deduzione}.  This
concludes the justification of
inequalities~\frmref{frm:stima-denominatori-f_l^(I;r,s)}, therefore
also estimates~\frmref{frm:stime-f_l^(I;r,s)-deduzione} and
\frmref{stime:f_l^(I;r,s)-lemmone} are completely verified.

Let us now consider the second stage of the $r$-th normalization step.
After having recalled
formul{\ae}~\frmref{frm:chi1r}--\frmref{frm:espansione-chi1r}, one can
easily justify the upper bound on the generating function
$\chi_{1}^{(r)}$ in formula~\frmref{stime:generatrici-lemmone}, by
using inequalities~\frmref{nonres}
and~\frmref{stime:f_l^(I;r,s)-lemmone} and the definition
in~\frmref{eq:def-M}. Let us remark that the upper bounds
in~\frmref{stime:f_l^(I;r,s)-lemmone} play now the role of inductive
estimates and they also include the case $r=1\,$, which allows
starting the induction.  For what concerns the functions
$f_{\ell}^{(\rmII;r,s)}$ appearing in the expansion of
$H^{(\rmII;r)}$, we can prove analogous estimates to those
in~\frmref{stime:f_l^(I;r,s)-lemmone}, just replacing the upper index
I with II. This can be verified, by starting from the recursive
definitions in~\frmref{frm:fII} and by patiently handling with many
different cases in a similar way to what we have done
for~\frmref{stime:f_l^(I;r,s)-lemmone}.

For what concerns the first part of the third stage of the $r$-th
normalization step, the upper bound on the generating function
$\chi_{2}^{(r)}$ in~\frmref{stime:generatrici-lemmone} can be
proved in a similar way to that for $\chi_{0}^{(r)}$.  By patiently
estimating all terms appearing in the recursive definitions
in~\frmref{frm:fIII}, we can provide the upper bounds
$$
\vcenter{\openup1\jot 
\halign{
$\displaystyle\hfil#$&$\displaystyle{}#\hfil$&$\displaystyle#\hfil$\cr
\|f_{\ell}^{(\rmIII;r,s)}\|_{1-d_{r-1}-3\delta_r}&\leq
\frac{\Ebarra M^{3s-3+\ell}}{2^{\ell}}\,
\frac{T_{r,s}^3}{\left(a_r\delta_r^2\right)^{\ell}}
\,\nu_{r,s}\exp(s\zeta_{r-1})
&\quad{\hbox{\rm for }0\le\ell\le 2\,,\ s\ge r\,,}
\cr
\|f_{\ell}^{(\rmIII;r,s)}\|_{1-d_{r-1}-3\delta_r}&\leq \frac{\Ebarra M^{3s}}{2^{\ell}}
\left(\frac{T_{r,s}}{a_m\delta_m^2}\right)^3
\nu_{r,s}\exp(s\zeta_{r-1})
&
\quad\hbox{for }{\vtop{\hbox{${\ell\ge 3\,,\ s\ge 1\,,}$}
\vskip-2pt\hbox{\hskip-5pt$\hbox{with }\ m=\min\{r,s\}\,.$}}}\cr
\cr
}}
\formula{stime:f_l^(III;r,s)-lemmone}
$$
Since all the estimates concerning both the second stage and the first
part of the third stage of the $r$-th normalization step are similar
to the ones considered in the first stage, we omit all the tedious
calculations necessary to fully
justify~\frmref{stime:f_l^(III;r,s)-lemmone}.

Let us now focus on the terms generated by the diagonalization of the
quadratic normal form part of order $\Oscr(\epsilon^r)\,$.  For $r=1\,$,
since $f_{2}^{(\rmIII;1,1)}=0$ (see~\frmref{eq:f_2^(III;1,1)=0}), then
equations~\frmref{frm:diagonalizza-con-telchi}--\frmref{frm:espansione-D2r}
imply that $\Dscr_2^{(1;j)}=0$ for $j\ge 1\,$; therefore,
$f_{\ell}^{(\rmIII;1,s)}=f_{\ell}^{(1,s)}$ for $0\le\ell\le 2$, $s>1$
or $\ell\ge 3$, $s\ge 0\,$, in view of the definitions
in~\frmref{frm:fr}. Therefore, for what concerns the first
normalization step, the upper bounds
in~\frmref{stime:f_l^(r,s)-lemmone} immediately follow from those
in~\frmref{stime:f_l^(III;r,s)-lemmone} and the fact that
$\zeta_{0}=\zeta_{1}=0$ (see~\frmref{eq:def-succ-zetagreco}).  Let us
now consider the generic case with $r\ge 2\,$. Comparing
formul{\ae}~\frmref{nonres.a}
and~\frmref{frm:diagonalizza-con-telchi}--\frmref{frm:diagonalizza-con-telchi-di-passo-j}
with the hypotheses of
lemma~\lemref{lem:stima-termini-trasformata-Lie}, one immediately
realizes that the smallness condition
$$
\epsilon^{\star}_{{\rm diag}}=\epsilon^{r-1}
\left(\frac{2e^2}{\delta_r^2}+\frac{2^9}{(1-d_{r-1}-3\delta_r)^2}\right)
\frac{\|f_{2}^{(\rmIII;r,r)}\|_{1-d_{r-1}-\delta_r}}{b_{r}\,R^2}\le
\frac{1}{2}\ ,
$$
must be satisfied.  Actually, the stronger inequality
$\epsilon^{\star}_{{\rm diag}}\le 2^{-(r+6)}$ holds true, as it can be
checked by
using~\frmref{frm:cond-eps-trasf-Diag}, \frmref{stime:f_l^(III;r,s)-lemmone}
and the definitions
in~\frmref{nonres}, \frmref{frm:dr-and-deltar}, \frmref{eq:def-M}.
The inequality bounding the effect of the operator $\EEE_{j}^{(r)}$ in
formula~\frmref{stime:generatrici-lemmone} is nothing but the first
estimate in~\frmref{frm:stima-finale-E_j-Z_j} with
$\epsilon^{\star}_{{\rm diag}}=2^{-(r+6)}$.  Applying
lemma~\lemref{lem:stima-termini-trasformata-Lie} to the third equation
in~\frmref{frm:fr}, we obtain
$$
\|f_{\ell}^{(r,s)}\|_{1-d_{r}} \le
\|f_{\ell}^{(\rmIII;r,s)}\|_{1-d_{r-1}-\delta_r} 
\,\exp\left(\frac{2^{-(r+6)}}{1-2^{-(r+6)}}\right)\ ,
$$
for $0\le\ell\le 2$, $s>r\ge 2$ or $\ell\ge 3$, $s\ge 1\,$.  Starting
from the estimates in~\frmref{stime:f_l^(III;r,s)-lemmone}, using
inequality above and the definition in~\frmref{eq:def-succ-zetagreco}
one can complete the justification
of~\frmref{stime:f_l^(r,s)-lemmone}.  Recall that we have already
considered the case $s=0$ at the beginning of the proof.

In order to complete the proof of the lemma, we have to evaluate the
variations of the frequencies induced by the $r$-th normalization
step.  Starting from equation~\frmref{chgfreq.veloci},
using~\frmref{frm:dr-and-deltar}, \frmref{stime:f_l^(III;r,s)-lemmone}
and the first Cauchy inequality in~\frmref{frm:stime-Cauchy}, we
obtain
$$
\max_{1\le j\le n_1}|\omega_{j}^{(r)}-\omega_{j}^{(r-1)}|\le
\epsilon^r\frac{\Ebarra}{3\,\rho} M^{3r-1}
\frac{T_{r,s}^3}{\left(a_r\delta_r^2\right)^{2}}
\nu_{r,r}\exp(r\zeta_{r-1})\ .
\formula{frm:stima-variazione-omeghini}
$$
Analogously, starting from~\frmref{chgfreq.lente},
using~\frmref{frm:dr-and-deltar}, \frmref{eq:def-succ-zetagreco},
\frmref{stime:f_l^(III;r,s)-lemmone}, the second
estimate (again with $\epsilon^{\star}_{{\rm diag}}=2^{-(r+6)}$)
in~\frmref{frm:stima-finale-E_j-Z_j} and twice the second Cauchy
inequality in~\frmref{frm:stime-Cauchy}, we get
$$
\max_{1\le j\le n_2}|\Omega_{j}^{(r)}-\Omega_{j}^{(r-1)}|\le
\epsilon^{r-1}\frac{8\,\Ebarra}{9\,R^2} M^{3r-1}
\frac{T_{r,s}^3}{\left(a_r\delta_r^2\right)^{2}}
\nu_{r,r}\exp(r\zeta_{r})\ .
\formula{frm:stima-variazione-omegoni}
$$
By using inequality $a_r\delta_r^2<1$ and the definition
in~\frmref{eq:def-M}, one can easily verify that both the
estimates~\frmref{frm:stima-variazione-omeghini}
and~\frmref{frm:stima-variazione-omegoni} are gathered
in~\frmref{frm:stima-variazione-frequenze}.
\endproof

\subsection{app:seq}{Estimates of some special numerical sequences}
We report here the proofs of
lemmas~\lemref{nrmlie.42}--\lemref{lem:nuest}.

\prooftx{of lemma~\lemref{nrmlie.42}}
Since $a_s\delta_s^2<1$ (see~\frmref{nonres}
and~\frmref{frm:dr-and-deltar}), it is enough to prove the second part
of the inequality stated in the lemma, i.e.,
$T_{r,s}/(a_s\delta_s^2)\le A^s e^{s\Gamma}$ for $r\ge 1\,,\ s\ge
1\,$.  Starting from the definition in~\frmref{nrmlie.26}, using
properties~(i) and~(ii) of lemma~\lemref{nrmlie.23}, the selection
rule~$\Smat$ and the fact that the sequence $\{a_s\delta_s^2\}_{s\ge
1}$ is decreasing, we get
$$
\frac{T_{r,s}}{a_s\delta_s^2}
= \frac{1}{a_s\delta_s^2}
   \max_{I\in\Jscr_{s,s}} \prod_{j\in I\,,\,j\ge 1} \frac{1}{a_j\delta_j^2}
\le\, \prod_{{j\in\{s\}\cup I^*_s}\,,\,{j\ge 1}}\, \frac{1}{a_j\delta_j^2}\ .
$$
It is convenient to evaluate the logarithm of
$T_{r,s}/(a_s\delta_s^2)\,$; thus, starting from the estimate above,
we can write
$$
\eqalign{
\log \frac{T_{r,s}}{a_s\delta_s^2} &\le -\log(a_s\delta_s^2)
-\sum_{k=1}^{\lfloor s/2\rfloor}\Bigl(\Bigl\lfloor\frac{s}{k}\Bigr\rfloor
-\Bigl\lfloor\frac{s}{k+1}\Bigr\rfloor\Bigr) \log(a_k\delta_k^2)
\cr
&\le-\sum_{k=1}^{s}\Bigl(\Bigl\lfloor\frac{s}{k}\Bigr\rfloor
-\Bigl\lfloor\frac{s}{k+1}\Bigr\rfloor\Bigr) (\log a_k + 2\log \delta_k)
\cr
&\le - s \sum_{k\ge 1} \frac{\log a_k + 2\log \delta_k}{k(k+1)}
= s\biggl(\Gamma -\sum_{k\ge 1} \frac{2\log \delta_k}{k(k+1)}\biggr)
\ ,
\cr
}
\formula{frm:deduzione-stima-sulle-T}
$$
where we used properties~(i) and~(ii) of lemma~\lemref{nrmlie.44}, the
fact that the sequence $\{a_s\delta_s^2\}_{s\ge 1}$ is decreasing and
the condition~$\tauvet$ in~\frmref{frm:Ctau}.  Thus,
using~\frmref{frm:dr-and-deltar}, one can easily get
$$
\eqalign{
-\sum_{k\ge 1} \frac{2\log \delta_k}{k(k+1)} &=
2\sum_{k\ge 1}\frac{\log\frac{8\pi^2}{3} + 2\log k}{k(k+1)}
\cr
&< 2\log\frac{8\pi^2}{3}+
4\left(\frac{\log 2}{6}+\int_{2}^{\infty}\frac{\log x\,\diff x}{x^2}\right)
\cr
&< 15\log 2\ ,
\cr
}
$$
where we explicitly used the relation $\sum_{k\ge 1} 1/[k(k+1)]=1$.
Replacing the estimate above
into~\frmref{frm:deduzione-stima-sulle-T}, we conclude the proof.
\endproof

\prooftx{of lemma~\lemref{lem:nuest}}
First, it is convenient to replace the definition
in~\frmref{frm:seqnu} for the sequence $\{\nu_{r,s}\}_{r\ge 0\,,\,s\ge
0}$ with a closed formula, avoiding to introduce
$\{\nu_{r,s}^{(\rmI)}\}_{r\ge 1\,,\,s\ge 0}$ and
$\{\nu_{r,s}^{(\rmII)}\}_{r\ge 1\,,\,s\ge 0}\,$.  Thus, let us remove
the symbol $\nu^{(\rmII)}$, by writing
$$
\eqalign{
\nu_{r,s}&=
\sum_{j=0}^{\lfloor s/r \rfloor}\bigl(\nu_{r,r}^{(\rmI)}+
\nu_{r,r}^{(\rmI)}\nu_{r,0}^{(\rmI)}\bigr)^{j}
\sum_{i=0}^{\lfloor s/r \rfloor-j} (\nu_{r,r}^{(\rmI)})^{i}\nu_{r,s-(i+j)r}^{(\rmI)}
=\sum_{j=0}^{\lfloor s/r \rfloor} (2\nu_{r,r}^{(\rmI)})^{j}
\sum_{i=j}^{\lfloor s/r \rfloor} (\nu_{r,r}^{(\rmI)})^{i-j}\nu_{r,s-ir}^{(\rmI)}
\cr
&=\sum_{i=0}^{\lfloor s/r \rfloor} (\nu_{r,r}^{(\rmI)})^{i}\nu_{r,s-ir}^{(\rmI)}
\sum_{j=0}^{i} 2^{j}
=\sum_{i=0}^{\lfloor s/r \rfloor} \left( 2^{i+1} -1 \right)
(\nu_{r,r}^{(\rmI)})^{i}\nu_{r,s-ir}^{(\rmI)}\ .
\cr
}
$$
Analogously, we can eliminate the occurrence of $\nu^{(\rmI)}$, by
writing
$$
\eqalign{
\nu_{r,s}&=\sum_{j=0}^{\lfloor s/r \rfloor} 2^j\left( 2^{j+1} -1 \right)\nu_{r-1,r}^{j}
\sum_{i=j}^{\lfloor s/r \rfloor} \nu_{r-1,r}^{i-j}\nu_{r-1,s-ir}
\cr
&=\sum_{i=0}^{\lfloor s/r \rfloor} \nu_{r-1,r}^{i}\nu_{r-1,s-jr}
\sum_{j=0}^{i}\left(2^{2j+1}-2^{j}\right)=
\sum_{i=0}^{\lfloor s/r \rfloor} \theta_i\nu_{r-1,r}^{i}\nu_{r-1,s-ir}\ ,
\cr
}
\formula{frm:deduzione-nuova-nu-ricorsiva}
$$
where we introduced the shorthand notation
$$
\theta_{i} = \sum_{j=0}^{i}\left(2^{2j+1}-2^{j}\right)=
\frac{2}{3}\left(2^{2(i+1)} -1\right) -2^{i+1} +1\ .
$$
From the definition above we can immediately verify that
$$
\theta_{0} = 1\ ,
\qquad
\theta_{1}=7\ ,
\qquad
\frac{1}{3}2^{2(j+1)}\leq \theta_{j}\leq\frac{2}{3}2^{2(j+1)}
\ \hbox{ for } j\geq 1\ .
\formula{frm:basics-succ-thetaj}
$$
Using such basic properties of the sequence $\{\theta_{j}\}_{j\ge 0}\,$,
one can easily gets
$$
\theta_{j+1} \leq 8 \theta_{j}\qquad\hbox{for } j\geq0\ .
\formula{frm:stimathetaj}
$$

Let us recall that combining~\frmref{frm:seqnu} with the recursive
equation~\frmref{frm:deduzione-nuova-nu-ricorsiva}, we can provide a
new and more compact definition of the sequence $\{\nu_{r,s}\}_{r\ge
0\,,\,s\ge 0}\,$, as
$$
\nu_{0,s} = 1
\quad\hbox{for } s\ge 0\ ,
\qquad
\nu_{r,s}=\sum_{j=0}^{\lfloor s/r \rfloor} \theta_j\nu_{r-1,r}^{j}\nu_{r-1,s-jr}
\quad\hbox{for } r\ge 1\,,\ s\ge 0\ .
\formula{frm:seqnu-nuova}
$$
As an immediate consequence, we remark that
$$
\nu_{0,s}\le\nu_{1,s}\le\ldots\le\nu_{s,s}=\nu_{s+1,s}=\ldots\ .
\formula{frm:succ-nu-proprieta-i}
$$
Moreover, since
$\nu_{r,r}=\theta_{0}\nu_{r-1,r}+\theta_{1}\nu_{r-1,r}\,$,
$\theta_{0}=1$ and $\theta_{1}=7$ we have
$$
\nu_{r,r}=8\nu_{r-1,r}\qquad\hbox{ for }r\ge 1\ .
\formula{frm:succ-nu-proprieta-ii}
$$
The following chains of inequalities allow us to justify other
useful properties of the sequence $\{\nu_{r,s}\}_{r\ge 0\,,\,s\ge
0}\,$.  Indeed, starting from~\frmref{frm:seqnu-nuova}, for $r\ge
2\,,\ s>r\,$, we can write
$$
\eqalign{
\nu_{r,s}&=\nu_{r-1,s}+
\nu_{r-1,r}\sum_{j=0}^{\lfloor s/r \rfloor-1}\theta_{j+1}\nu_{r-1,r}^{j}\nu_{r-1,s-r-jr}
\cr
&\leq\nu_{r-1,s}+8\nu_{r-1,r}
\sum_{j=0}^{\lfloor s/r \rfloor-1}\theta_{j}\nu_{r-1,r}^{j}\nu_{r-1,s-r-jr}
\cr
&\leq\nu_{r-1,s}+8\nu_{r-1,r}\nu_{r-1,s-r}
\leq\nu_{r-1,s}+\nu_{r,r}\nu_{s-r,s-r}\ ,
\cr
}
\formula{frm:succ-nu-proprieta-iv}
$$
where we used inequality~\frmref{frm:stimathetaj} and
equation~\frmref{frm:succ-nu-proprieta-ii}.  In a similar way, for
$r=1$ (and, again, $s>r$) we can provide a more accurate estimate,
i.e.,
$$
\eqalign{
\nu_{1,s}&=\nu_{0,s}+\nu_{0,1}\sum_{j=0}^{s-1}\theta_{j+1}\nu_{0,1}^{j}\nu_{0,s-1-j}
\cr
&\leq(1+\theta_{1})\nu_{0,s-1}+
8\sum_{j=1}^{s-1}\theta_{j}\nu_{0,1}^{j}\nu_{0,s-1-j}
\leq8\nu_{0,s-1}\leq\nu_{s-1,s-1}\ ,
\cr
}
\formula{frm:succ-nu-proprieta-iii}
$$
where some particular values of the sequences involved have been
inserted (namely, $\theta_{1}=7$ and $\nu_{0,s}=1$ for $s\ge 0$). We
aim to control all the sequence $\{\nu_{r,s}\}_{r\ge 0\,,\,s\ge 0}\,$;
thus, looking at formula~\frmref{frm:succ-nu-proprieta-i}
one immediately realizes that it is enough to provide an upper bound
on the diagonal elements, for which we can write
$$
\eqalign{
\nu_{r,r}&=8\nu_{r-1,r}\leq8\nu_{r-2,r}+8\nu_{r-1,r-1}\nu_{1,1}
\leq\ldots
\cr
&\leq8\nu_{1,r}+8\left( \nu_{2,2}\nu_{r-2,r-2}+\ldots+\nu_{r-1,r-1}\nu_{1,1}\right)
\leq8\sum_{j=1}^{r-1}\nu_{j,j}\nu_{r-j,r-j}\ ,
\cr
}
\formula{frm:nurr}
$$
for $r\ge 2$, where we orderly used all the properties described by
formul{\ae}~\frmref{frm:succ-nu-proprieta-ii}--\frmref{frm:succ-nu-proprieta-iii}.

Using the fact that $\nu_{1,1}=8$ (see~\frmref{frm:seqnu-nuova}
and~\frmref{frm:succ-nu-proprieta-ii}) and the recursive inequality
for the diagonal elements in~\frmref{frm:nurr}, by induction one can
easily verify that
$$
\nu_{r,r}\leq\frac{64^r}{8}\lambda_{r}
\qquad
\hbox{for } r\ge 1\ ,
$$
being $\{\lambda_r\}_{r\ge 1}$ the Catalan sequence, whose definition
is recalled in~\frmref{frm:def-Catalan}. By combining the information
provided by formula~\frmref{frm:succ-nu-proprieta-i}, the inequality
above and the well known upper bound $\lambda_{r}\leq4^{r-1}$, one can
fully justify the statement.
\endproof

\subsection{app:measure}{On the Geometry of the Resonant Regions}

\prooftx{of proposition~\proref{prop:geometrico}}
The proof proceeds by induction.  However, in order to highlight the
key points, we first describe in detail the first two steps.  By
hypotheses, $\Omega^{(0)}(\omega)$ is an analytic function on the
complex extended domain $\Wscr^{(0)}_{h_0}$ and its Jacobian is
uniformly bounded in $\Wscr^{(0)}_{h_0}\,$, namely
$\big|\partial\Omega^{(0)}/\partial\omega\big|_{\infty;\Wscr^{(0)}_{h_0}}\leq
J_0\,$, where $\big|\cdot\big|_{\infty;\Wscr^{(0)}_{h_0}}$ is defined
in~\frmref{def:cost-Lipschitz}. Thus, starting from the
inequality at point~(a') of
lemma~\lemref{lem:analiticita-espansione-Ham}, if $h_0$ is so small
that
$$
h_0 \leq \min\left\{\frac{1}{K+2J_0\epsilon}
\frac{\gamma}{2K^{\tau}}\cdot\frac{\bar b}{4 J_0}\right\}\ ,
\formula{diseq:h0-piccolo-quanto-basta}
$$
then the non-resonance conditions~\frmref{nonres}
and~\frmref{nonres.a} are satisfied in the complexified domain
$\Wscr^{(0)}_{h_0}$ for $0<|k|\leq K\,$, $|l|\leq 2$ and
$0\le i< j\le n_2\,$.  More precisely,
$$
\left| k\cdot\omega +\epsilon l\,\cdot\Omega^{(0)}(\omega)\right|\geq
\frac{(2-1/2)\gamma}{K^\tau}
\quad {\rm and} \quad
\left| \Omega^{(0)}_i(\omega)-\Omega^{(0)}_j(\omega)\right|\geq
\left(2-\frac{1}{2}\right) \bar b\ .
\formula{nonres-cond-diof-step1}
$$
For $r=0$, the inequalities in~\frmref{nonres-cond-diof} immediately
follow from the the previous ones, recalling that $\phi^{(0)}={\rm
Id}$.  Moreover, the frequencies are not modified by the first
normalization step (see~\frmref{chgfreq-passo1}), therefore we set
$$
\Wscr^{(1)}=\Wscr^{(0)}
\quad \hbox{and}\quad J_1=J_0\ .
$$
Let us now require the new radius of the complex extension be so
small that
$$
h_1 \le \min\left\{h_0\,,
\ \frac{1}{\max\{K\,,\,1/\sigma\}+\epsilon J_1}\cdot
\frac{\gamma}{4(2K)^{\tau}}\right\}\ .
\formula{def:h1}
$$
Using condition above and the first non-resonant condition in~(a') of
lemma~\lemref{lem:analiticita-espansione-Ham}, the first inequality
in~\frmref{nonres-cond-diof-step1} is replaced by
$$
\left| k\cdot\omega +\epsilon l\,\cdot\Omega^{(1)}(\omega)\right|\geq
\frac{(2-1/2)\gamma}{(2K)^\tau}\ ,
\formula{nonres-cond-diof-step2}
$$
for $\omega\in\Wscr^{(1)}_{h_1}\,$, $0<|k|\leq 2K$ and $|l|\leq
2\,$. This concludes the proof in the case $r=1\,$.

The first actual change of the frequencies might occurs at the end of the
second perturbation step and the transformed fast frequencies reads
$$
\omega^{(2)}(\omega^{(0)}) = \omega^{(2)}(\omega^{(1)}) = \omega^{(1)}+
\delta\omega^{(2)}(\omega^{(1)})
= \left( {\rm Id} +\delta\omega^{(2)} \right)(\omega^{(1)})
$$
where $\max_{1\le j\le n_1}\sup_{\omega\in \WWW^{(1)}_{h_{1}}}
\big|\delta\omega^{(2)}_j(\omega)\big|\le
\sigma\big(\epsilon\AAA\big)^2$ in view of the assumption
in~\frmref{frm:stima-variazione-frequenze-per-parte-geom}.  If
$$
\mu_1 = \frac{4\sigma(\epsilon\AAA)^2}{h_1} < 1\ ,
\formula{def:mu1}
$$
in view of lemma~D.1 in~\dbiref{Poschel-1989}, the function $({\rm
Id}+\delta\omega^{(2)})$ admits an analytic inverse
$\phi^{(2)}:\WWW^{(1)}_{h_1/4} \to \WWW^{(1)}_{h_1/2}$ and, in the
domain $\WWW^{(1)}_{h_1/4}\,$, the following estimates hold true
$$
\max_{{\scriptstyle{1\le j\le n_2}}}\,\sup_{\omega\in \WWW^{(1)}_{h_1/4}}\,
\left|\phi^{(2)}_j(\omega)-\omega_j\right|
\leq \sigma(\epsilon\AAA)^2\ ,
\qquad
\left|\frac{\partial\left(\phi^{(2)}-{\rm Id}\right)}{\partial\omega}
\right|_{\infty;\WWW^{(1)}_{h_1/4}}
\leq \mu_1\ ,
\formula{diseq:stime-phi2}
$$
where the norm $\big|\cdot\big|_{\infty;\WWW^{(1)}_{h_1/4}}$ on the
Jacobian of the function $\phi^{(2)}-{\rm
Id}:\,\WWW_{h_1/4}\to\complessi^{n_1}$ is defined in an analogous way
to~\frmref{def:cost-Lipschitz}.  The growth of the
Lipschitz constants for the sequence of functions
$\big\{\phi^{(r)}-{\rm Id}\big\}_{r\ge 0}\,$ is controlled by setting
$$
\bar J_0=\bar J_1=0\ ,
\qquad
\bar J_2=e^{\mu_1}-1\ ,
\formula{def:Jbarra-da-0-a-2}
$$
so that, in particular, we have $\big|{\partial\left(\phi^{(2)}-{\rm
Id}\right)}/{\partial\omega} \big|_{\infty;\WWW^{(1)}_{h_1/4}}\leq \bar J_2\,$.

Recall now that the step $r=2$ includes also the preparation of the
next step $r=3$, namely cutting out the resonant regions
$$
\Rscr_{k,l}^{(2)}=\left\{\omega\in\Wscr^{(2)}:
\big| k\cdot\omega+\epsilon l\,\cdot\Omega^{(2)}\circ\phi^{(2)}(\omega)\big|
\leq 2\gamma/(3K)^{\tau}\right\}\ .
$$

Thus we need an upper bound on both the sup-norm of
$\Omega^{(2)}\circ\phi^{(2)}$ and the Lipschitz constant of its
Jacobian.  The new transversal frequencies can be written
as
$$
\epsilon\Omega^{(2)}(\omega^{(1)})=
\epsilon\Omega^{(1)}(\omega^{(1)})+
\epsilon\Delta\Omega^{(2)}(\omega^{(1)})\ ,
$$
where we have $\max_{1\le j\le n_1}\sup_{\omega\in \WWW^{(1)}_{h_1}}
\big|\epsilon\Delta\Omega^{(2)}_j(\omega)\big|\leq (\epsilon\AAA)^2$,
in view of the hypothesis
in~\frmref{frm:stima-variazione-frequenze-per-parte-geom}.  Moreover,
in the domain $\WWW^{(1)}_{h_1/4}$ the new transversal
frequencies are functions of the transformed fast frequencies, namely
$$
\Omega^{(2)}\big(\phi^{(2)}(\omega^{(2)})\big) =
\Omega^{(1)}\big(\phi^{(2)}(\omega^{(2)})\big) +
\Delta\Omega^{(2)}\big(\phi^{(2)}(\omega^{(2)})\big)\ .
\formula{frm:Omegoni-passo2-da-passo1}
$$
Using this formula we can bound the Jacobian of the function
$\Omega^{(2)}\circ\phi^{(2)}$.  To this end, we need the preliminary
estimate
$$
\eqalign{
\left|\frac{\partial\big(\Omega^{(2)}\circ\phi^{(2)}-\Omega^{(1)}\big)}
        {\partial\omega}\right|_{\infty;\WWW^{(1)}_{h_1/4}}
&\leq\left|\frac{\partial\Omega^{(1)}}{\partial\omega}
\right|_{\infty;\WWW^{(1)}_{h_1/2}}
\left|\frac{\partial\big(\phi^{(2)}-{\rm Id}\big)}{\partial\omega}
\right|_{\infty;\WWW^{(1)}_{h_1/4}}
\cr
&\phantom{\leq}+
\left|\frac{\partial\Delta\Omega^{(2)}}{\partial\omega}
\right|_{\infty;\WWW^{(1)}_{h_1/2}}
\,\left|\frac{\partial\phi^{(2)}}{\partial\omega}\right|_{\infty;\WWW^{(1)}_{h_1/4}}
\cr
&\leq J_1 \mu_1 + \frac{2(\epsilon\Ascr)^2}{\epsilon h_1}
(1+\mu_1)
\leq J_1 \mu_1 + \frac{\mu_1}{2\epsilon\sigma}(1+\mu_1)\ ,
}
$$
where we used formul{\ae}~\frmref{def:mu1}--\frmref{diseq:stime-phi2},
the estimate
$\big|{\partial\Omega^{(1)}}/{\partial\omega}\big|_{\infty;\WWW^{(1)}_{h_1/2}}\le
J_1$ and Cauchy inequality to estimate
$\big|{\partial\Delta\Omega^{(2)}}/{\partial\omega}\big|_{\infty;\WWW^{(1)}_{h_1/2}}\,$.
Thus, we can ensure that the Jacobian of the transformed
transversal frequencies is bounded as
$$
\left|\frac{\partial\big(\Omega^{(2)}\circ\phi^{(2)}\big)}
{\partial\omega}\right|_{\infty;\WWW^{(1)}_{h_1/4}}
\leq \left(J_1  + \frac{\mu_1}{2\epsilon\sigma}\right)(1+\mu_1)=\mathrel{\mathop:} J_2\ .
\formula{eq:J2}
$$
Using again formula~\frmref{frm:Omegoni-passo2-da-passo1}, we can
bound the deterioration of the non-resonance conditions involving the
transversal frequencies. In fact, for $l\in\interi^{n_2}$, $|l|\le
2\,$, we have 
$$
\eqalign{
&\sup_{\omega\in \WWW^{(1)}_{h_1/4}}
\left|\epsilon l\cdot\left[\Omega^{(2)}\big(\phi^{(2)}(\omega)\big)-
\Omega^{(1)}(\omega)\right]\right|
\cr
&\qquad\leq 2\epsilon \max_j
\sup_{\omega\in \WWW^{(1)}_{h_1/4}}\left|\Omega^{(1)}_j\big(\phi^{(2)}(\omega)\big)-
\Omega^{(1)}_j(\omega)\right|
+2 \max_j \sup_{\omega\in \WWW^{(1)}_{h_1/4}}
\left|\epsilon\Delta\Omega^{(2)}_j\big(\phi^{(2)}(\omega)\big)\right|
\cr
&\qquad\leq 2\left|\epsilon\frac{\partial\Omega^{(1)}}{\partial\omega}
\right|_{\infty;\WWW^{(1)}_{h_1/2}}
\,\max_j\,\sup_{\omega\in \WWW^{(1)}_{h_1/4}}\left|\phi^{(2)}_j(\omega)-\omega_j\right|
+2\max_j\,\sup_{\omega\in \WWW^{(1)}_{h_1/2}}
\left|\epsilon\Delta\Omega^{(2)}_j(\omega)\right|
\cr
&\qquad\leq 2\epsilon J_1 \sigma(\epsilon\Ascr)^2+2(\epsilon\Ascr)^2
= 2(1+\epsilon J_1\sigma)(\epsilon\Ascr)^2
= \mu_1 \left(\frac{1}{\sigma}+\epsilon J_1\right) \frac{h_1}{2}\ .
}
\formula{diseq:stima-var-freq-sec}
$$
Thus, for $\omega\in\Wscr^{(1)}_{h_1/4}\,$, $0<|k|\leq 2K$ and
$|l|\leq 2\,$, we obtain the non-resonance estimate
$$
\eqalign{
\left| k\cdot\omega
+\epsilon l\,\cdot\Omega^{(2)}\big(\phi^{(2)}(\omega)\big)\right|
&\ge
\frac{(2-1/2)\gamma}{(2K)^\tau}
-\sup_{\omega\in \WWW^{(1)}_{h_1/4}}
\left|\epsilon l\cdot\left[\Omega^{(2)}\big(\phi^{(2)}(\omega)\big)-
\Omega^{(1)}(\omega)\right]\right|
\cr
&\ge \frac{(2-1/2)\gamma}{(2K)^{\tau}}-
\mu_1 \left(\frac{1}{\sigma}+\epsilon J_1\right)\frac{h_1}{2}
\ge
\frac{(2-1/2-1/4)\gamma}{(2K)^{\tau}}\ ,
}
\formula{nonres-cond-diof-step2-vecchio-blocco}
$$
where we started from inequality~\frmref{nonres-cond-diof-step2}
(holding true on all the complex domain $\Wscr^{(1)}_{h_1/4}$), we
used~\frmref{diseq:stima-var-freq-sec}, \frmref{def:mu1} and the
definition of $h_1$ in~\frmref{def:h1}. Moreover, requiring also
$$
\epsilon\Ascr \leq \frac{1}{(1+\epsilon J_1\sigma)\Ascr}\,\frac{\bar b}{8}\ ,
\formula{eps-piccolo-per-non-ris-trasv-step2}
$$
one can easily obtain the lower bound
$$
\left| \Omega^{(2)}_i\big(\phi^{(2)}(\omega)\big)-
\Omega^{(2)}_j\big(\phi^{(2)}(\omega)\big)\right|
\geq
\left(2-\frac{1}{2}-\frac{1}{4}\right) \bar b
\formula{nonres-cond-trasversa-step2}
$$
uniformly with respect to $\omega\in\Wscr^{(1)}_{h_1/4}\,$, when
$i\neq j\,$. It is now time to consider the new subset of resonant
regions $\Rscr_{k,l}^{(2)}$ for $2K<|k|\leq 3K\,$, $|l|\leq 2\,$.
First, we remove them from the domain, by defining $\Wscr^{(2)}$
according to~\frmref{def:WWWr}--\frmref{def:strisciarisonantekl}.
Having required that the new radius of the complex extension is
so small that
$$
h_2 \le \min\left\{\frac{h_1}{4}\,,
\ \frac{1}{\max\{3K/2\,,\,1/\sigma\}+\epsilon J_2}\,
\frac{\gamma}{8(3K)^{\tau}}\right\}\ ,
\formula{def:h2}
$$
for $2K<|k|\leq 3K\,$, $|l|\leq 2\,$ we have
$$
\eqalign{
\inf_{\omega\in\Wscr^{(2)}_{h_{2}}}
\left| k\cdot\omega
+\epsilon l\,\cdot\Omega^{(2)}\big(\phi^{(2)}(\omega)\big)\right|
&\ge
\inf_{\omega\in\Wscr^{(2)}}
\left| k\cdot\omega
+\epsilon l\,\cdot\Omega^{(2)}\big(\phi^{(2)}(\omega)\big)\right|
\cr
&\phantom{\ge}-3Kh_{2}-2\epsilon J_2h_{2}
\ge\frac{(2-1/2-1/4)\gamma}{(3K)^{\tau}}\ .
\cr
}
\formula{nonres-cond-diof-step2-nuovo-blocco}
$$
% porta giu' l'inf!
Here we used the
definitions~\frmref{def:WWWr}--\frmref{def:strisciarisonantekl} for
the real set $\Wscr^{(2)}$ and subtract the contribution due to the
complex extension; moreover, we also used
formul{\ae}~\frmref{def:h2}, \frmref{eq:J2} (recall that
$\Wscr^{(2)}_{h_{2}}\subseteq\Wscr^{(1)}_{h_{1}/4}\,$, in view
of~\frmref{def:WWWr} and~\frmref{def:h2}).  The inequalities
in~\frmref{nonres-cond-diof} are easily justified in view
of~\frmref{nonres-cond-diof-step2-vecchio-blocco},
\frmref{nonres-cond-trasversa-step2}
and~\frmref{nonres-cond-diof-step2-nuovo-blocco};  This concludes the
proof of the statement for $r=2\,$.

Iterating the procedure for a generic step $r>2$ it is now
straightforward, provided the sequence of restriction of the frequency
domain is suitably selected.  Let us give some details.  For $r>2\,$,
we restart from the relation
$$
\omega^{(r)}(\omega^{(0)}) =
\omega^{(r)}\circ\phi^{(r-1)}\circ\omega^{(r-1)}(\omega^{(0)})=
\left( {\rm Id} + \delta\omega^{(r)}\right)\big(\omega^{(r-1)}(\omega^{(0)})\big)\ ,
$$
where $\max_{1\le j\le n_1}\sup_{\omega\in \WWW^{(r-1)}_{h_{r-1}}}
\big\{\big|\delta\omega^{(r)}_j(\omega)\big|\big\}\le
\sigma\big(\epsilon\AAA\big)^r$ in view of assumption~\frmref{frm:stima-variazione-frequenze-per-parte-geom}.  Thus, using again lemma~D.1 in~\dbiref{Poschel-1989}, we obtain $\omega^{(r-1)}(\omega^{(0)})$ from
$\omega^{(r)}(\omega^{(0)})\,$ via the function
$\flusso^{(r)}:\WWW^{(r-1)}_{h_{r-1}/4} \to \WWW^{(r-1)}_{h_{r-1}/2}$.
The function $\phi^{(r)}$, namely the inverse of
$\omega^{(r)}(\omega^{(0)})$, is obtained by composition, i.e.,
$\phi^{(r)}=\phi^{(r-1)}\circ\flusso^{(r)}=
\flusso^{(2)}\circ\ldots\circ\flusso^{(r)}$ and, by construction, we have
$\phi^{(r)}\big(\WWW^{(r-1)}_{h_{r-1}/4}\big)
\subset\phi^{(r-1)}\big(\WWW^{(r-1)}_{h_{r-1}}\big)\,$. Replacing $\phi^{(2)}$ with $\flusso^{(r)}$,
formul{\ae}~\frmref{def:mu1}--\frmref{nonres-cond-diof-step2-nuovo-blocco}
can be suitably adapted to the case $r>2\,$, so as to prove the
inequalities corresponding
to~\frmref{nonres-cond-diof-step2-vecchio-blocco},
\frmref{nonres-cond-trasversa-step2}
and~\frmref{nonres-cond-diof-step2-nuovo-blocco}.  For this purpose,
for $r\ge 2\,$, it is convenient to impose the following
conditions
$$
\eqalignno{
\mu_{r-1}&=\frac{4\sigma(\epsilon\Ascr)^r}{h_{r-1}}\le
\frac{\min\big\{1\,,\,\epsilon\sigma\big\}}{2^{r}}
\ ,
&{{\rm (i)}} \cr
\bar J_r &= \bar J_{r-1}  (1+\mu_{r-1}) + \mu_{r-1}
\le e^{\tilde\mu_{1}} -1
\leq \epsilon\sigma\ ,
&{{\rm (ii)}} \cr
J_r &= \left(J_{r-1}  + \frac{\mu_{r-1}}{2\epsilon\sigma}\right)(1+\mu_{r-1})
\le \left(J_0+\frac{\tilde\mu_{1}}{2\epsilon\sigma}\right)
e^{\tilde\mu_{1}}
\leq 2 J_0 + 1\ ,
&{{\rm (iii)}} \cr
(\epsilon\Ascr)^{r-1} &\leq
\frac{\bar b}{2^{r+1}\Ascr(1+\epsilon J_{r-1}\sigma)}\ ,
&{{\rm (iv)}} \cr
h_r &\leq \min\left\{\frac{h_{r-1}}{4}\,,
\ \frac{1}{\max\big\{\frac{(r+1)K}{2}\,,\,\frac{1}{\sigma}\big\}+
\epsilon J_{r}}\,
\frac{\gamma}{2^{r+1}\big((r+1)K\big)^{\tau}}\right\}\ ,
&{{\rm (v)}} \cr
}
$$
where $\tilde\mu_{r}=\sum_{s=r}^{\infty}\mu_{s}\,$. Let us remark that
the estimates~(ii)--(iii) are straightforward consequences of
condition~(i).  The sequence $\{h_r\}_{r\ge 0}$ in~\frmref{def:hr} has
been chosen so as to satisfy all the smallness
conditions~\frmref{diseq:h0-piccolo-quanto-basta}, \frmref{def:h1}
and~(v), which are required along this proof.  This is seen because $\epsilon
J_r\sigma\le\epsilon (2J_0+1)\sigma<1$ in view of point~(iii),
hypothesis $\epsilon<\epsilon^{*}_{{\rm ge}}$ and $\epsilon^{*}_{{\rm
ge}}\le 1/[(2J_0+1)\sigma]$ due to~\frmref{def:soglia-geometrica}.

The smallness
conditions~\frmref{def:mu1}, \frmref{eps-piccolo-per-non-ris-trasv-step2},
(i) and~(iv) on $\epsilon$ are satisfied in view of
definition~\frmref{def:soglia-geometrica} since
$\epsilon<\epsilon^{*}_{{\rm ge}}\,$.  This concludes the proof.
\endproof

\acknowledgements{A.~G. and U.~L. have been partially supported by the
research program ``Teorie geometriche e analitiche dei sistemi
Hamiltoniani in dimensioni finite e infinite'', PRIN 2010JJ4KPA\_009,
financed by MIUR.  The work of M.~S. is supported by an FSR Incoming
Post-doctoral Fellowship of the Acad\'emie universitaire Louvain,
co-funded by the Marie Curie Actions of the European Commission.}

\references

\bye